\documentclass[11pt]{article}

\usepackage{amsmath, amsfonts}
\RequirePackage{natbib}
\usepackage{bbm}
\usepackage{graphicx,psfrag,epsf}
\usepackage{float}
\usepackage{subfig}
\usepackage{caption}
\usepackage{subcaption}
\usepackage{enumerate}
\usepackage[inline, shortlabels]{enumitem}
\usepackage{geometry}

\newcommand{\fdp}{\textnormal{FDP}}
\newcommand{\fdr}{\textnormal{FDR}}

\newcommand{\fdphat}{\widehat{\textnormal{FDP}}}
\newcommand{\cR}{\mathcal R}
\newcommand{\cU}{\mathcal U}

\newcommand{\F}{\mathfrak F}
\newcommand{\EE}[1]{\mathbb E\left[#1\right]}
\newcommand{\PP}[1]{\mathbb P\left[#1\right]}
\newcommand{\EEst}[2]{\mathbb E\left[\left. #1 \right| #2 \right]}

\newcommand{\ind}{\mathbbm 1}

\newcommand{\Par}{\text{Par}}

\newcommand{\hd}{\mathcal{H}_{d}}

\newcommand{\chd}{\text{Ch}}

\newcommand{\bmath}{\boldsymbol}

\newtheorem{proposition}{Proposition}
\newtheorem{condition}{Condition}
\newtheorem{assumption}{Assumption}
\newtheorem{definition}{Definition}
\newtheorem{theorem}{Theorem}

\newtheorem{procedure}{Procedure}

\newcommand{\bm}{\mathbf}

\numberwithin{equation}{section}

\usepackage{titlesec}
\usepackage{chngcntr}

\makeatother
\usepackage{xr}
\title{Leveraging the group structure of hypotheses for more powerful multiple testing with FDR control for the filtered rejection set}
\author{Marina Bogomolov\thanks{marinabo@technion.ac.il} \\
 Faculty of Data and Decision Sciences, Technion - Israel Institute of Technology, \\ Haifa 3200003, Israel
\and
Shinjini Nandi\thanks{shinjini.nandi@montana.edu} \\
Department of Mathematical Sciences, Montana State University, \\
Bozeman, MT 59717, U.S.A.}
\date{}

\begin{document}

\maketitle

\begin{abstract}
Modern biological studies often involve testing many hypotheses organized in a group or a hierarchical structure, such as a directed acyclic graph (DAG). In these studies, researchers often wish to control the false discovery rate (FDR) after filtering the discoveries to obtain interpretable results. For addressing this goal, Katsevich, Sabatti, and Bogomolov (2023, Journal of the American Statistical Association, 118(541), 165-176) developed a general method, Focused BH, that guarantees FDR control for the filtered rejection set for a pre-specified filter, under certain assumptions. We propose improving the power of Focused BH by adapting it to group or hierarchical structures of hypotheses using data-dependent weights.  The general method incorporating such weights is referred to as Weighted Focused BH (WFBH). For DAG-structured hypotheses, we propose a variant of WFBH, which can gain power by being adaptive to the DAG structure, and by exploiting the logical relationships among the hypotheses. We prove that WFBH with weights that were proposed to adapt the Benjamini-Hochberg procedure to different group structures, as well as its proposed variant for testing DAG-structured hypotheses, control the post-filtering FDR under certain assumptions. Through simulations, we demonstrate that the latter variant is robust to deviations from these assumptions and can be considerably more powerful than comparable methods. Finally, we elucidate its practical use by applying it to real datasets from microbiome and gene expression studies.
\end{abstract}

\section{Introduction}
\label{s:intro}
Many scientific studies address a large set of hypotheses that have a pre-specified structure, such as a group structure or a hierarchical structure. For example, consider gene expression studies where the goal is to find genes that are associated with a certain disease. Researchers often consider the hierarchical structure of the gene ontology (GO, \cite{ashburner2000gene}), where nodes are annotated with gene sets with similar biological functions, and child nodes are annotated to subsets of gene sets corresponding to their parent nodes. The GO structure is a directed acyclic graph (DAG), see Section \ref{sec:weights} for its definition, as well as other DAG-related definitions. The self-contained hypothesis for each gene set \citep{goeman2007analyzing} states that all genes in the gene set are not associated with the disease, and these hypotheses have a hierarchical structure imposed by the GO. Other examples of studies where hierarchical structures of hypotheses have been addressed include genome-wide association studies (GWAS, see, e.g., \cite{sesia2020multi}) and \cite{bogomolov2020hypotheses}), microbiome studies (see, e.g., \cite{SankaranHolmes14} and \cite{li2022bottom}), and fMRI studies (see, e.g., \cite{BB14} and \cite{schildknecht2016more}). 

When testing hypotheses with such structures, researchers often impose constraints on the structure of the rejected hypotheses, possibly as well as their $p$-values, 
to obtain an interpretable set of discoveries.  For illustration, consider a gene expression study with hypotheses that have a structure induced by the GO. These hypotheses satisfy the following assumption that addresses their logical relationships, often referred to as the strong heredity principle: 
\begin{assumption}\label{logic-rel}
    If a null hypothesis is false, then all its ancestor hypotheses are also false null hypotheses.
\end{assumption}
Therefore, it is natural to require that if a certain hypothesis is rejected, then all its ancestor hypotheses are also rejected. In other words, the set of rejected hypotheses is required to be a rooted sub-DAG of the original DAG. 
%
%
However, a researcher may claim that, given a discovery of a certain gene set, discoveries of all its supersets corresponding to its ancestors are redundant. In order to avoid such redundancies in the rejection set, one can impose the constraint that all the rejected hypotheses are \textit{outer nodes} \citep{yekutieli2006approaches, Y08}, i.e., are not ancestors of other rejected hypotheses. Focusing on the outer nodes is also natural in phenome-wide association studies \citep{denny2010phewas, filtering}, while requiring that a rejection set is a rooted sub-DAG of the original DAG is natural in genetic interaction studies (see \cite{smoothed} for details). 
The constraints regarding the rejected hypotheses can be with respect to their $p$-values as well as their structure; see \cite{brzyski2017controlling} and \cite{filtering} for such constraints in GWAS, where the hypotheses have a spatial structure. 

All of the studies discussed above are large-scale studies that aim to find promising hypotheses for follow-up studies, so controlling the false discovery rate (FDR) is common in such studies. 
Obviously, the rejection set of the popular FDR-controlling Benjamini-Hochberg procedure (BH, \cite{BH95}) may not satisfy the constraints required by the researcher to obtain an interpretable set of discoveries. 
A natural approach is filtering the rejection set of an FDR-controlling procedure to obtain its largest subset satisfying the constraints and reporting only rejected hypotheses in this subset as discoveries. However, this approach can result in a violation of the FDR for the reported discoveries; see \cite{meijer2016multiple} and \cite{filtering}   for illustrations. 
\cite{filtering} proposed a general solution that reconciles  filtering with FDR control: Their proposed Focused BH procedure (FBH hereafter) receives the filter which is intended to be applied on the rejection set as an input, and incorporates 
this filter in such a way that FDR control is guaranteed for the filtered rejection set that it outputs. However, this guarantee for general filters and structures naturally comes at the cost of reduced power: 
FBH always rejects a subset of hypotheses rejected by the BH procedure. 

In this paper, we propose incorporating data-adaptive weights in FBH, which can leverage the structure of the hypotheses to potentially increase its power. Specifically, we propose the Weighted Focused BH (WFBH hereafter), which can be viewed as FBH applied to weighted $p$-values, where the weights are data-dependent. In Section \ref{sec:prel} we present the method with general data-dependent weights and prove its FDR control under certain conditions on the weights, the filter, and the $p$-values. This result shows that the weights of \cite{nandi1} and \cite{nandi2} can be incorporated in FBH to capture various group structures studied in these papers. We next focus on testing hypotheses with a DAG structure (Section \ref{sec:hier}), and propose to adapt the data-adaptive weights of \cite{nandi1} and \cite{nandi2} for capturing the group structure within a DAG. 
For DAGs that satisfy Assumption \ref{logic-rel}, we suggest exploiting the logical relationships by applying WFBH on the \textit{smoothed} $p$-values, 
using the techniques proposed by  \cite{smoothed}. 
Our FDR control results address WFBH with our proposed weights for
DAGs, when applied on the original or smoothed $p$-values, for general classes of filters, including those enforcing the DAG-related structural constraints discussed above.  The effects of incorporating the proposed data-adaptive weights and smoothing in FBH 
are 
studied in a simulation study (Section \ref{sec:sim}) and real data examples (Section \ref{sec:real}). The paper ends with a discussion in Section \ref{s:discuss}. 

%

\section{Weighted Focused BH}
\label{sec:prel}

\subsection{Preliminaries}
\label{sec:prel:1}
\begin{figure}
    \centering
\includegraphics[width=0.8\linewidth]{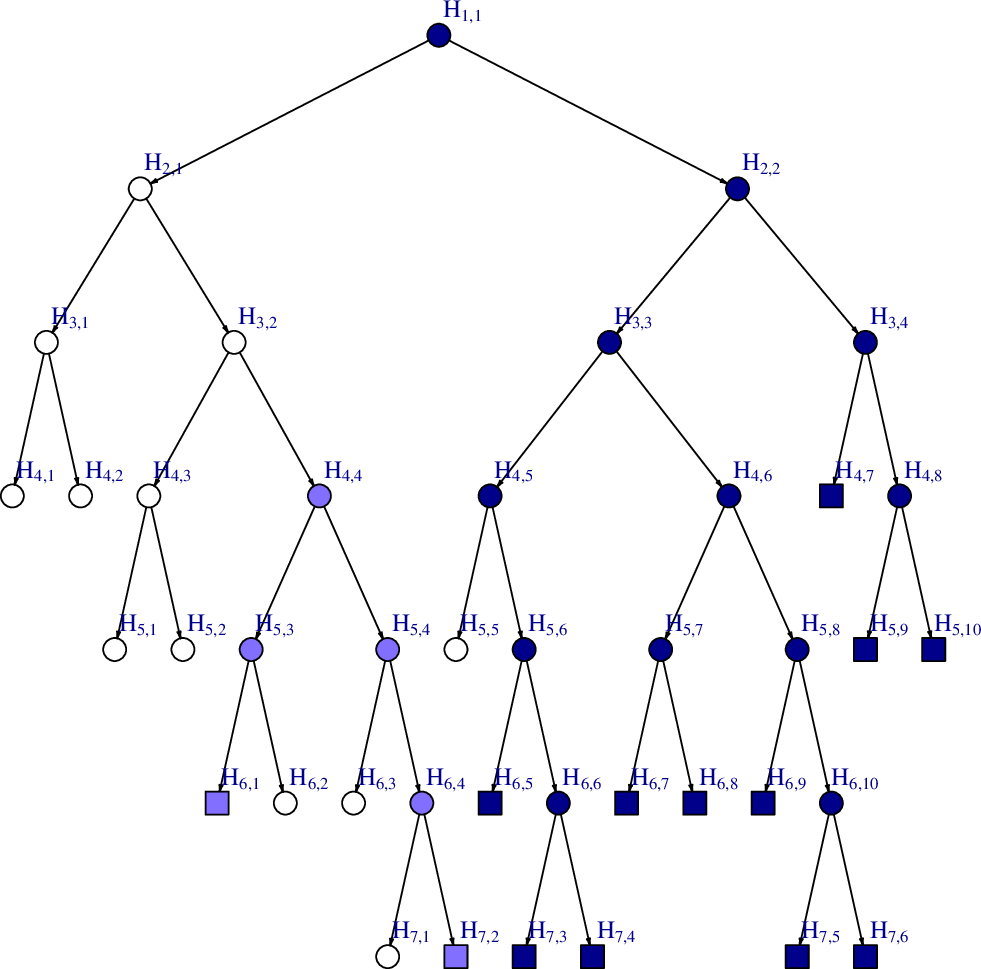}
    \caption{Rejections (before and after filtering) made by the BH procedure at level 0.2 on a tree structure consisting of $41$ nodes, in a study addressed in Section \ref{sec:real}. 
    Each node represents a hypothesis, the $i$-th hypothesis at depth $d$ is denoted by $H_{d,i}$. The $p$-values are given in Figure \ref{fig:chlm}.  
     The colored nodes are rejected  by BH at level 0.2, the slate-blue nodes are eliminated by the DAG-structured filter $\F_{\text{DS}}$ when applied on the rejection set of BH, while the dark blue nodes constitute the subset of the rejections remaining after applying $\F_{\text{DS}}.$
     The square nodes represent rejections remained after applying the outer nodes filter $\F_{\text{out}}$ on the rejection set of BH. Our proposed variant of WFBH for testing hypotheses on a DAG  leverages the group structure of hypotheses at each depth (see Section \ref{sec:hier}). For example, the hypotheses at depth $d=3$ are divided into two groups:  $\{H_{3,1}, H_{3,2}\}$ and $\{H_{3,3}, H_{3,4}\},$ and the hypotheses at depth $d=4$ are divided into four groups: $\{H_{4,1}, H_{4,2}\},$ $\{H_{4,3}, H_{4,4}\},$ $\{H_{4,5}, H_{4,6}\},$ and $\{H_{4,7}, H_{4,8}\}.$}
    \label{fig:filtering}
\end{figure}

Consider a set of $m$ null hypotheses $H_1, \ldots, H_m,$ with a vector of corresponding $p$-values $\bmath{p}=(p_1, \ldots, p_m).$ 
Define $\mathcal{H}=\{1, \ldots, m\},$ and let $\mathcal{H}_0\subseteq \mathcal{H}$ be the subset of indices of true null hypotheses. 
We restate the definition of a filter from \cite{filtering}.
\begin{definition}[\cite{filtering}]
Given a vector of $p$-values $\bmath{p}$ and a subset of indices $\mathcal{R}\subseteq \mathcal{H},$ a filter $\F$ is a map
$\F: (\mathcal{R}, \bmath{p})\mapsto\mathcal{U}\subseteq\mathcal{H},$
with the property that $\mathcal{U}\subseteq \mathcal{R}.$
\end{definition}
When the map $\F$ depends only on the input set $\mathcal{R},$ we call it a fixed filter. Examples of fixed filters which can be applied on a rejection set of a multiple testing procedure applied on hypotheses with a DAG structure are the \textit{DAG-structured filter} $\F_{\text{DS}},$ defined as 
\begin{align}\F_{\text{DS}}: \mathcal{R} \mapsto \mathcal{U}=\{j\in \mathcal{R}: \text{all the ancestors of $j$ are also
in $\mathcal{R}$}\},\label{filter-DAG}
\end{align}
and the \textit{outer nodes filter}, $\F_{\text{out}},$ defined as
\begin{align}\F_{\text{out}}: \mathcal{R} \mapsto \mathcal{U}=\{j\in \mathcal{R}: \text{no descendants of $j$ are also in $\mathcal{R}$}\}
\label{filter-outer}.\end{align}
Note that applying $\F_{\text{DS}}$ or $\F_{\text{out}}$ to $\mathcal{R}^*(\bm p),$ which is a rejection set of a certain multiple testing procedure, results in its largest subset which is a rooted sub-DAG of the original DAG, or which contains only the outer nodes, respectively. 
For illustration, consider the tree in Figure \ref{fig:filtering} for the real data example addressed in Section \ref{sec:real}. 
An example of a non-fixed filter is a \textit{screening filter}, $\F(\mathcal{R}, \bm p)=\mathcal{R}\cap \mathcal{S}(\bm p),$ where $\mathcal{S}: \bm p\mapsto \mathcal{S}_0\subseteq \mathcal{H}$ is a \textit{screening} function. See \cite{filtering} and references therein for examples of applications of screening filters, as well as more complex filters. 

\subsection{The Weighted Focused BH algorithm}\label{sec:algor}
Consider general weight functions $\hat{w}_i:[0,1]^m\rightarrow [0,\infty),\,\, i=1, \ldots,m.$ Each hypothesis $H_i$ is associated with a weight $\hat{w}_i\equiv \hat{w}_i(\bm p),$ and we consider the weighted $p$-values  $\hat{w}_1p_1, \ldots, \hat{w}_mp_m.$    WFBH is obtained by applying the original FBH of \cite{filtering} to these weighted $p$-values. 
\begin{procedure}[WFBH]\label{main-alg} Given as input the vector of $p$-values $\bm p,$ 
the filter $\F$, the target FDR level $q$, and the weight functions $\hat{w}_i:[0,1]^{m}\rightarrow [0, \infty),\, i=1, \ldots, m,$ perform the following steps:
   \begin{enumerate}
        \item 
      Compute 
       the data-adaptive weights $\hat{w}_i=\hat{w}_i(\bm p)$ for $i=1, \ldots, m;$
          \item  For $t \in \{0, \hat{w}_1p_1, \dots, \hat{w}_mp_m\},$ compute $$\fdphat(t) \equiv \frac{m \cdot t}{|\F(\{i: \hat{w}_ip_i \leq t\}, \bm p)|};$$
          \item Compute $t^* \equiv \max\{t \in \{0, \hat{w}_1p_1, \dots, \hat{w}_mp_m\}: \fdphat(t) \leq q\};$	
          \item Find the base rejection set $\mathcal{R}^*=\{i: \hat{w}_ip_i\leq t^*  \};$ 
          \item Compute the filtered rejection set $\mathcal{U}^*=\F(\mathcal{R}^*, \bm p)$ and report $\mathcal{U}^*$ as the set of discoveries.
   \end{enumerate}
\end{procedure}
Our general theoretical results regarding WFBH require that the weight functions satisfy the following condition.
\begin{condition}\label{Cond-weight}
For each $i\in\{1, \ldots, m\},$ the weight function $\hat{w}_i(\bm p)$ is nondecreasing in each coordinate of $\bm p,$ and the weight functions satisfy:
     \begin{align*}
     \sum_{i\in \mathcal{H}_0} \EE{\frac{1}{\hat{w}_i(\bm p_{0,i})}}\leq m,
     \end{align*}
where 
$\bm p_{0,i}$ is the vector of $p$-values $\bm p$ with $p_i$ replaced by 0.
\end{condition}
Examples of weight functions that satisfy Condition \ref{Cond-weight} include non-informative weight functions $\hat{w}_i(\bm p)\equiv 1, \,i=1, \ldots, m,$ or, more generally, any constant functions $\hat{w}_i(\bm p)\equiv w_i, i=1, \ldots, m,$ where $\sum_{i=1}^m 1/w_i=m,$ and the constants $w_1, \ldots, w_m$ are nonnegative. 
For a general filter $\F(\mathcal{R}, \bm p),$ WFBH with unity weights reduces to FBH with the same filter. For a trivial filter $\F(\mathcal{R}, \bm p)=\mathcal{R},$ WFBH with constant weight functions described above 
reduces to the weighted BH procedure of \cite{genovese2006false}, who suggested choosing weights based on prior knowledge for improving the power of BH.

Non-constant functions satisfying Condition \ref{Cond-weight} under independence of $p$-values include $\hat{w}_i(\bm p)=$ $\hat{\pi}_0(\bm p), \,i=1, \ldots, m,$ where $\hat{\pi}_0(\bm p)$ is an estimator for the proportion of nulls, satisfying  $\EE{1/\hat{\pi}_0(\bm p_{0,i})}\leq \pi_0^{-1}$  under independence for every $i\in \mathcal{H}_0,$ where $\pi_0=|\mathcal{H}_0|/m.$  It follows from the results of \cite{benjamini2006adaptive} and \cite{blan2009} that several choices of $\hat{\pi}_0$ satisfy the above property, e.g.,  Storey's estimator, $\hat{\pi}_0^{Storey}$ (\cite{Storey2002, SetS04}), leading to the following data-adaptive weights,
\begin{align}
\hat{w}_i(\bm p)=
\hat{\pi}_0^{Storey}\equiv\frac{1+\sum_{i=1}^m \ind\{p_i>\lambda\}}{m(1-\lambda)},\,\,\lambda \in (0,1),\,\,i=1, \ldots, m,
\label{pi-Storey}
\end{align}
as well as other null-proportion estimators given in Corollary 13 of \cite{blan2009}. WFBH with the weights given in 
(\ref{pi-Storey}), after a minor modification in Step 3, reduces to Storey-Focused BH, an adaptive variant of FBH defined by \cite{filtering}, which in turn reduces to the adaptive BH procedure of \cite{SetS04} (referred to as Storey-BH hereafter) for the trivial filter $\F(\mathcal{R}, \bm p)=\mathcal{R}.$ Finally, \cite{nandi1} and \cite{nandi2}  showed that the data-adaptive weights that they developed for capturing different group structures 
also satisfy Condition \ref{Cond-weight} under independence. WFBH with these weights and a trivial filter reduces to the corresponding  data-adaptive weighted BH methods of \cite{nandi1} and \cite{nandi2}. Thus, WFBH can be viewed as a generalization of FBH incorporating data-adaptive weights, or as a generalization of weighted BH with data-adaptive weights, allowing filtering operations for fixed-in-advance filters. We show next theoretical results for general data-adaptive weights and filters, recovering some known results for FBH and weighted BH.

\subsection{General FDR control results}\label{sec:gen}
We begin with an assumption regarding the $p$-values and a definition of a monotonic filter.
\begin{assumption}[validity of $p$-values]\label{ass:valid}
    For each $j\in \mathcal{H}_0,$ the distribution of $p_j$ is stochastically lower bounded by a uniform distribution on $[0,1],$  i.e., $\PP{p_j \leq t} \leq t$ for all $t \in [0,1].$
    \end{assumption}
\begin{definition}[\cite{filtering}]
A filter $\F$ is monotonic if for any $\bm p^1 \leq \bm p^2$ (where the inequality is understood component-wise) and $\cR^1 \supseteq \cR^2$ we have $|\F(\cR^1, \bm p^1)| \geq |\F(\cR^2, \bm p^2)|.$
\end{definition}
Recall the filters in Section \ref{sec:prel:1}. The DAG-structured filter $\F_{\text{DS}}$ 
is monotonic for any DAG, while the outer nodes filter $\F_{\text{out}}$ 
is monotonic only for a DAG that is a tree (see Section \ref{sec:weights} for a definition). The latter result was shown by \cite{filtering}. 
The following is our result for WFBH with general data-dependent weights. 
\begin{theorem}
    \label{main_theorem}
	Suppose that the $p$-values are valid and independent. 
Assume that the weight functions incorporated in WFBH satisfy Condition \ref{Cond-weight}, and the filter $\F$ is monotonic. 
Then 
WFBH controls the FDR at level $q, $ i.e., it satisfies \begin{align*}
\EE{\frac{|\mathcal{U^*}\cap \mathcal{H}_0|}{|\mathcal{U^*}|}}\leq q,
\end{align*}
 where we define $0/0\equiv0.$
\end{theorem}
The proof of Theorem \ref{main_theorem} is given in Appendix B. 
 Theorem \ref{main_theorem} for WFBH with non-informative unity weight functions 
 recovers certain results of \cite{filtering} regarding FBH. The result of Theorem \ref{main_theorem} for WFBH with weight functions defined in (\ref{pi-Storey})  is similar to the result of \cite{filtering} regarding Storey-Focused BH with a monotonic filter.  
 By applying Theorem \ref{main_theorem} with a trivial filter and either constant weight functions described in Section \ref{sec:algor}, 
 or non-constant weight functions 
 suggested by \cite{nandi1} and \cite{nandi2}, we recover the results of \cite{genovese2006false}, \cite{nandi1}, and \cite{nandi2} regarding their weighted BH methods. 
Theorem \ref{main_theorem} extends these results by allowing incorporating general data-adaptive weights satisfying Condition \ref{Cond-weight} in conjunction with a wide class of filters. This theorem is useful for obtaining FDR control results for WFBH with data-adaptive weights that we propose for DAGs in the next section. 
\section{A variant of WFBH for testing hypotheses on a DAG}\label{sec:hier}
\subsection{Data-adaptive weights capturing the group structure embedded by a DAG}\label{sec:weights}
We consider the case where $H_1, \ldots, H_m$ have a DAG structure. 
We start by  defining a DAG and introducing some DAG-related notations and definitions, following \cite{DAGGER}. A DAG  is a hierarchical structure in which each node represents a distinct hypothesis and the directed edges between the nodes indicate the relationships between them. If there is an edge between two nodes, the source node is the parent while the receiver is the child. Naturally a node may have multiple children and multiple parents. Nodes that have no parents are defined as roots and nodes that have no children are called leaves. A special case of the DAG structure is the tree, where every node except the roots has only one parent. 
 Each hypothesis corresponds to a node of a DAG, so we use the terms hypotheses and nodes interchangeably, i.e., node $a$ is equivalent to $H_a,$ for $a\in\{1, \ldots, m\}.$ 
The depth of node $a,$ denoted as $\text{Depth}(a),$ is defined as the longest path from any root to node $a$ plus 1. All roots are set to have depth 1, a node which has only root nodes as its parents  has depth 2, and in general, $\text{Depth}(a)=1+\max_{b\in\text{Par}(a)}\text{Depth}(b),$ where $\text{Par}(a)$ denotes the set of indices of parents of node $a.$ Similarly, $\chd(a)$ denotes the set of indices of children of node $a$. If there is a path from $H_a$ to $H_b,$ we say that $H_a$ is an ancestor of $H_b,$ and $H_b$ is a descendant of $H_a.$

Let $\mathcal{H}_d$ be the subset of indices of hypotheses with depth $d,$ and let $D$ be the maximal depth of any node in the DAG, so $\cup_{d=1}^D \mathcal{H}_d=\{1, \ldots,m\}.$ For convenience, for each $d\in\{1, \ldots, D\},$ we give a pair of new indices to all the hypotheses with depth $d,$ so that the set of all the hypotheses with depth $d$ is 
$\{H_{d,1}\ldots, H_{d, |\mathcal{H}_d|}\}.$ Thus, for any depth $d,$ we have a one-to-one correspondence between the indices  in $\mathcal{H}_d$ and the pairs of indices $\{(d,i), i=1, \ldots, |\mathcal{H}_d|\},$ so we have two equivalent forms for   referring to each hypothesis. 

We are now ready to define the data-adaptive weights that we suggest to incorporate in WFBH for testing hypotheses on a DAG.
We address separately the hypotheses at each depth $d$ of a DAG. The set of hypotheses with depth $d$ have a natural group structure based on their parent hypotheses: Hypotheses which have the same parent belong to the same group, see Figure \ref{fig:filtering} for an illustration. If a DAG is a tree, i.e., each non-root hypothesis has a single parent, then for each depth $d$, we have non-overlapping groups of hypotheses. 
Otherwise, some hypotheses may have more than one parent, which results in overlapping groups. 
We denote by $n_d$ the number of groups at depth $d,$ for each $d\in\{1, \ldots, D\}.$ The root hypotheses form a single group, so $n_1=1.$ For $d>1,$ 
$$n_d=\sum_{{\ell}=1}^{d-1}\sum_{a\in \mathcal{H}_\ell}\ind\{|\chd(a)\cap\mathcal{H}_{d}|>0\}.$$ 
For a given depth $d,$ the data-dependent weight associated with $H_{d,i}$ is denoted by $\widetilde{w}_{d,i}$ for $i=1, \ldots, |\mathcal{H}_d|.$ For unifying the notations, we imagine an artificial additional dummy hypothesis $H_{m+1}$ which is a parent of all the root hypotheses, so $\text{Par}(1,i)=m+1$ for  $i=1, \ldots, |\mathcal{H}_1|,$ and $\chd(m+1)\cap \mathcal{H}_1=\mathcal{H}_1.$ The inverses of the data-dependent weights for hypotheses with depth $d\in\{1, \ldots, D\}$ are defined below:
\begin{align}\widetilde{w}_{d,i}^{-1} = 
\frac{1}{|\Par(d,i)|}\sum_{a \in \Par(d,i)}\widetilde{w}^{-1}(\chd(a)\cap\hd),\,\, i=1, \ldots, |\mathcal{H}_d|,\label{weight-fin}\end{align} 
with
\begin{align*}
\widetilde{w}(\chd(a)\cap\hd)  =\left\{
\begin{array}{ll}
    \widehat{\pi}(\chd(a)\cap \hd)\cdot K_{a,d}, & \text{if } |\chd(a)\cap \hd|>c \\
    K_{a,d}, & \text{if } |\chd(a)\cap \hd|\leq c,
\end{array}
\right.
\end{align*}
where $$K_{a,d}=\frac{ |\chd(a)\cap \hd|}{|\mathcal{H}_d|}\cdot n_{d},$$ $c$ is a pre-defined threshold for the size of the group where the proportion of true null hypotheses will be estimated,
and
$\widehat{\pi}(\chd(a)\cap \hd)$ is Storey's estimator for the proportion of true null hypotheses in $\chd(a)\cap \hd,$ i.e.,
\begin{align*}\widehat{\pi}(\chd(a)\cap \hd) = \frac{1+\sum\limits_{j \in \chd(a) \cap \hd}\ind\{p_j>\lambda\} }{(1-\lambda)\cdot | \chd(a) \cap \hd|},\,\,\,\lambda\in(0,1).
\end{align*} 
If $c=0,$ these data-dependent weights reduce to those developed by \cite{nandi1} and \cite{nandi2} for capturing the group structure of the hypotheses with depth $d$. 
We incorporate the threshold $c$ for avoiding the estimation of the proportion of true null hypotheses in small groups that can be encountered in a DAG, where Storey's estimator is not reliable.
A researcher can use any threshold $c,$ as long as the choice is not data-dependent. 

Consider several special cases. If all the groups at a certain depth $d$ are non-overlapping (which holds for every depth if the DAG is a tree), and are of equal size $n,$ then for each $i\in \{1, \ldots, |\mathcal{H}_d|\}$, $\tilde{w}_{d,i}$ is equal to Storey's 
estimate for the proportion of nulls in the group to which $H_{d,i}$ belongs, provided that the group size is at least $c$, and $\tilde{w}_{d,i}=1$ otherwise. Thus, the weight associated with each root hypothesis is Storey's estimate for the proportion of nulls among the roots if the number of roots is greater than $c.$ 
For $H_{d,i}$ with multiple parents, $\widetilde{w}_{d,i}$ is influenced by the weights of all the groups to which it belongs. 
The weighting scheme is advantageous for hypotheses with depth $d$  if most of the signals at this depth are concentrated in certain groups, and there is not much overlap among the groups, so $n_d/|\mathcal{H}_d|$ is small. 
If the latter does not hold, i.e.,  many hypotheses with depth $d$ have common parents, it can happen that the proposed weights for hypotheses with depth $d$ are all greater than one, in which case it is worth replacing the proposed weights by the non-informative unity weights for improving the power of the procedure. 
To decide whether hypotheses with a given depth $d$ will be assigned with the proposed data-adaptive weights or with unity weights, one can compute the minimum possible weight for a hypothesis with depth $d,$ and assign unity weights to all hypotheses with this depth only if this minimum weight is greater than one. 
For example, for depth $d$ with groups of two or more hypotheses, assuming $\lambda=0.5,$ the minimum possible value of $\tilde{w}_{d,i}$ is $2n_d/|\mathcal{H}_d|.$ 
              Formally, we consider the following assumption regarding the weights $\{\hat{w}_{d,i},\, d=1, \ldots, D, i=1, \ldots, |\mathcal{H}_d|\}$ 
       that are 
       incorporated into WFBH for testing 
       hypotheses on a DAG. 
    \begin{assumption}\label{ass-weights-hier}
The weights $\{\widehat{w}_{d,i},\, d=1, \ldots, D, i=1, \ldots, |\mathcal{H}_d|\}$ are given by
\begin{align*}
\widehat{w}_{d,i}  =\left\{
\begin{array}{ll}
    \widetilde{w}_{d,i} & \text{if } d\in\mathcal{D}_w \\
    1, & \text{if } d\in\{1, \ldots, D\}\setminus\mathcal{D}_w,
\end{array}
\right.
\end{align*}
where $\widetilde{w}_{d,i}$ is given by (\ref{weight-fin}), and $\mathcal{D}_w\subseteq \{1, \ldots, D\}$ is a subset of depths assigned with data-dependent weights, selected using a rule which does not rely on the data (such as the rule given above). 
\end{assumption}
     The following result is useful for obtaining FDR control results for WFBH applied on a DAG, with weights that satisfy Assumption \ref{ass-weights-hier}. See Appendix C for the proof. 
\begin{proposition}\label{prop-weights}
Consider a set of weights $\{\hat{w}_{d,i},\,d=1, \ldots, D, i=1, \ldots, |\hd| \}$ that satisfies Assumption \ref{ass-weights-hier}. Assume that for each $d\in\mathcal{D}_w,$  the $p$-values in the set $\{p_{d,i}: i=1, \ldots,  |\mathcal{H}_d|\}$ are valid and independent. Then  Condition \ref{Cond-weight} is satisfied by $\{\hat{w}_{d,i},\, d=1, \ldots, D, \,i=1, \ldots, |\hd|\}.$
\end{proposition}
\subsection{Exploiting logical relationships via smoothing}\label{sec:smooth}
\cite{smoothed} proposed a general technique, called \textit{smoothing},  for improving the power of multiple testing methods that are applied on a DAG that satisfies the logical relationships in Assumption \ref{logic-rel}. We focus here on all-descendant smoothing, where the $p$-value of each node is combined with the $p$-values of all its descendant nodes, resulting in a vector of smoothed $p$-values $\tilde{\bm p}=(\tilde{p}_{1}, \ldots, \tilde{p}_m).$ These smoothed $p$-values are valid if Assumption \ref{logic-rel} holds and the combining method gives a valid global null $p$-value under the dependency between the combined $p$-values. 
The idea behind all-descendant smoothing  is improving the power of the test of each parent hypothesis by borrowing statistical strength from the tests of its descendants. \cite{smoothed} suggested applying a multiple testing method on smoothed $p$-values, and showed that this can lead to a substantial power gain for several methods, while preserving their error rate control. 
We show a similar result for WFBH when it is applied on smoothed $p$-values which are obtained using Simes' combining function \citep{Simes1986} or any function satisfying the following condition: 
\begin{condition}\label{ass-smoothing} 
Let $p_1, \ldots, p_n$ 
be a vector of $p$-values 
and  
$G: \mathbb{R}\mapsto \mathbb{R}$ be a monotone increasing function. Let $U_i,\,i=1, \ldots, n$  be independent and identically distributed as $\text{Un}[0,1].$ 
The combining function $T:[0,1]^n \rightarrow [0, \infty)$ 
satisfies one of the following conditions:
    \begin{itemize}
        \item [(a)] $T(p_1, \ldots, p_n)=F(G(\sum_{i=1}^n H^{-1}(p_i))),$
        where 
        $H: \mathbb{R}\mapsto[0,1]$ is a monotone increasing function with the first-order derivative $H^{'}$ being log-concave,  and $F(x)\equiv\PP{G(\sum_{i=1}^n H^{-1}(U_i))\leq x}.$ 
        \item [(b)] $T(p_1, \ldots, p_n)=F(G(p_{(i)}),$ 
    where  $i\in \{1, \ldots, n\}$ is arbitrary, $p_{(i)}$ is the $i$-th order statistic of $p_1, \ldots, p_n,$ 
    and $F(x)\equiv\PP{G(U_{(i)})\leq x},$ where  $U_{(i)}$ is the $i$-th order statistic of $U_1, \ldots, U_n.$
    \end{itemize}
\end{condition}
Combining functions satisfying part (a) of Condition \ref{ass-smoothing} include Fisher's \citep{fisher1932statistical} and 
Stouffer's \citep{stouffer1949american} functions, 
as well as several functions addressed in \cite{vovk2020combining}. Part (b) of Condition \ref{ass-smoothing} is satisfied by Tippett's \citep{tippett1931methods} and R\"uger's \citep{ruger1978maximale} methods; see \cite{smoothed} for details.

\subsection{FDR control results }\label{sec:theory:hier}
Our theoretical result below addresses a vector $\bm p'=(p'_1, \ldots, p'_m),$ which is the vector of original 
$p$-values (i.e. $\bm p'=\bm p$), or the vector of smoothed $p$-values (i.e. $\bm p'=\tilde{\bm p}$) obtained by all-descendant smoothing. 
\begin{theorem}\label{thm:hier}
    Assume that the hypotheses have a DAG structure and their $p$-values $p_1, \ldots, p_m$ are valid and independent.  WFBH with a monotonic filter and weight functions satisfying Assumption \ref{ass-weights-hier} controls the FDR at level $q$ when it is applied on  $\bm p'$ if either of the following sets of additional assumptions are satisfied:
    \begin{itemize}
        \item[(i)] $\bm p'=\bm p,$ i.e., the method is applied to the original $p$-values; 
        \item[(ii)] The DAG is a tree satisfying the logical relationships in Assumption \ref{logic-rel};  $\bm p'=\tilde{\bm p}$ where the smoothing function is Simes' function; 
        \item [(iii)] The same assumptions as in item (ii) with the smoothing function replaced by any combining function satisfying Condition \ref{ass-smoothing}, and an additional assumption that $p_i\sim \text{Un}[0,1]$ for each $i\in \mathcal{H}_0.$
    \end{itemize}
\end{theorem}
The result of Theorem \ref{thm:hier} under the additional assumption of item (i) follows from Theorem \ref{main_theorem} in conjunction with Proposition \ref{prop-weights}. 
The assumption that the DAG is a tree in items (ii) and (iii) can be weakened. Moreover, it can be relaxed for the case where $\mathcal{D}_w=\emptyset,$ in which WFBH reduces to FBH. 
In addition, Theorem \ref{thm:hier} can be adjusted to address intersection trees, where the hypothesis for each node is an intersection of its leaf descendants or of certain other hypotheses.  We give the variant of Theorem \ref{thm:hier} with weaker assumptions in the last two items, as well as the results for FBH and intersection trees in Appendices D and F, 
along with all the proofs. Our results extend the results of \cite{filtering} for FBH, in addition to addressing its weighted variant that we propose. The proofs of the above results are based on a lemma given in Appendix D, which extends Lemma 2 of \cite{pfilter2}, and can be useful for obtaining FDR control for other methods.

Consider two filters that we addressed in Section \ref{sec:prel:1}: the DAG-structured filter $\F_{\text{DS}}$ and the outer nodes filter $\F_{\text{out}},$ given in (\ref{filter-DAG}) and (\ref{filter-outer}), respectively.  As noted in Section \ref{sec:gen}, 
$\F_{\text{DS}}$ is monotonic for any DAG, while $\F_{\text{out}}$ is monotonic only for a tree. 
Hence, Theorem \ref{thm:hier} shows that when the original $p$-values are independent, WFBH with the proposed weights and the DAG-structured filter $\F_{\text{DS}}$ guarantees FDR control for any DAG when applied to the original $p$-values, and for a tree when applied to smoothed $p$-values, under the additional assumptions in items (ii) and (iii). When $\F_{\text{DS}}$ is replaced by $\F_{\text{out}}$, Theorem \ref{thm:hier} gives such FDR control guarantees only for a tree.
For 
allowing arbitrary dependencies across parent and child $p$-values, as well as arbitrary filters and smoothing methods (as long as they give valid smoothed $p$-values), 
one can revert to Weighted Reshaped Focused BH, a more conservative variant of the proposed method; 
 see Appendices A and E. 
However, our simulations show that WFBH with the proposed weights and $\F_{\text{DS}}$ appears to preserve its FDR control under positive dependence, provided that $\lambda$ is set to the target FDR level, which agrees with the results of \cite{blan2009} regarding Storey-BH and of \cite{nandi1, nandi2}  regarding their weighted BH methods. 

\subsection{Comparison to existing methods for testing hypotheses on a DAG} 
 It was shown by \cite{filtering} that FBH with a fixed filter, such as $\F_{\text{DS}},$ can be recast as a variant of Structured BH, a general method for controlling FDR under structural constraints. This method was independently discovered and theoretically studied by \cite{pfilter2} and \cite{filtering}. However, to our knowledge, the variant of Structured BH that requires the rejection set to respect the strong heredity principle, 
 which is equivalent to FBH with $\F_{\text{DS}},$ was neither empirically studied nor compared to competitors in simulation studies. Therefore, in our simulation study we address FBH and WFBH with weights satisfying Assumption \ref{ass-weights-hier}, both with the DAG-structured filter $\F_{\text{DS}}$. 
 We next discuss other methods respecting the strong heredity principle of a DAG.

For family-wise error rate (FWER) controlling methods, see, e.g., \cite{meijer2015multiple, meijer2016multiple, goemanmansmann, Meinshausen08}. 
These methods exploit the logical relationships among the hypotheses and preserve FWER control after applying arbitrary filters, which can be chosen even after obtaining the rejection set. Such flexibility is not allowed for general FDR-controlling procedures which respect the strong heredity principle for a DAG or a tree. For example, the procedures of \cite{lynch2014control},  and the DAGGER method of \cite{DAGGER}, which reduces to the method of \cite{lynch2016procedures} when the DAG is a tree, 
control the FDR for the entire set of discoveries,  
however, they do not control the FDR for the subset of outer node discoveries. 
The procedure of \cite{Y08} for testing hypotheses on a tree is somewhat different: its testing level can be adjusted to control
the FDR for the entire set of discoveries, or only for the subset of outer nodes. The FDR-controlling procedure of \cite{lei2021general} is even more flexible: it can satisfy arbitrary structural constraints on the rejection set, including the constraint of respecting the strong heredity principle, 
however, it cannot respect constraints on the $p$-values of the rejected hypotheses. 
WFBH is more flexible than Yekutieli's procedure and the procedure of \cite{lei2021general}, in the sense that it can account for arbitrary filters, including those incorporating screening functions and thereby enforcing constraints on the $p$-values of the rejected hypotheses. 
However, WFBH  does not possess the flexibility of 
FWER-controlling procedures, since it requires the filter to be pre-specified and given as an input to the procedure. Naturally, the more stringent error control and the flexibility of FWER-controlling procedures for testing hypotheses on a DAG usually come with a power cost, as illustrated in simulation studies of \cite{DAGGER} and \cite{filtering}.


The algorithmic nature of WFBH is different from that of the DAGGER and Yekutieli's methods. 
The latter methods are sequential top-down methods, starting from testing the roots and proceeding to testing child hypotheses only if their parent hypotheses were rejected; therefore, they allow the $p$-values to be collected sequentially as well. In contrast, WFBH 
requires all the $p$-values to be available in advance, as well as the variant of the method of \cite{lei2021general} respecting the strong heredity principle, which we refer to as the DAG-structured method of \cite{lei2021general} hereafter. The lack of a possibility to obtain the data sequentially can be viewed as a limitation of these methods.  
However, when these methods are applicable and the signals are weak at low depths of the DAG, these methods can have a power advantage over sequential methods, as shown by \cite{filtering} and \cite{lei2021general}. The reason is that 
in these cases, sequential methods can stop at low depths, thereby missing many signals at high depths, while WFBH and the DAG-structured method of  \cite{lei2021general} can borrow statistical strength from signals at high depths of the DAG for increasing their power.

Although Yekutieli's method is sequential while WFBH is not, they are similar in the sense that they are adaptive to the  proportions of signals within groups of hypotheses with the same parents, becoming more liberal when the signals are concentrated within such groups. 
The adaptiveness of Yekutieli's method is due to the fact that it tests each group with a rejected parent separately, using the BH procedure, while the adaptiveness of WFBH is due to up-weighting or down-weighting the $p$-values in each group with respect to the within-group  estimated proportion of nulls. The other FDR-controlling methods described above do not respect the group structure of hypotheses within the same depth, and thus lack this group-adaptivity property. Finally, we note that the methods discussed above also differ with respect to theoretical FDR control guarantees. While all methods are valid under independence, WFBH and DAGGER are also valid under dependencies induced by descendant smoothing, under certain conditions. 
\section{Simulation study}\label{sec:sim}
We consider various DAG structures, and study the performance of FBH and WFBH with the DAG-structured filter $\F_{\text{DS}}$ 
and with weights satisfying Assumption \ref{ass-weights-hier}. We refer to these variants as FBH and WFBH hereafter. 
Our goals are: (a) to study the effects of incorporating the suggested data-adaptive weights and smoothing techniques on the power of FBH; 
(b) to compare the power performance of FBH and WFBH to their competitors, which test hypotheses on a DAG with FDR control and respect the strong heredity principle; (c) to study whether WFBH can maintain its FDR control under positive dependence.
In this section, we address the following DAG structures and choices of $\mathcal{D}_w$ for  weights: 
\begin{enumerate}[label=(\arabic*)] 
\item Wide tree: \label{widetree} A DAG with $550$ nodes distributed over two depths, $50$ nodes are roots, each of which has $10$ child leaf nodes  at depth 2. We set $\mathcal{D}_w=\{1,2\}.$ 
\item Bipartite graph 1: \label{bip1}A DAG with $550$ nodes at two depths -  $200$ roots and $350$ leaves. Each root is randomly assigned to 10 leaves that are set to be its children. We set $\mathcal{D}_w=\{1\}.$
\end{enumerate}
We set $\lambda=0.5$ for the weights, and choose $\mathcal{D}_w=\{1\}$ for bipartite graph 1 because the smallest possible weight for the leaf hypotheses is $2(200/350)>1.$ A proportion $p$ of randomly selected leaves is set to be non-null, and is varied in $(0,1).$ The remaining hypotheses in the graph are set to be non-null only if they have at least one  non-null child hypothesis, so it is guaranteed that Assumption 1 is satisfied for each of the graphs. 
For every $d\in\{1, \ldots, D\}$ and $i\in\{1, \ldots, |\mathcal{H}_d|\},$  $p_{d,i}=1-\Phi(X_{d,i}),$ where $X_{d,i}\sim N( \mu_{d,i}, 1).$ 
If $H_{d,i}$ is true, then $\mu_{d,i}=0;$ otherwise, 
the value of $\mu_{d,i}$ is determined according to the following three setups, addressing the depth of $H_{d,i}:$ 
\begin{enumerate*}[label=(\roman*), itemjoin=\quad]
   \item Global alternate:
$\mu_{d,i}=2;$ 
   \item Decremental alternate: $\mu_{d,i} = 2 + 1.5(d-1);$ 
    \item Incremental alternate:
    $\mu_{d,i} = 2 + 1.5(D-d)$. 
\end{enumerate*}
In the decremental setup, the signal strength increases with depth, making it easier to reject child hypotheses rather than their parents. 
In the incremental setup 
the opposite is true: the signal becomes weaker as we proceed from the roots to the leaves.


For the tree structure \ref{widetree}, the competitors we consider are the DAGGER method \citep{DAGGER} and Yekutieli's procedure \citep{Y08}, referred to as Yk hereafter, targeting full-tree FDR control. For the bipartite structure \ref{bip1}, which is not a tree, Yk is not applicable;  therefore, for this structure, we consider only the DAGGER as a competitor. 
We do not consider additional competitors addressed in \cite{DAGGER}, because in most of their simulation settings DAGGER was the most powerful. 
The target FDR level for all procedures is set to $0.05,$ so
the level of Yk is adjusted to $0.05/(2\times 1.44)$  according to \cite{Y08}. 

\begin{figure}
\centering
  \includegraphics[width=1\linewidth]{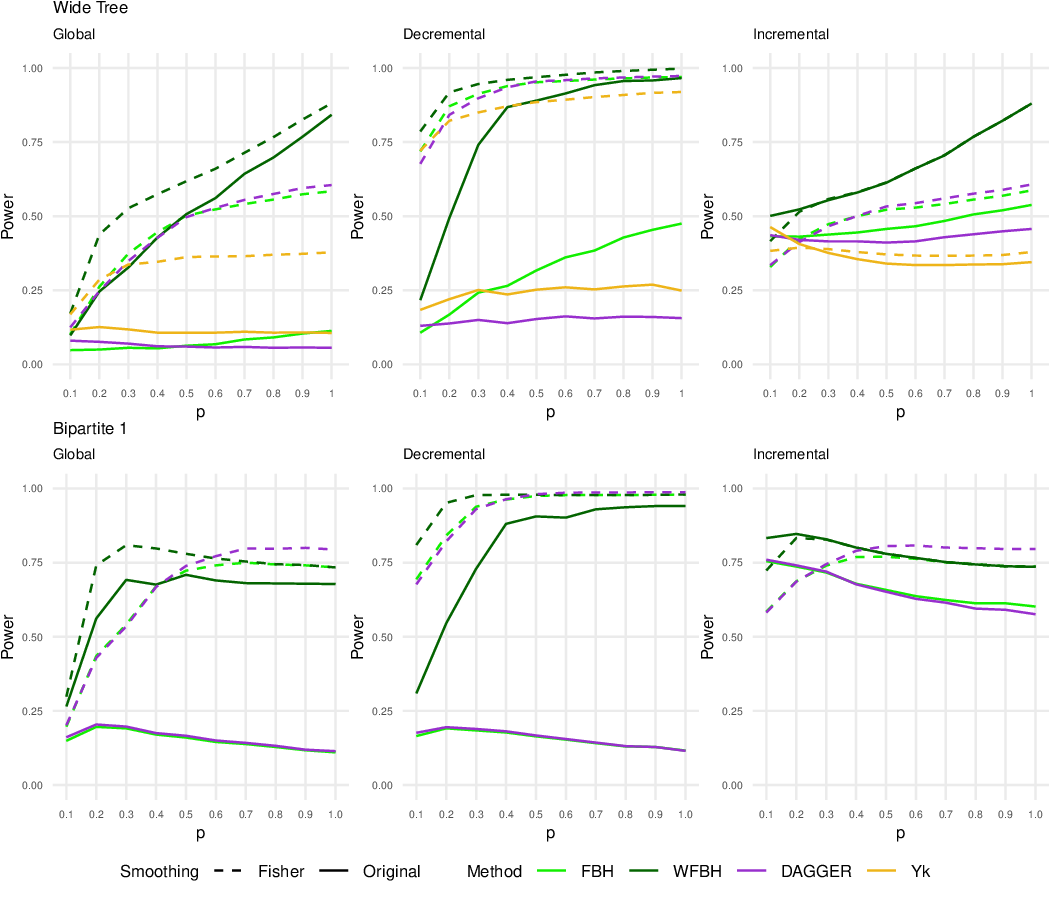}
\caption{Comparison of estimated average power (based on 200 iterations) of WFBH, FBH, DAGGER, and Yk (the method of \cite{Y08}) applied on the original independent $p$-values, and on Fisher's smoothed $p$-values based on all-descendant smoothing. The top panel shows the comparisons for the wide tree  (see \ref{widetree} in Section \ref{sec:sim}) across the three setups: global, decremental and incremental. 
The bottom panel shows the comparisons for the bipartite graph 1 (see \ref{bip1} in Section \ref{sec:sim}) for the same three setups. Since Yekutieli's method is not suitable for DAGs that are not trees, it is omitted and only FBH, WFBH and DAGGER are compared.}
\label{fig_pow_ind_simple}
\end{figure}

Figure \ref{fig_pow_ind_simple} demonstrates the estimated average power (obtained based on 200 iterations) of FBH, WFBH and their competitors as a function of $p \in (0,1).$ 
 We demonstrate the power of all the methods when applied to independent original $p$-values and their corresponding Fisher's smoothed versions, obtained using all-descendant smoothing. In this setting, all methods have theoretical FDR control guarantees when applied to the original $p$-values; however, only the following methods have such guarantees also when applied to smoothed $p$-values: WFBH for a tree;  FBH and DAGGER for both graphs. 


Since DAGGER and Yk are sequential top-down methods,  the decremental setup is less favorable for their power than the incremental, where the signal is the strongest at the roots. 
In the decremental setup, when the methods are applied to the original $p$-values, WFBH significantly outperforms the sequential methods for both structures, however, its power gain is much smaller when all the methods are applied to smoothed $p$-values. 
Note that smoothing is the most advantageous in the decremental setup, due to the fact that the signals corresponding to non-null parent hypotheses are weaker than those corresponding to non-null child hypotheses.
For global and incremental setups, in the wide tree structure, WFBH applied to smoothed $p$-values is considerably more powerful than its competitors applied to the same $p$-values, for all values of $p.$ This is not so for the bipartite graph in these setups: While WFBH with smoothed $p$-values outperforms its competitors for small and moderate values of $p,$ DAGGER with smoothed $p$-values is the most powerful for higher values of $p.$ 
Further results, addressing FDR and average power under various scenarios, including positively dependent $p$-values, for structures \ref{widetree} and \ref{bip1}, as well as two additional graph structures, are provided in Appendix G.

\section{Real Data Analysis}\label{sec:real}
\begin{figure}
\centering
\includegraphics[width=\linewidth]{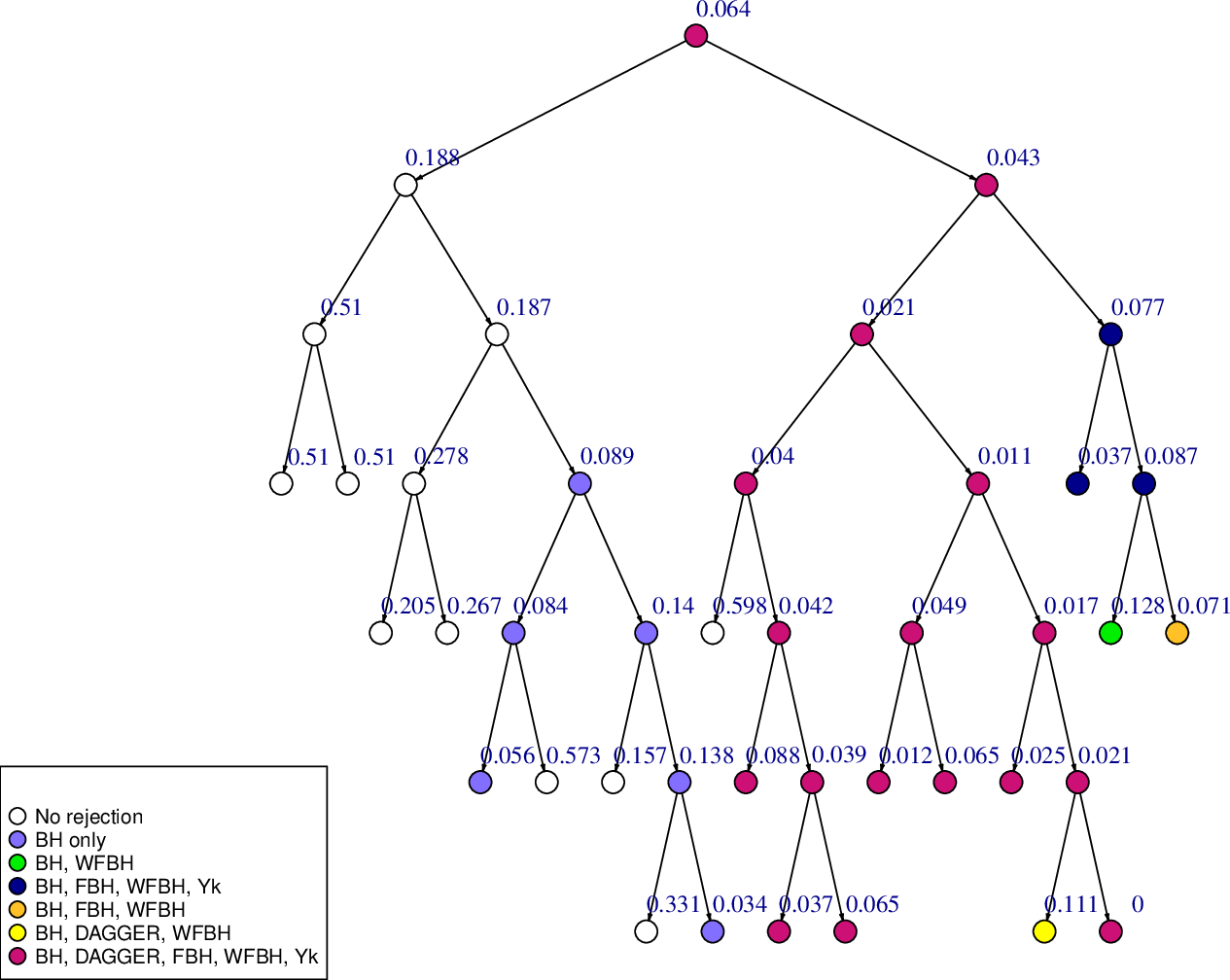}
\caption{Phylogenetic tree structure for the  phylum Chlamydiae, along with $p$-values for all the hypotheses (see Section \ref{sec:real} for details). The colored nodes indicate discoveries made by different methods with target full-tree FDR level of $q=0.2$ for all the methods. Yekutieli's method is applied at level $q/2=0.1$ The FBH and WFBH methods are applied with DAG-structured filter, and the weights for WSBH are obtained according to Assumption \ref{ass-weights-hier} with $\mathcal{D}_w=\{1,\ldots, 7\}$ and $\lambda=q$.  
All the colored nodes are
rejected by BH; all the colored nodes except the slate-blue nodes are remained after applying the DAG-structured filter
on the rejection set of BH, and these are the rejections of 
WFBH. The rejection sets of other procedures are obtained by excluding the following hypotheses from the rejection set of WFBH:  the two leaves with $p$-values
0.111 and 0.128 for FBH; the three leaves with $p$-values
0.128, 0.071, 0.111  for Yekutieli's method; the hypotheses in the sub-tree
originating from node at depth 2 with $p$-value 0.077 for DAGGER. 
} 
\label{fig:chlm}
\vspace{0.5in}
\centering
\begin{tabular}{ccccccc}
  \hline
 & DAGGER & WFBH & FBH & Yekutieli & BH \\ 
  \hline
 & 18 & 23 & 21 & 20 & 29 \\ 
   \hline
\end{tabular}
\vspace{0.2in}
\captionof{table}{
The number of discoveries made by different methods for the tree structure corresponding to phylum Chlamydiae, described in Figure \ref{fig:chlm}.}
\end{figure}

We 
address a subset of data from \cite{Caporaso2011} examining the abundance of microbial communities across different natural environments. The data, along with the phylogenetic tree that classifies the microbial community at different taxonomic resolutions, are available through the Bioconductor package `phyloseq' \citep{mcmurdie2012phyloseq}. 
The null hypothesis for each node of the phylogenetic tree states that the corresponding taxonomic classification of microbes is uniformly abundant across all environments. 
Abundances of individual microbes are aggregated to higher level terms in the phylogenetic tree, and an F-test is applied to obtain a $p$-value for each node. In this section, we address the phylogenetic tree corresponding to Chlamydiae phylum, consisting of 41 nodes, using the R package `strucSSI' of \cite{SankaranHolmes14}, who analyzed this tree using Yk at level $q=0.1,$ which targets full-tree FDR control at level 0.2 (up to an inflation factor which is close to one in settings like ours, according to \cite{Y08}). We complement the analysis of \cite{SankaranHolmes14} by applying  the other methods we considered in Section \ref{sec:sim}, along with BH, with target FDR level 0.2. Figure \ref{fig:chlm} illustrates the phylogenetic tree for this phylum, and the location of discoveries for different methods. In this example BH disregards the strong heredity principle, and WFBH with $\lambda=0.2$ leads to the highest number of rejections among all the methods respecting this principle. 
Analysis of a larger tree corresponding to the Actinobacteria phylum and  an additional real data example addressing the GO graph are deferred to Appendix H.   

\section{Discussion}
\label{s:discuss}
We have introduced Weighted Focused BH, a variant of Focused BH which can incorporate general data-adaptive weights satisfying Condition \ref{Cond-weight}. This method is equivalent to applying Focused BH on weighted $p$-values, 
and is therefore expected to be more powerful than the original Focused BH if the weights for false null hypotheses are expected to be smaller than the other weights.  

Following \cite{nandi1, nandi2} who suggested data-adaptive weights  that 
satisfy this property in cases where hypotheses have different group structures and most signals are concentrated within certain groups, we suggested similar weights to capture the group structure embedded by a DAG. In our simulation study we showed that incorporating such weights in Focused BH 
can lead to a significant power gain when the target is FDR control under the constraint that the strong heredity principle is respected. 
Moreover, we showed that this method 
seems to preserve its FDR control in cases where the assumptions required for our theoretical results are not satisfied. 

The choice of weights for the weighted BH procedure of \cite{genovese2006false}  has been extensively studied, and various fixed and data-adaptive weights have been proposed to capture the structure of hypotheses or domain-specific knowledge, possibly using auxiliary covariates %
\citep{hu2010false,  RetD06,
roeder2009genome, li2013using, li2019multiple,   ignatiadis2021covariate}. Using similar ideas for constructing the weights for Weighted Focused BH for addressing different structures, filters, and applications, can lead to specialized variants of Weighted Focused BH with good power properties.
The question of the existence of optimal weights for Weighted Focused BH with a certain filter, under a certain model for $p$-values, can be an interesting direction for future research. 
The ideas and techniques could be borrowed from research regarding optimal weights for weighted BH of \cite{genovese2006false} under different models for $p$-values, see, e.g., \cite{roquain2009optimal} and \cite{durand2019adaptive}.

The weighting scheme of \cite{nandi1} 
can be useful for applications with a spatial structure, like GWAS, where the false null hypotheses are likely to appear in spatial clusters which can be defined in advance or constructed based on data which are independent of $p$-values for the hypotheses of interest. The original Focused BH with a  filter accounting for such clusters was shown to be useful in GWAS by \cite{filtering}. Investigating the FDR and power of the Weighted Focused BH with the same filter and the weights of \cite{nandi1} under dependencies encountered in GWAS is a possible direction for future research. 

\section*{Acknowledgements}
The research of MB was partially supported by the Israel Science Foundation grant no. 1726/19.\\
Computational efforts were performed on the Tempest High Performance Computing System, operated and supported by University Information Technology Research Cyberinfrastructure \\ (RRID:SCR\_026229) at Montana State University.
\appendix
\newtheorem{atheorem}{Theorem}[section]
\renewcommand{\theatheorem}{\Alph{section}\arabic{atheorem}}

\newtheorem{adefinition}{Definition}[section]
\renewcommand{\theadefinition}{\Alph{section}\arabic{adefinition}}

\newtheorem{alemma}{Lemma}[section]
\renewcommand{\thealemma}{\Alph{section}\arabic{alemma}}

\section*{Appendices}
\titleformat{\section}
  {\normalfont\Large\bfseries}   
  {Appendix \Alph{section}:}     
  {1em}                          
  {}                             
\renewcommand{\thesection}{\Alph{section}} 
\counterwithin{figure}{section}
\counterwithin{table}{section}
\section{Weighted Reshaped Focused BH}\label{Reshaped}
    Weighted Reshaped Focused BH is obtained by Procedure 1 with the following changes in Steps 2 and 3: $\fdphat(t)$  is replaced by $$\displaystyle\fdphat^{\beta}(t)\equiv  \frac{m \cdot t}{\beta(|\F(\{j: \hat{w}_jp_j \leq t\}, \bm p)|)},$$ where 
\begin{align}\beta_{\nu}(r)=\int_{0}^{r}xd\nu(x),\label{beta}\end{align}
 		where $\nu$ is an arbitrary probability distribution on $(0,\infty).$ Following \cite{pfilter2} we refer to a function of form (\ref{beta}) as a reshaping function.
        Note that $\beta(r)\leq r$ for all $r>0,$ so for all $t\geq0,$ $\fdphat^{\beta}(t)\geq \fdphat(t),$  which yields that the new variant of the procedure incorporating the function $\beta$ is more conservative than the original variant. 
An example of a function of form (\ref{beta}) is $\beta(r)=r/(\sum_{i=1}^m1/i),$ which was originally incorporated in the BH procedure by \cite{BY01} for theoretical FDR guarantees under arbitrary dependence. For other examples of such functions and multiple testing procedures incorporating those functions, see \cite{blanchard2008two}. 
In fact, Weighted Reshaped Focused BH is equivalent to applying Reshaped Focused BH (defined in Section C.2 of the supplementary material of \cite{filtering}) on the weighted $p$-values $\hat{w}_1p_1, \ldots, \hat{w}_mp_m.$ 

For presenting our theoretical results regarding Weighted Reshaped Focused BH, we need to introduce several notations.
For each $i\in\{1, \ldots, m\},$ we define $Desc(i)$ as the set of indices of descendants of $H_i.$ Define $\bm p_{i, Desc(i)}=p_i\cup \{p_j: j\in Desc(i)\},$ i.e., $\bm p_{i, Desc(i)}$ is the set containing $p_i$ and the $p$-values for the descendants of $H_i$. If $H_i$ has no descendants, then $\bm p_{i, Desc(i)}=p_i.$ According to all-descendant smoothing, $\tilde{p}_i=T(\bm p_{i, Desc(i)}),$ where $T$ is the combining function. Let $\mathcal{D}_t\subseteq\{1, \ldots, D\}$ be the set of depths that satisfy the following property: For each $d\in\mathcal{D}_t,$ and any two hypotheses $H_i$ and $H_j$ with depth $d,$ the set of descendants of $H_i$ does not overlap with the set of descendants of $H_j.$ For a tree, $\mathcal{D}_t=\{1, \ldots, D\}.$ For each $d\in \mathcal{D}_t,$  define $\bm p_{-i}^{d, Desc}=\{\bm p_{j, Desc(j)}: j\in \mathcal{H}_d, j\neq i\},$ i.e., $\bm p_{-i}^{d, Desc}$ is the set of all the $p$-values for hypotheses with depth $d$ along with the $p$-values for their descendants, excluding $p_i$ and the $p$-values for its descendants.

\begin{atheorem}\label{thm-gen-dep}
   Assume that the original $p$-values $(p_1, \ldots, p_m)$ are valid. Consider Weighted Reshaped Focused BH with weights that satisfy Assumption 3 and an arbitrary filter $\F$, applied on the vector $\bm p'.$  This method controls the FDR at level $q$ if either of the following two sets of additional assumptions are satisfied:  
    \begin{itemize}
        \item [(i)]$\bm p'=\bm p;$  for each $d\in \mathcal{D}_w,$  the $p$-values in the set $\{p_{d,i}, i=1, \ldots, |\mathcal{H}_d|\}$ are independent (while the dependency among the $p$-values for hypotheses with different depths can be arbitrary). 
        \item [(ii)] $\bm p'=\tilde{\bm p},$ where each smoothed $p$-value $\tilde{p}_{d,i}$ is obtained using any $p$-value combination method which gives a valid $p$-value for $H_i\cap (\cap_{j\in Desc(i)} H_j)$ under the dependency structure among the $p$-values in the set $\bm p_{i, Desc(i)};$  $\mathcal{D}_w\subseteq \mathcal{D}_t,$ and for each $d\in \mathcal{D}_w$ and each $i\in \mathcal{H}_d,$ the $p$-values in the set $\bm p_{i, Desc(i)}$ are independent of the $p$-values in the set $\bm p_{-i}^{d, Desc}.$
\end{itemize}
\end{atheorem} 
The proof of Theorem \ref{thm-gen-dep} is given in Appendix \ref{sec:proof:gen}. Let us consider the last assumption of item (ii), i.e., that $\mathcal{D}_w\subseteq \mathcal{D}_t,$ and for each $d\in \mathcal{D}_w$ and  $i\in \mathcal{H}_d,$ the $p$-values in the set $\bm p_{i, Desc(i)}$ are independent of the $p$-values in the set $\bm p_{-i}^{d, Desc}.$ If the DAG is a tree with independent $p$-values, then this assumption is satisfied for any $\mathcal{D}_w.$
In this case,  the more conservative variant of the procedure allows using an arbitrary filter and an arbitrary  combining method which gives a valid global null $p$-value under independence,  in contrast to the original procedure, which has theoretical guarantees only under certain additional conditions on these components of the procedure (see items (ii) and (iii) of Theorem 2). For another example where the last  assumption of item (ii) is satisfied, consider a DAG consisting of several rooted sub-DAGs, where a sub-DAG originating from any given root has no common nodes with any of the sub-DAGs originating from other roots. Assume that the $p$-values for hypotheses belonging to any given sub-DAG are independent of all the other $p$-values, while the $p$-values belonging to the same sub-DAG are allowed to be arbitrarily dependent. In this case, for each root hypothesis $H_i$ (with $i\in\mathcal{H}_1$), the $p$-values in the set $\bm p_{i, Desc(i)}$ are independent of the $p$-values in the set $\bm p_{-i}^{d, Desc},$ therefore, if  the data-dependent weights are assigned only to the root hypotheses, i.e., $\mathcal{D}_w=\{1\},$  the last assumption of item (ii) is satisfied. 

\section{Proof of Theorem \ref{main_theorem}} \label{sec:supp_proofs}
\subsection*{Definitions and lemma for the proof of Theorem \ref{main_theorem}}
Consider a set of $m$ null hypotheses $H_1, \ldots, H_m,$ with a vector of corresponding 
$p$-values $\bmath{p}=(p_1, \ldots, p_m).$ We address a general multiple testing procedure based on $p$-values, which is a map
$\bm p \mapsto \cU^*(\bm p),$ where $\cU^*(\bm p)\subseteq\{1, \ldots, m\}$ is the set of indices of rejected hypotheses.


\begin{adefinition} \label{def:thresh_collection}
    Let $\hat{w}_i:[0,1]^m \rightarrow (0, \infty)$ be weight functions for $i=1, \ldots, m,$ and let $q \in (0,1)$ be a target FDR level. Then, the function
	\begin{equation}
	\Delta(\bm p, r, i) = \frac{rq}{m\hat{w}_i(\bm p)}
\label{thresh_collection}
	\end{equation}
	is an adaptively weighted threshold collection with respect to $ q, m, \hat{w}_i, i=1, \ldots, m$.\label{adaptive_thresh}
\end{adefinition}
This definition is similar to the definition of an adaptive threshold collection of \cite{blan2009}, which was also used by \cite{filtering}. 
The definition of an adaptive threshold collection of \cite{blan2009} 
for the identity shape function is given by (\ref{thresh_collection}) with $\hat{w}_i(\bm p)=1/G(\bm p)$ for $i=1, \ldots, m,$  where $G:[0,1]^m\rightarrow(0, \infty)$ is an estimator for $\pi_0^{-1},$ the inverse of the proportion of true null hypotheses. 
The next definition addresses the self-consistency property with respect to the adaptively weighted threshold collection, which is analogous to the self-consistency definitions of \cite{blanchard2008two, blan2009} with respect to other threshold collections. 

\begin{adefinition} \label{def:SC}
	A multiple testing procedure $\bm p \mapsto \cU^*(\bm p)$ is self-consistent with respect to adaptively weighted threshold collection $\Delta(\bm p, r,i)$ if 
	\begin{equation}
	p_i \leq \Delta(\bm p, |\cU^*(\bm p)|, i) \quad \text{for all } i \in \cU^*(\bm p).
	\label{SC}
	\end{equation}
\end{adefinition}
 We are now ready to state a lemma that is useful for the proof of Theorem 1. 

\begin{alemma} \label{lem:FDR_control}
	Assume that the $p$-values $p_1, \ldots, p_m$ are valid and independent. Let $\cU^*$ be a self-consistent multiple testing procedure with respect to  adaptively weighted threshold collection~\eqref{thresh_collection} with weight functions $\hat{w}_i(\bm p)$ satisfying Condition 1. 
    Then $\cU^*$ applied on $\bm p$ controls the FDR at level $q$ if the mapping $\bm p \mapsto |\cU^*(\bm p)|$ is coordinate-wise non-increasing. 
\end{alemma}
 The result of Lemma \ref{lem:FDR_control} generalizes the result of Theorem 11 of \cite{blan2009}, and reduces to this result when $\hat{w}_i(\bm p)=1/G(\bm p)$ for $i=1, \ldots, m,$ where $G:[0,1]^m\rightarrow (0, \infty)$ is a coordinate-wise non-increasing function, and $\EE{G(\bm p_{0,i})}\leq \pi_0^{-1}$ for every $i\in \mathcal{H}_0$ under independence. As discussed in Section 2.2, these weight functions satisfy Condition 1 under independence. 
\subsection*{Proof of Lemma \ref{lem:FDR_control}}
     
	 The proof is almost identical to the proof of Theorem 11 in \cite{blan2009}. 
     Let $\mathcal U^*$ be a self-consistent procedure with respect to thresholds of form (\ref{thresh_collection}) with weights $\hat{w}_1(\bm p), \ldots, \hat{w}_m(\bm p) $ that satisfy Condition 1 under independence. According to (\ref{SC}), for each $i\in \mathcal{U}^*(\bm p)$ 
    it holds
    \begin{align*}
        p_i\leq \frac{|\cU^*(\bm p)|q}{m\hat{w}_i(\bm p)}.
    \end{align*}
    Therefore, the FDR of $\mathcal{U}^*,$ denoted by $\fdr(\mathcal{U}^*),$ satisfies:
    \begin{align*}
        &\fdr(\mathcal{U^*})=\EE{\frac{\sum_{i \in \mathcal H_0} \ind(i \in \cU^*(\bm p))}{|\cU^*(\bm p)|}} = \sum_{i \in \mathcal H_0} \EE{\frac{\ind(i \in \cU^*(\bm p))}{|\cU^*(\bm p)|}} \leq \\&\sum_{i \in \mathcal H_0} \EE{\frac{\ind(p_i\leq  q|\cU^*(\bm p)|/(m\hat{w}_i(\bm p))}{|\cU^*(\bm p)|}}.
        \end{align*}
        For each $i\in \{1, \ldots, m\},$ let $\bm p_{-i}$ be the vector of $p$-values excluding $p_i.$ Since the functions $\hat{w}_i(p), i=1, \ldots, m$ satisfy Condition 1, each of them is coordinate-wise nondecreasing in $\bm p,$ therefore
        \begin{align}
            \sum_{i \in \mathcal H_0} \EE{\frac{\ind(p_i \leq  q|\cU^*(\bm p)|/(m\hat{w}_i(\bm p))}{|\cU^*(\bm p)|}}
        &\leq \sum_{i \in \mathcal H_0} \EE{\frac{\ind(p_i \leq  q|\cU^*(\bm p)|/(m\hat{w}_i(\bm p_{0,i}))}{|\cU^*(\bm p)|}}\notag\\&=
        \sum_{i \in \mathcal H_0} \EE{\EE{  \frac{\ind(p_i \leq  q|\cU^*(\bm p)|/(m\hat{w}_i(\bm p_{0,i}))}{|\cU^*(\bm p)|} \mid  \bm p_{-i}}}\notag\\&\leq\frac{q}{m}\sum_{i \in \mathcal{H}_0}\EE{\frac{1}{\hat{w}_i(\bm p_{0,i})}}\leq q.\label{last}
    \end{align}
    The first inequality in (\ref{last}) follows from Lemma 3.2 of \cite{blanchard2008two}, which  was restated as Lemma 27 in \cite{blan2009}. It is given below.
    \begin{alemma}[\cite{blanchard2008two}]\label{indep}
        Let $g:[0,1]\rightarrow (0, \infty)$ be a nonincreasing function. Let $U$ be a random variable which satisfies $\PP{U\leq u}\leq u$ for all $u\in[0,1].$ Then, for any constant $c>0,$ we have
        \begin{align*}
            \EE{\frac{\ind(U\leq cg(U))}{g(U)}}\leq c.
        \end{align*}
    \end{alemma}
   Let us fix $j\in \mathcal{H}_0.$ Since  the $p$-values are valid, $\PP{p_j\leq u}\leq u$ for all $u\in[0,1].$ By independence of $p_j$ and $\bm p_{-j},$ we obtain that the conditional distribution of $p_j$ given $\bm p_{-j}$ is the same as the marginal distribution of $p_j.$ In addition, for fixed $\bm p_{-j},$ the function $g(U)=|\mathcal{U}^*(\bm p_{-j},U))|$ is nonincreasing in $U$, because $|\mathcal{U^*}(\bm p)|$ is nonincreasing in each $p_j$. Finally, $q/(m\hat{w}_i(\bm p_{0j}))$ depends only on $\bm p_{-j},$ and therefore, $q/(m\hat{w}_i(\bm p_{0j}))$ is a constant for fixed $\bm p_{-j}.$ 
   Hence, by applying Lemma \ref{indep} with $U=p_j,$ $g(U)=|\mathcal{U}^*(\bm p_{-j},U)|,$ and $c=q/(m\hat{w}_i(\bm p_{0j})),$ we obtain the first inequality in (\ref{last}). The second inequality in (\ref{last}) follows from the assumption that the weight functions satisfy Condition 1. This completes the proof. 

\subsection*{Proof of Theorem 1}
We first prove that WFBH is self-consistent with respect to the threshold collection given in (\ref{thresh_collection}),  i.e. $\Delta(\bm p, r, i)=rq/(m\hat{w}_i(\bm p)).$ Let $i\in \mathcal{U}^*(\bm p).$ Recall that the filtered rejection set of WFBH is $\mathcal{U}^*(\bm p)=\F(
\{j: \hat{w}_jp_j\leq t^*\}, \bm p
),$ and
by definition of a filter, 
$\F(
\{j: \hat{w}_jp_j\leq t^*\}, \bm p
)\subseteq \{j: \hat{w}_jp_j\leq t^*\},$ therefore \begin{align}\hat{w}_ip_i\leq t^*.\label{self:1}\end{align} By definition of WFBH, 
\begin{align*}\fdphat(t^*)=\frac{mt^*}{|\mathcal{U}^*(\bm p)|}\leq q,\end{align*}
which yields that
\begin{align}
    t^*\leq \frac{|\mathcal{U}^{*}(\bm p)|q}{m}.
\label{self:2}\end{align}
Combining (\ref{self:1}) and (\ref{self:2}), we obtain that 
\begin{align*}
   \hat{w}_ip_i\leq t^*\leq \frac{|\mathcal{U}^*(\bm p)|q}{m},
\end{align*}
implying that $p_i\leq |\mathcal{U}^*(\bm p)|q/\{m\hat{w}_i(\bm p)\}=\Delta(\bm p, |\mathcal{U}^*(\bm p)|, i).$ Thus we proved that WFBH is self-consistent with respect to thresholds of form (\ref{thresh_collection}).

Now we shall prove that the mapping $\bm p\mapsto |\mathcal{U}^*(\bm p)|$ with weights satisfying Condition 1 and a monotonic filter $\F$ is coordinate-wise non-increasing. Our proof is almost identical to the proof of \cite{filtering} that the mapping corresponding to Focused BH with a monotonic filter is coordinate-wise  non-increasing. Suppose $\bm p^1 \leq \bm p^2$, where the inequality is understood coordinate-wise, and let $t_1^*$ and $t_2^*$ be the corresponding $p$-value cutoffs for WFBH.
	Let 
	\begin{align*}
	t^{**}_2 = \max\{t \in \{0, \hat{w}_1(\bm p^1)p^1_1, \dots, \hat{w}_m(\bm p^1)p^1_m\}: t \leq t^*_2\}.
	\end{align*}
	By definition of $t_2^{**}$, it holds that $\{j:\hat{w}_j(\bm p^1)p_j^1\leq t_2^{**}\} = \{j:\hat{w}_j(\bm p^1)p_j^1\leq t_2^{*}\}.$ In addition, note that if the weight functions $\hat{w}_i(\bm p)$ satisfy  Condition 1, then they are coordinate-wise non-decreasing, so $\hat{w}_j(\bm p^1)\leq \hat{w}_j(\bm p^2)$ for $j=1, \ldots, m.$ Therefore, $\{j:\hat{w}_j(\bm p^2)p_j^2\leq t_2^{*}\}\subseteq \{j:\hat{w}_j(\bm p^1)p_j^1\leq t_2^{*}\}.$ Hence, we obtain $\{j:\hat{w}_j(\bm p^2)p_j^2\leq t_2^{*}\}\subseteq \{j:\hat{w}_j(\bm p^1)p_j^1\leq t_2^{**}\}.$ 
    We denote the estimator of $\fdp(t_2^{**})$ in Step 2 of Procedure 1, based on the vector $\bm p^1,$ by $\fdphat(t^{**}_2, \bm p^1).$ We obtain
	\begin{align}
\fdphat(t^{**}_2, \bm p^1) = \frac{m \cdot t^{**}_2}{|\F(\{j:\hat{w}_j(\bm p^1)p_j^1\leq t_2^{**}\}, \bm p^1)|}&\leq \frac{m \cdot t^{**}_2}{|\F(\{j:\hat{w}_j(\bm p^2)p_j^2\leq t_2^{*}\}, \bm p^2)|}\label{mono1}\\&\leq \frac{m \cdot t^{*}_2}{|\F(\{j:\hat{w}_j(\bm p^2)p_j^2\leq t_2^{*}\}, \bm p^2)|}\leq q.
\label{monotonicity_argument1}
	\end{align}
    The inequality in (\ref{mono1}) follows by the monotonicity of $\F,$ and the second inequality in (\ref{monotonicity_argument1}) follows from the definition of $t_2^*.$
    From (\ref{monotonicity_argument1})  we obtain 
    $$\fdphat(t_2^{**}, \bm p^1)\leq q,$$ so by definition of $t_1^*,$ it follows that $t_1^*\geq t_2^{**}.$ Therefore, $$\{j: \hat{w}_j(\bm p^2)p^2_j\leq t_2^{*}\}\subseteq \{j:\hat{w}_j(\bm p^1)p^1_j\leq t_2^*\}= \{j:\hat{w}_j(\bm p^1)p^1_j\leq t_2^{**}\}\subseteq \{j:\hat{w}_j(\bm p^1)p^1_j\leq t_1^{*}\}.$$ By monotonicity of $\F,$ it follows that $$|\F(\{j:\hat{w}_j(\bm p^1)p^1_j\leq t_1^*\}, \bm p_1)| \geq |\F(\{j:\hat{w}_j(\bm p^2)p^2_j\leq t_2^*\}, \bm p_2)|.$$ Therefore, the mapping $\bm p \mapsto |\cU^*(\bm p)|$ with weights satisfying Condition 1 and a monotonic filter  is indeed 
    nonincreasing in each component of $\bm p$.
 We conclude that WFBH with a monotonic filter and weights satisfying Condition 1 satisfies the assumptions of Lemma \ref{lem:FDR_control}; therefore, the result of Theorem 1  follows from Lemma \ref{lem:FDR_control}. 
   \section{Proof of Proposition 1}
   Let $\{\hat{w}_{d,i}, d=1, \ldots, D, i=1, \ldots, |\mathcal{H}_d|\}$ be a set of weight functions that satisfy Assumption 3 in the main text. We need to prove that they satisfy Condition 1. We first prove that these functions are coordinate-wise non-decreasing in  $\bm p.$ Indeed, this  obviously holds for $\hat{w}_{d,i}$ with $d\neq \mathcal{D}_w$ and
   any $i\in \{1, \ldots, |\mathcal{H}_d|\},$ because in this case $\hat{w}_{d,i}\equiv 1.$
   This is also true for $\hat{w}_{d,i}$ with $d\in \mathcal{D}_w$  and
   any $i\in \{1, \ldots, |\mathcal{H}_d|\},$ 
    because $\hat{\pi}(\chd(a)\cap \mathcal{H}_d)$ is coordinate-wise non-decreasing in each $p$-value in the set $\{p_j: j\in \chd(a)\cap \mathcal{H}_d\},$ for any node $a$ such that $|\chd(a)\cap \mathcal{H}_d|>0.$ 
         
          For the rest of the proof, for convenience, we convert the indices  $\{(d,i), d=1, \ldots, D, i=1, \ldots, |\mathcal{H}_d|\}$ to the indices $\{1, \ldots,m\}$, according to the one-to-one correspondence between them, as discussed in Section 3.1 in the main manuscript. After the conversion,  $\{\hat{w}_{d,i}, d=1, \ldots, D, i=1, \ldots, |\mathcal{H}_d|\}= \{\hat{w}_1, \ldots, \hat{w}_m\}.$ It remains to prove that these functions satisfy the following inequality:
$$\sum_{d=1}^D\sum_{i\in \mathcal{H}_d^0}\EE{\frac{1}{\hat{w}_{i}(\bm p_{0,i})}}\leq m,
$$
where $\mathcal{H}_d^0=\mathcal{H}_d\cap \mathcal{H}_0,$ i.e. $\mathcal{H}_d^0$ is the set of indices of true null hypotheses with depth $d.$ 
  Following \cite{DAGGER}, for each $d\in\{1, \ldots, D\},$
we define $\mathcal{H}_{1:d}=\cup_{\ell=1}^d\mathcal{H}_{\ell},$ i.e., 
$H_{1:d}$ is the subset of $\mathcal{H}$ that contains indices of hypotheses with depths not greater than $d.$
Recall that we added an artificial dummy parent hypothesis to all root nodes, so now each hypothesis has a parent in $\mathcal{H}_{1:d-1},$ where $\mathcal{H}_{1:0}$ consists of the single artificial hypothesis that is a parent for all root nodes. For each depth $d\in\mathcal{D}_w$ and 
node $a\in \mathcal{H}_{1:d-1}$ such that $|\text{Ch}(a)\cap \mathcal{H}_d|>0,$ for  $j\in \chd(a)\cap \mathcal{H}_d$ we define
$$\hat{\pi}_{0,j}(\chd(a)\cap \mathcal{H}_d)=\frac{1+\sum_{i\in \chd(a)\cap \mathcal{H}_d, i\neq j}\ind\{p_i>\lambda\}}{(1-\lambda)\cdot |\chd(a)\cap \mathcal{H}_d|},\,\, \,\lambda\in(0,1),$$
which is
Storey's estimator for the proportion of null hypotheses in $\chd(a)\cap\mathcal{H}_d$ obtained when $p_j$ is replaced by 0. 
        Fix $d\in \mathcal{D}_w.$ We obtain the following:
\begin{align}
    &\sum_{i\in \mathcal{H}_d^0}\EE{\frac{1}{\hat{w}_i(\bm p_{0,i})}}\leq\notag\\&\sum_{a\in \mathcal{H}_{1:d-1}}\ind\{|\chd(a)\cap \mathcal{H}_d|>c\}\sum_{i \in \mathcal{H}_d^{0}\cap \chd(a)}\EE{\frac{1}{\hat{\pi}_{0,i}(\chd(a)\cap \mathcal{H}_d)}}\cdot\frac{|\mathcal{H}_d|}{|\chd(a)\cap\mathcal{H}_d|\cdot n_d}
+\label{sum1}\\&\sum_{a\in \mathcal{H}_{1:d-1}}\ind\{0<|\chd(a)\cap \mathcal{H}_d|\leq c\}\sum_{i \in \mathcal{H}_d^{0}\cap \chd(a)}\frac{|\mathcal{H}_d|}{|\chd(a)\cap\mathcal{H}_d|\cdot n_d}.\label{sum2}
\end{align}
Consider first the sum in (\ref{sum1}). Since the $p$-values in $\chd(a)\cap \mathcal{H}_d$ are independent, based on Corollary 13 of \cite{blan2009} 
we obtain that for each $i\in \chd(a)\cap\mathcal{H}_d^0:$
\begin{align}\EE{\frac{1}{\hat{\pi}_{0,i}(\chd(a)\cap \mathcal{H}_d)}}\leq \frac {1}{\pi_0(\chd(a)\cap \mathcal{H}_d)},\label{cor-blan}\end{align}
where $\pi_0(\chd(a)\cap \mathcal{H}_d)$ is the proportion of true null hypotheses in $\chd(a)\cap \mathcal{H}_d,$ i.e., 
$$\pi_0(\chd(a)\cap \mathcal{H}_d)=\frac{|\chd(a)\cap \mathcal{H}_d^0|}{|\chd(a)\cap\mathcal{H}_d|}.$$
Using (\ref{cor-blan}), we obtain an upper bound for the sum in (\ref{sum1}):
\begin{align}&\sum_{a\in \mathcal{H}_{1:d-1}}\ind\{|\chd(a)\cap \mathcal{H}_d|>c\}\sum_{i \in \mathcal{H}_d^{0}\cap \chd(a)}\EE{\frac{1}{\hat{\pi}_{0,i}(\chd(a)\cap \mathcal{H}_d)}}\cdot\frac{|\mathcal{H}_d|}{|\chd(a)\cap\mathcal{H}_d|\cdot n_d}\leq\notag\\&\sum_{a\in \mathcal{H}_{1:d-1}}\ind\{|\chd(a)\cap \mathcal{H}_d|>c\}\sum_{i \in \mathcal{H}_d^{0}\cap \chd(a)}\frac{|\chd(a)\cap \mathcal{H}_d|}{|\chd(a)\cap\mathcal{H}_d^0|}\cdot\frac{|\mathcal{H}_d|}{|\chd(a)\cap\mathcal{H}_d|\cdot n_d}=\notag\\&\frac{|\mathcal{H}_d|}{n_d}\sum_{a\in \mathcal{H}_{1:d-1}}\ind\{|\chd(a)\cap \mathcal{H}_d|>c\}\sum_{i \in \mathcal{H}_d^{0}\cap \chd(a)}\frac{1}{|\chd(a)\cap\mathcal{H}_d^0|}=\notag\\&\frac{|\mathcal{H}_d|}{n_d}\sum_{a\in \mathcal{H}_{1:d-1}}\ind\{|\chd(a)\cap \mathcal{H}_d|>c\}\label{upper1}
\end{align}
Let us now obtain an upper bound for the sum in (\ref{sum2}). 
\begin{align}
    &\sum_{a\in \mathcal{H}_{1:d-1}}\ind\{0<|\chd(a)\cap \mathcal{H}_d|\leq c\}\sum_{i \in \mathcal{H}_d^{0}\cap \chd(a)}\frac{|\mathcal{H}_d|}{|\chd(a)\cap\mathcal{H}_d|\cdot n_d}=\notag\\&
    \frac{|\mathcal{H}_d|}{n_d}\sum_{a\in \mathcal{H}_{1:d-1}}\ind\{0<|\chd(a)\cap \mathcal{H}_d|\leq c\}\sum_{i \in \mathcal{H}_d^{0}\cap \chd(a)}\frac{1}{|\chd(a)\cap\mathcal{H}_d|}\leq\notag\\& \frac{|\mathcal{H}_d|}{n_d}\sum_{a\in \mathcal{H}_{1:d-1}}\ind\{0<|\chd(a)\cap \mathcal{H}_d|\leq c\}.\label{upper2}
\end{align}
Using the upper bounds in (\ref{upper1}) and (\ref{upper2}) for the sums in (\ref{sum1}) and (\ref{sum2}) respectively, we obtain 
\begin{align}
\sum_{i\in \mathcal{H}_d^0}\EE{\frac{1}{\hat{w}_i(\bm p_{0,i})}}\leq &\frac{|\mathcal{H}_d|}{n_d}\sum_{a\in \mathcal{H}_{1:d-1}}\ind\{|\chd(a)\cap \mathcal{H}_d|>c\}+\notag\\&\frac{|\mathcal{H}_d|}{n_d}\sum_{a\in \mathcal{H}_{1:d-1}}\ind\{0<|\chd(a)\cap \mathcal{H}_d|\leq c\}=\notag\\&\frac{|\mathcal{H}_d|}{n_d}\sum_{a\in \mathcal{H}_{1:d-1}}\ind\{|\chd(a)\cap \mathcal{H}_d|>0\}=|\mathcal{H}_d|\label{inter-imp}
\end{align}
where the last definition follows from the definition of $n_d.$ Now, fix $d\in\{1, \ldots, D\}\setminus \mathcal{D}_w.$ For each $i\in \mathcal{H}_d^0,$ $\hat{w}_i(\bm p_{0,i})=1,$ therefore $$\sum_{i\in \mathcal{H}_d^0}\EE{\frac{1}{\hat{w}_i(\bm p_{0,i})}}=|\mathcal{H}_d^0|\leq |\mathcal{H}_d|.$$ Thus, for each $d\in\{1, \ldots, D\},$ the following holds: 
\begin{align*}\sum_{i\in \mathcal{H}_d^0}\EE{\frac{1}{\hat{w}_i(\bm p_{0,i})}}\leq|\mathcal{H}_d|.\end{align*}
It follows that $$\sum_{d=1}^D\sum_{i\in \mathcal{H}_d^0}\EE{\frac{1}{\hat{w}_{i}(\bm p_{0,i})}}\leq \sum_{d=1}^D|\mathcal{H}_d|=m,
$$
which completes the proof.

\section{Proof of Theorem 2 and its extensions}\label{sec:proof-2}
The following lemma is needed to prove the extension of Theorem 2, which we present below.
This lemma is similar to Lemma 2 in \cite{pfilter2}, recovering some of its parts and adding a novel result. 
\begin{alemma}\label{main-lemma}
Consider  hypotheses $H_1, \ldots, H_m$ with corresponding $p$-values $p_1, \ldots, p_m.$ Assume that these $p$-values are valid. Let $\mathcal{H}_0\subseteq \{1, \ldots, m\}$ be a subset of indices of true null hypotheses, and let $\mathcal{G}\subseteq \mathcal{H}_0.$ Let $\bm p_{\mathcal{G}}=\{p_i: i\in \mathcal{G}\}$ be the set of $p$-values with indices in $\mathcal{G}$.
\begin{enumerate}
    \item[(i)] If the $p$-values $p_1, \ldots, p_m$ are independent,  and $f:[0,1]^m\rightarrow [0,\infty)$ is a coordinate-wise non-increasing function, then we have 
    \begin{align*}
    \EE{\frac{T(\bm p_\mathcal{G})\leq f(\bm p)\}}{f(\bm p)}}\leq 1,
 \end{align*}
  where $T$ is Simes' combining function.
  \item[(ii)] If the  $p$-values $p_1, \ldots, p_m$ are independent, $p_i$ is distributed as $\text{Un}[0,1]$ for each $i\in \mathcal{H}_0,$ 
  and $f:[0,1]^m\rightarrow [0,\infty)$ is a coordinate-wise non-increasing function, then we have 
    \begin{align*}
    \EE{\frac{\ind\{T(\bm p_{\mathcal{G}})\leq f(\bm p)\}}{f(\bm p)}}\leq 1,
 \end{align*}
  where $T:[0,1]^{|\mathcal{G}|}\rightarrow [0, \infty)$ is any combining function that satisfies Condition 2, e.g. it can be Fisher's, Stouffer's, or Tippett's combining function.
  \item [(iii)]  If the $p$-values are arbitrarily dependent, then for any constant $c>0,$ any function $\beta$ of form (\ref{beta}) and any function $f:[0,1]^m\rightarrow [0,\infty),$  we have
    \begin{align*}
    \EE{\frac{\ind\{T(\bm p_{\mathcal{G}})\leq c\beta(f(\bm p))\}}{cf(\bm p)}}\leq 1,
 \end{align*}
  where 
  $T:[0,1]^{|\mathcal{G}|}\rightarrow [0, \infty)$ is any combining function which guarantees that under the dependency of the $p$-values in $\bm p_{\mathcal{G}},$ $T(\bm p_{\mathcal{G}})$ is a valid $p$-value for the global null $\cap_{i\in \mathcal{G}}H_i,$ i.e. $\PP{T(\bm p_{\mathcal{G}})\leq x}\leq x$ for all $x\in[0,1].$ 
\end{enumerate}
\end{alemma}
Item (i) follows immediately from items (a) and (b) of Lemma 2 in \cite{pfilter2}, and item (iii) gives the result of item (c) of the same lemma. Item (ii) is novel, its proof is based on the results of \cite{smoothed}.  The proof of item (ii) is given at the end of this Appendix. 
We give below a result that includes the result of Theorem 2.
\begin{atheorem}\label{thm:hier:SM}
    Assume that the hypotheses have a DAG structure and 
    their $p$-values $p_1, \ldots, p_m$ are valid and independent.  WFBH with a monotonic filter and weight functions satisfying Assumption 3 controls the FDR at level $q$ when it is applied on  $\bm p'$ if either of the following sets of additional assumptions are satisfied:
    \begin{itemize}
        \item[(i)] $\bm p'=\bm p,$ i.e., the method is applied to the original $p$-values;
        \item[(ii)] The hypotheses satisfy the logical relationships in Assumption 1; the set of depths $\mathcal{D}_w$ addressed in Assumption 3 satisfies $\mathcal{D}_w\subseteq\mathcal{D}_t$ (see Appendix~\ref{Reshaped} for the definition of $\mathcal{D}_t),$ $\bm p'=\tilde{\bm p}$ where the smoothing function is Simes' function; 
                \item [(iii)] The same assumptions as in item (ii) with the smoothing function replaced by any combining function satisfying Condition 2, and an additional assumption that $p_i\sim \text{Un}[0,1]$ for each $i\in \mathcal{H}_0.$
    \end{itemize}
\end{atheorem}
In contrast to Theorem 2, this result does not require the assumption that the DAG is a tree in items (ii) and (iii), and requires instead that $\mathcal{D}_w\subseteq \mathcal{D}_t.$ As discussed in Appendix~\ref{Reshaped}, if the DAG is a tree, then $\mathcal{D}_t=\{1, \ldots, D\},$ therefore we have $\mathcal{D}_w\subseteq \mathcal{D}_t$ for any $\mathcal{D}_w.$ This shows that the assumptions of items (ii) and (iii) of Theorem \ref{thm:hier:SM} are weaker than the assumptions of items (ii) and (iii) of Theorem 2, respectively. An example of a DAG that is not a tree and a choice of $\mathcal{D}_w$ satisfying $\mathcal{D}_w\subseteq \mathcal{D}_t$ is a DAG consisting of several non-overlapping rooted sub-DAGs with $\mathcal{D}_w=\{1\},$ which we considered in Appendix~\ref{Reshaped}.

As noted in Section 3.3, Weighted Focused BH incorporating weights satisfying Assumption 3 with $\mathcal{D}_w=\emptyset$ reduces to the original Focused BH procedure. Thus, Theorem \ref{thm:hier:SM} gives FDR control results for Focused BH as well as for its weighted variant. In fact, the result for the Focused BH under the additional assumptions in item (iii) is novel. Note that since $\mathcal{D}_w=\emptyset$ for Focused BH, the assumption regarding $\mathcal{D}_w$ trivially holds for any DAG. The assumption that the original $p$-values are independent can be weakened for Focused BH when applied to smoothed $p$-values based on Simes' combining method. We summarize below the results for the Focused BH when applied to smoothed $p$-values. 
\begin{atheorem}\label{thm:Focused}
     Assume that $H_1, \ldots, H_m$ have a DAG structure, and satisfy the logical relationships in Assumption 1. Assume that their $p$-values $p_1, \ldots, p_m$ are valid.  Focused BH with a monotonic filter controls the FDR at level $q$ when applied to the vector of smoothed $p$-values  $\tilde{\bm p}$ if either of the following sets of additional assumptions are satisfied:
     \begin{itemize}
         \item[(i)] The $p$-values $p_1, \ldots, p_m$ are PRDS on $\mathcal{H}_0$ 
         (see Definition \ref{def:PRDS}); the smoothing function is Simes' function.
         \item[(ii)] The $p$-values $p_1, \ldots, p_m$ are independent; $p_i\sim \text{Un}[0,1]$ for each $i\in\mathcal{H}_0,$ the smoothing function is any combining function satisfying Condition 2.
     \end{itemize}
\end{atheorem}
The proof of Theorem \ref{thm:Focused} under the additional assumptions of item (i) is almost identical to the proof of Theorem 3 of \cite{filtering}, which addresses an intersection DAG (see Appendix~\ref{sec:intersec} for a definition). Both proofs rely on item (b) of Lemma 2 of \cite{pfilter2}. As noted above, the result under the additional assumptions of item (ii) is novel, and follows from the result of Theorem \ref{thm:hier:SM} under the additional assumptions of item (iii) for $\mathcal{D}_w=\emptyset.$ In fact, this result can be obtained directly, by using  Lemma 1 and Theorems 1 and 2 of \cite{smoothed} in conjunction with item (i) of Theorem 1 of \cite{filtering}. Specifically, the former three results show that  under the additional assumptions of item (ii) of Theorem \ref{thm:Focused}, the smoothed $p$-values satisfying Condition 2 are valid and are PRDS on $\mathcal{H}_0,$ 
and 
item (i) of Theorem 1 of \cite{filtering} shows that Focused BH with a monotonic filter controls the FDR when applied to valid $p$-values satisfying the above PRDS property.

\subsection*{Proof of Theorem \ref{thm:hier:SM}} 
The proof of Theorem \ref{thm:hier:SM} under the additional assumption of item (i) follows from Theorem 1 and Proposition 1. We shall now prove the result of Theorem \ref{thm:hier:SM} under the additional assumptions in items (ii) and (iii).
Let us show that the smoothed $p$-values $\tilde{p}_1, \ldots, \tilde{p}_m$ addressed in items (ii) and (iii) are valid. In item (ii), we assume that the smoothed $p$-values are obtained using Simes' method, i.e., $\tilde{p}_i=T_{Simes}(\bm p_{i, Desc(i)}),$ where $T_{Simes}$ is Simes' combining function. Let $H_i$ be a true null hypothesis. We need to show that $\PP{\tilde{p}_i\leq x}\leq x$ for all $x\in[0,1], $ i.e., that $\tilde{p}_i$ is superuniform. Since Assumption 1 is satisfied, all descendants of $H_i$ are also true null hypotheses, so the global null $H_i\cap (\cap_{j\in Desc(i)} H_j)$ is also true. The original $p$-values are assumed to be valid and independent; therefore the $p$-values in the set $\bm p_{i, Desc(i)}$ are superuniform and independent. It follows from the result of \cite{Simes1986} that $\tilde{p}_i=T_{Simes}(\bm p_{i, Desc(i)})$ is superuniform, i.e., it is a valid $p$-value for $H_i.$ This shows the validity of smoothed $p$-values addressed in item (ii). The validity of smoothed $p$-values addressed in item (iii) follows from Lemma 1 of \cite{smoothed}. 

For the rest of the proof, we address a vector of smoothed $p$-values $\tilde{\bm p},$ where the smoothing function is $T,$ which is either Simes' combining function addressed in item (ii), or any function satisfying Condition 2, addressed in item (iii). Let $\{\hat{w}_{d,i}, d=1, \ldots, D, i=1, \ldots, |\mathcal{H}_d|\}$ be weight functions satisfying Assumption 3.
Define $\mathcal{D}_c=\{1, \ldots, D\}\setminus \mathcal{D}_w,$ that is, $\mathcal{D}_c$ is the set of depth indices that were assigned with non-informative unity weights. 
For each depth $d\in\{1, \ldots, D\},$ let $\tilde{\bm p}^d$ be the vector of smoothed $p$-values for the hypotheses with depth $d,$ i.e. $\tilde{\bm p}^d=(\tilde{p}_{d,1},\ldots, \tilde{p}_{d,|\mathcal{H}_d|}).$ Similarly to the proof of Proposition 1, for convenience, we convert the pairs of indices $(d,i)$ to the original indices in $\mathcal{H},$ using the one-to-one correspondence between them. After conversion, we have the weight functions $\hat{w}_1, \ldots, \hat{w}_m,$ and the vector $\tilde{\bm p}^d$ that contains the smoothed $p$-values for hypotheses with indices in $\mathcal{H}_d.$ Note that for each $d\in\mathcal{D}_c$ and $i\in \mathcal{H}_d$ we have $\hat{w}_{i}(\bm p)= 1,$ while for each $d\in \mathcal{D}_w$ and $i\in\mathcal{H}_d,$ it holds: 
 (1) $\hat{w}_{i}(\tilde{\bm p})=\hat{w}_{i}(\tilde{\bm p}^d),$ that is, $\hat{w}_{i}$ depends on $\tilde{\bm p}$ only through $\tilde{\bm p}_d;$ (2) $\hat{w}_{i}$ is a coordinate-wise 
 non-decreasing function of $\tilde{\bm p}^d,$ therefore  $\hat{w}_{i}(\tilde{\bm p}^d_{0,i})\leq \hat{w}_{i}(\tilde{\bm p}^d),$ where 
 $\tilde{\bm p}^d_{0,i}$ is the vector $\tilde{\bm p}^d$ with $\tilde{p}_{i}$ replaced by 0. 
 In addition, according to the proof of Theorem 1, Weighted Focused BH is self-consistent with respect to the threshold collection given in (\ref{thresh_collection}), i.e. $\Delta(\bm p, r, i)=rq/(m\hat{w}_i(\bm p)).$ Using these facts, we begin to evaluate the FDR of the Weighted Focused BH when applied to $\tilde{\bm p}:$

 \begin{align}
     \text{FDR}&=\sum_{d=1}^D\sum_{i\in \mathcal{H}_d^0}\EE{\frac{\ind\{i\in\mathcal{U}^{*}(\tilde{\bm p})\}}{|\mathcal{U}^{*}(\tilde{\bm p})|}}\notag\\&\leq\sum_{d\in \mathcal{D}_c}\sum_{i\in \mathcal{H}_d^0}\EE{\frac{\ind\{\tilde{p}_{i}\leq q|\mathcal{U}^{*}(\tilde{\bm p})|/m\}}{|\mathcal{U}^{*}(\tilde{\bm p})|}}+\sum_{d\in \mathcal{D}_w}\sum_{i\in \mathcal{H}_d^0}\EE{\frac{\ind\{\tilde{p}_{i}\leq q|\mathcal{U}^{*}(\tilde{\bm p})|/(m\hat{w}_{i}(\tilde{\bm p}^d))\}}{|\mathcal{U}^{*}(\tilde{\bm p})|}}\notag\\&\leq\sum_{d\in \mathcal{D}_c}\sum_{i\in \mathcal{H}_d^0}\EE{\frac{\ind\{\tilde{p}_{i}\leq q|\mathcal{U}^{*}(\tilde{\bm p})|/m\}}{|\mathcal{U}^{*}(\tilde{\bm p})|}}+\sum_{d\in \mathcal{D}_w}\sum_{i\in \mathcal{H}_d^0}\EE{\frac{\ind\{\tilde{p}_{i}\leq q|\mathcal{U}^{*}(\tilde{\bm p})|/(m\hat{w}_{i}(\tilde{\bm p}_{0,i}^d))\}}{|\mathcal{U}^{*}(\tilde{\bm p})|}}\label{thm2-term2-item2}
 \end{align}
For each $d\in\{1, \ldots, D\}$ and $i\in\mathcal{H}_d,$ we define the following functions:
$$f_i(\bm p)=q|\mathcal{U}^{*}(\tilde{\bm p})|/m, \,\,f^w_{i}(\bm p)=q|\mathcal{U}^{*}(\tilde{\bm p})|/\{m\hat{w}_i(\tilde{\bm p}^d_{0,i})\}.$$ 
These are functions of the original $p$-values, because for each $i\in\{1, \ldots, m\},$ $\tilde{p}_i=T(\bm{p}_{i, Desc(i)}),$ i.e.,  each smoothed $p$-value is a function of the original $p$-values. 
Using these notations and (\ref{thm2-term2-item2}), we obtain:
\begin{align}
\text{FDR}&\leq \frac{q}{m}\sum_{d\in \mathcal{D}_c}\sum_{i\in \mathcal{H}_d^0}\EE{\frac{
\ind\{T(\bm p_{i, Desc(i)})\leq f_{i}(\bm p)\}}{f_{i}(\bm p)}}+\frac{q}{m}\sum_{d\in \mathcal{D}_w}\sum_{i\in \mathcal{H}_d^0}\EE{\frac{
\ind\{T(\bm p_{i, Desc(i)})\leq f^w_{i}(\bm p)\}}{\hat{w}_{i}(\tilde{\bm p}^d_{0,i})f_{i}^w(\bm p)}}\label{thm2-2items}
 \end{align}
Let us fix $d\in D_c$ and $i\in \mathcal{H}_d^0.$ We shall first show that the function $f_i(\bm p)$ is coordinate-wise non-increasing in $\bm p.$ We showed in the proof of Theorem 1 that $|\mathcal{U}^*|$ is 
nonincreasing in each of its input $p$-values when the filter is monotonic and each of the weight functions is nondecreasing in each of its input $p$-values. The filter is assumed to be monotonic in both items (ii) and (iii), and the weight functions satisfying Assumption 3 are nondecreasing in each of their input $p$-values.
Therefore, $|\mathcal{U}^*(\tilde{\bm p})|$ is a coordinate-wise non-increasing function of $\tilde{\bm p}.$ In addition, each smoothed $p$-value is a coordinate-wise non-decreasing function of $\bm p$ under the assumptions of items (ii) and (iii), because Simes' combining function, as well as any combining function satisfying Condition 2 is nondecreasing in each of its input $p$-values. Hence,  $|\mathcal{U}^*(\tilde{\bm p})|$  is  a coordinate-wise non-increasing function of $\bm p,$ which shows that $f_i(\bm p)$ is indeed a   coordinate-wise non-increasing function of $\bm p.$ In addition, as we noted above, since the DAG is assumed to satisfy Assumption 1 and $H_i$ is a true null hypothesis,  all its descendants are also true null hypotheses, which yields $\{i\}\cup Desc(i)\subseteq \mathcal{H}_0.$ Therefore, we obtain 
\begin{align}\EE{\frac{
\ind\{T(\bm p_{i, Desc(i)})\leq f_{i}(\bm p)\}}{f_{i}(\bm p)}}\leq 1\label{ineq-unity}\end{align}
under the additional assumptions of item (ii) of Theorem \ref{thm:hier:SM} using item (i) of Lemma \ref{main-lemma}, and under the additional assumptions in item (iii) of Theorem  \ref{thm:hier:SM} using item (ii) of Lemma \ref{main-lemma}. 
Applying the inequality in (\ref{ineq-unity}) for each $d\in \mathcal{D}_c$ and $i\in \mathcal{H}_d^0,$ we obtain an upper bound for the first term in (\ref{thm2-2items}):
\begin{align}
   \frac{q}{m}\sum_{d\in \mathcal{D}_c}\sum_{i\in \mathcal{H}_d^0}\EE{\frac{
\ind\{T(\bm p_{i, Desc(i)})\leq f_{i}(\bm p)\}}{f_{i}(\bm p)}}\leq  \frac{q}{m}\sum_{d\in \mathcal{D}_c}|\mathcal{H}_d^0|\leq \frac{q}{m}\sum_{d\in \mathcal{D}_c}|\mathcal{H}_d|.\label{upper-first-hier}
\end{align}
Now let us fix $d\in \mathcal{D}_w$ and $i\in \mathcal{H}_d^0.$ It is assumed that $\mathcal{D}_w\subseteq \mathcal{D}_t,$ therefore $\bm p_{-i}^{d, Desc}$ is defined, and it holds:
\begin{align}
    \EE{\frac{
\ind\{T(\bm p_{i, Desc(i)})\leq f^w_{i}(\bm p)\}}{\hat{w}_{i}(\tilde{\bm p}^d_{0,i})f_{i}^w(\bm p)}}&=\EE{\EEst{\frac{\ind\{T(\bm p_{i, Desc(i)})\leq f_{i}^w(\bm p)\}}{\hat{w}_{i}(\tilde{\bm p}^d_{0,i})f_{i}^w(\bm p)}}{\bm p_{-i}^{d, Desc}}}\notag\\
&=\EE{\frac{1}{\hat{w}_{i}(\tilde{\bm p}^d_{0,i})}\EEst{\frac{
\ind\{T(\bm p_{i, Desc(i)})\leq f_{i}^w(\bm p)\}}{f_{i}^w(\bm p)}}{\bm p_{-i}^{d, Desc}}}\label{thm2-cond}
\end{align}
The equality in (\ref{thm2-cond}) holds because the vector $\tilde{\bm p}_{0,i}^d$ is a function only of the $p$-values  in the set $\bm p_{-i}^{d, Desc(i)},$
therefore after fixing $\bm p_{-i}^{d, Desc},$ $\hat{w}_{i}(\tilde{\bm p}^d_{0,i})$ is a constant. 
Let us now consider the conditional expectation \begin{align*}\EEst{\frac{
\ind\{T(\bm p_{i, Desc(i)})\leq f_{i}^w(\bm p)\}}{f_{i}^w(\bm p)}}{\bm p_{-i}^{d, Desc}}\label{cond-exp}.\end{align*} 
We shall show that we can apply items (i) and (ii) of Lemma \ref{lem:FDR_control} to obtain an upper bound for this conditional expectation, under the additional assumptions in items (ii) and (iii) of Theorem \ref{thm:hier:SM}, respectively. We showed above that $|\mathcal{U}^*(\tilde{\bm p})|$ is a coordinate-wise non-increasing function of $\bm p.$ Using the facts that $\hat{w}_{i}$ is nondecreasing in each of its input $p$-values, and that each smoothed $p$-value is a coordinate-wise non-decreasing function of the original $p$-values $\bm p,$  
we obtain that $\hat{w}_{i}(\tilde{\bm p}_{0,i}^d)$ is a coordinate-wise non-decreasing function of $\bm p$. Therefore, $f_{i}^w(\bm p)$ is a coordinate-wise non-increasing function of $\bm p,$ so after fixing $\bm p_{-i}^{d, Desc},$ the function $f_{i}^w(\bm p)$ is 
nonincreasing in each of the remaining $p$-values that were not fixed. 
Now, we shall show that the conditional distribution of $T(\bm p_{i, Desc(i)})$ given $\bm p_{-i}^{d, Desc}$ is the same as its marginal distribution. Let us first show that the $p$-values in the set $\bm p_{i, Desc(i)}$ are independent of the $p$-values in the set $\bm p_{-i}^{d, Desc}.$ Indeed, since $d\in \mathcal{D}_t,$ these two sets of $p$-values are disjoint, therefore, they are independent due to the assumption that the original $p$-values $p_1, \ldots, p_m$ are independent. Hence, we obtain that $T(\bm p_{i, Desc(i)})$ is independent of $\bm p_{-i}^{d, Desc},$ which yields that the conditional distribution of $T(\bm p_{i, Desc(i)})$ given $\bm p_{-i}^{d, Desc}$ is the same as its marginal distribution.
Finally, as we showed above, due to Assumption 1 we have $\{i\}\cup Desc(i)\subseteq \mathcal{H}_0.$ Using these facts, we obtain 
\begin{equation}\EEst{\frac{
\ind\{T(\bm p_{i, Desc(i)})\leq f_{i}^w(\bm p)\}}{f_{i}^w(\bm p)}}{\bm p_{-i}^{d, Desc}}\leq 1\label{cond-exp-1}\end{equation}
under the additional assumptions of item (ii) of Theorem \ref{thm:hier:SM} using item (i) of Lemma \ref{main-lemma}, and under the additional assumptions in item (iii) of Theorem  \ref{thm:hier:SM} using item (ii) of Lemma \ref{main-lemma}. 
Using (\ref{thm2-cond}) and  (\ref{cond-exp-1}) for each $d\in \mathcal{D}_w$ and $i\in \mathcal{H}_d^0,$ we obtain an upper bound for the second term in (\ref{thm2-2items}):
\begin{align}&\frac{q}{m}\sum_{d\in \mathcal{D}_w}\sum_{i\in \mathcal{H}_d^0}\EE{\frac{
\ind\{T(\bm p_{i, Desc(i)})\leq f^w_{i}(\bm p)\}}{\hat{w}_{i}(\tilde{\bm p}^d_{0,i})f_{i}^w(\bm p)}}=\notag\\&\frac{q}{m}\sum_{d\in \mathcal{D}_w}\sum_{i\in \mathcal{H}_d^0}\EE{\frac{1}{\hat{w}_{i}(\tilde{\bm p}^d_{0,i})}\EEst{\frac{
\ind\{T(\bm p_{i, Desc(i)})\leq f_{i}^w(\bm p)\}}{f_{i}^w(\bm p)}}{\bm p_{-i}^{d, Desc}}}\leq\notag\\&\frac{q}{m}\sum_{d\in \mathcal{D}_w}\sum_{i\in \mathcal{H}_d^0}\EE{\frac{1}{\hat{w}_{i}(\tilde{\bm p}^d_{0,i})}}. \label{upper-2}
\end{align}
Let us show that it is enough to prove that for each $d\in \mathcal{D}_w:$ 
\begin{align}\sum_{i\in\mathcal{H}_d^0}\EE{\frac{1}{\hat{w}_{i}(\tilde{\bm p}_{0,i})}}\leq |\mathcal{H}_d|\label{ineq-weights-last}.\end{align}
Indeed, (\ref{ineq-weights-last}) combined with (\ref{upper-2}) yields that 
\begin{align}&\frac{q}{m}\sum_{d\in \mathcal{D}_w}\sum_{i\in \mathcal{H}_d^0}\EE{\frac{
\ind\{T(\bm p_{i, Desc(i)})\leq f^w_{i}(\bm p)\}}{\hat{w}_{i}(\tilde{\bm p}^d_{0,i})f_{i}^w(\bm p)}}\leq \frac{q}{m}\sum_{d\in \mathcal{D}_w}|\mathcal{H}_d|.\label{upper-cond-2}
\end{align}
This gives an upper bound for the second term in (\ref{thm2-2items}). Combining (\ref{upper-cond-2}) with an upper bound (\ref{upper-first-hier}) for the first term in (\ref{thm2-2items}), we obtain an upper bound for FDR:
\begin{align*}
    \text{FDR}\leq \frac{q}{m}\sum_{d\in \mathcal{D}_c}|\mathcal{H}_d|+\frac{q}{m}\sum_{d\in \mathcal{D}_w}|\mathcal{H}_d|=\frac{q}{m}\sum_{d=1}^D|\mathcal{H}_d|=q.
\end{align*}
Thus, it remains to prove (\ref{ineq-weights-last}). 
%
Let us show that the assumptions of Proposition 1 regarding the $p$-values given as input to the weight functions are satisfied, i.e. that the smoothed $p$-values in the set $\{\tilde{p}_{d,i}, i=1, \ldots, |\mathcal{H}_d|\}$ are valid and independent for each depth $d\in\mathcal{D}_w.$ We showed above that under the additional assumptions of  items (ii) and (iii), the smoothed $p$-values are valid, and that the $p$-values in the set $\bm p_{i, Desc(i)}$ are independent of the $p$-values in the set $\bm p_{-i}^{d, Desc}$ for each $d\in\mathcal{D}_w$ and $i\in \mathcal{H}_d.$ The latter implies that the smoothed $p$-values in the set $\{\tilde{p}_{d,i}, i=1, \ldots, |\mathcal{H}_d|\}$ are independent. 
Thus,  using the same arguments that lead to (\ref{inter-imp}) in the proof of Proposition 1, where we replace the original $p$-values by smoothed $p$-values, we obtain (\ref{ineq-weights-last}), which completes the proof. 
\subsection*{Proof of item (ii) of Lemma \ref{main-lemma}}
Let $f:[0,1]^m\rightarrow[0, \infty)$ be a coordinate-wise non-increasing function. Let us fix $\mathcal{G}\subseteq\mathcal{H}_0,$ and a combining function 
$T:[0,1]^{|\mathcal{G}|}\rightarrow [0, \infty)$ satisfying Condition 2 for $n=|\mathcal{G}|.$ We need to prove that
\begin{align}
    \EE{
    \frac{\ind\{T(\bm p_{\mathcal{G}})\leq f(\bm p)\}}{f(\bm p)}}\leq 1.\label{desired-res-lemma}
\end{align}
We  first assume that $T$ satisfies part (a) of Condition 2, i.e., $T(\bm p_{\mathcal{G}})=F(G(\sum_{i\in \mathcal{G}}H^{-1}(p_i))),$
       where $G:\mathbb{R}\rightarrow \mathbb{R}$ and $H: \mathbb{R}\mapsto[0,1]$ are monotone increasing functions,
         $H^{'}$ is log-concave,  and $F(x)\equiv\PP{G(\sum_{i\in\mathcal{G}}H^{-1}(U_i))\leq x},$ where $\{U_i,\,i\in\mathcal{G}\}$ are independent and identically distributed as $U[0,1].$  In this case, the proof is based on Theorem 1.2 in the Supplementary Material of \cite{smoothed}. 
This result is given in the following. 
\begin{atheorem}[\cite{smoothed}]\label{thm-smooth}
    Let $X_1, \ldots, X_n$ be random variables on $\mathbb{R}$
or $\mathbb{Z},$ and $\mathcal{S}\subset\{1, \ldots,n\}$ such that: (i) $X_i$ has a log-concave density for each $i\in \mathcal{S},$ (ii) $\{X_i: i\in \mathcal{S}\}$ are mutually independent, and (iii) $\{X_i: i\in \mathcal{S}\}$ are independent of $\{X_i: i \notin \mathcal{S}\}.$ Let 
$$Y_1=g(\sum_{i\in \mathcal{S}}a_iX_i),$$
where $a_i\geq 0$ for all 
$i \in \mathcal{S}$ and $g$ is 
nondecreasing. For any $j=2, \ldots, M, $ let 
$$Y_j=f_j(X_1, \ldots, X_n)$$ for some coordinate-wise non-decreasing function $f_j.$ Then $(Y_1, \ldots, Y_M)$ are positive regression dependent with respect to $Y_1.$  
\end{atheorem}
 The notion of positive regression dependence on each one from a subset (PRDS) was defined by \cite{BY01}. 
\begin{adefinition}[PRDS, \cite{BY01}]\label{def:PRDS}
    $(X_1, \ldots, X_m)\in\mathbb{R}^m$ is positive regression dependent on each one from a subset $I\subseteq\{1, \ldots, m\}$ (in short, PRDS on $I$), if for any nondecreasing set $D\in\mathbb{R}^m$ and for each $i\in I,$ $\PP{(X_1, \ldots, X_m)\in D\,|\,X_i=x}$ is nondecreasing in $x.$
\end{adefinition}
Recall that a set $D\in\mathbb{R}^{m}$ is called nondecreasing if for all $x,y\in \mathbb{R}^m$ such that $x\leq y,$ i.e., $x_i\leq y_i$ for $i=1, \ldots, m,$ $x\in D$ implies that $y\in D.$ 
 The result of 
 Theorem \ref{thm-smooth} means that for any nondecreasing set $D\in \mathbb{R}^M,$ $\PP{(Y_1, \ldots, Y_M)\in D\mid Y_1=t}$ is nondecreasing in $t.$ 
The proof of Theorem \ref{thm-smooth} is based on a result of \cite{efron1965increasing}, an interested reader is referred to \cite{smoothed} for the proof.

The proof of item (ii) of Lemma \ref{main-lemma} is similar to the proof of Theorem 1 of \cite{smoothed}. Define $X_i=H^{-1}(p_i),$ $i=1, \ldots, m.$ Since it is assumed that $p_i$ is distributed as $\text{Un}[0,1]$ for each $i\in \mathcal{H}_0$ 
and $\mathcal{G}\subseteq \mathcal{H}_0,$ we obtain that $p_i$ is distributed as  $\text{Un}[0,1]$ for each $i\in \mathcal{G}.$ Using this fact, and the fact that $H$ is monotone increasing, we obtain for any $x\in \mathbb{R}:$
\begin{align*}
    \PP{X_i\leq x}=\PP{p_i\leq H(x)}=H(x).
\end{align*}
Hence, the density function of $X_i$ is $H',$ which is log-concave. In addition, the combined $p$-value $T(\bm p_{\mathcal{G}})$ is given by $F(G(\sum_{i\in \mathcal{G}}X_i),$ where $F\circ G$ is a non-decreasing function, since $G$ is increasing, and $F$ is nondecreasing. Finally, for each $i\in\mathcal{G},$ $p_i=H(X_i)$ is a coordinate-wise non-decreasing function of $(X_1, \ldots, X_m),$ and $X_1, \ldots, X_m$ are independent because $p_1, \ldots, p_m$ are assumed to be independent.  
Thus, by applying Theorem \ref{thm-smooth} with $n=m,$ $M=m+1,$ $\mathcal{S}=\mathcal{G},$ $g=F\circ G,$ $a_i=1$ for $i\in \mathcal{G},$ and $f_j(X_1, \ldots, X_m)=H(X_{j-1})$ for $j= 2, \ldots, m+1,$ 
we obtain that the vector $(T(\bm p_{\mathcal{G}}), p_1, \ldots, p_{m})$ is PRDS with respect to $T(\bm p_{\mathcal{G}}).$ 
Using this result, we shall obtain the inequality in (\ref{desired-res-lemma}) from Lemma 1(b) of \cite{pfilter2}, or equivalently from Proposition 3.6 of \cite{blanchard2008two}. 
We state this lemma below.
\begin{alemma}[\cite{pfilter2}]\label{grouplemma}
    Let $\bm p=(p_1, \ldots, p_n)$ be a vector of $p$-values corresponding to the hypotheses $H_1, \ldots, H_n$. Let $H_i$ be a true null hypothesis with $p$-value $p_i.$ 
    If $h:[0,1]^n\rightarrow [0, \infty)$ is a coordinate-wise non-increasing function, and $\bm p$ is PRDS with respect to $p_i,$ 
    then  $\EE{\ind\{p_i\leq h(\bm p)\}/h(\bm p)}\leq 1.$%
\end{alemma}
Define a vector $\bm p'=(T(\bm p_{\mathcal{G}}), p_1, \ldots, p_{m})\in[0,1]^{m+1}$ and a function $h':[0,1]^{m+1}\rightarrow[0, \infty)$ as follows: $h'(x_1, \ldots, x_{m+1})=f(x_2, \ldots, x_{m+1}),$ so that $h'(\bm p')=f(p_1, \ldots, p_m)=f(\bm p).$ Since $f$ is nonincreasing in each of its arguments,   $h'$ is also nonincreasing in each of its arguments. Since it is assumed that the $p$-values for true null hypotheses are distributed uniformly on $[0,1],$ and $\mathcal{G}\subseteq\mathcal{H}_0,$ we obtain from Lemma 1 of \cite{smoothed} that the distribution of $T(\bm p_{\mathcal{G}})$ is stochastically lower bounded by a uniform distribution on $[0,1].$ Applying Lemma \ref{grouplemma} with $(p_1 \ldots, p_n)=(T(\bm p_{\mathcal{G}}), p_1, \ldots, p_{m}),$ $h=h',$ and with index $i=1,$ we obtain (\ref{desired-res-lemma}), which completes the proof for the case where the combining function $T$ satisfies part (a) of Condition 2. 
    
    Now we assume that $T$ satisfies part (b) of  Condition 2, i.e., 
    $T(\bm p_{\mathcal{G}})=F(G(p_{(i);\mathcal{G}})),$ where $p_{(i);\mathcal{G}}$ is the $i$-th order statistic of $\bm p_{\mathcal{G}},$ $G: \mathbb{R}\mapsto \mathbb{R}$ is a monotone increasing function, and $F(x)=\PP{G(U_{(i)})\leq x},$ where $U_1, \ldots, U_{|\mathcal{G}|}$ are independent and identically distributed as $\text{Un}[0,1],$ 
    and $U_{(i)}$ is the $i$-th order statistic of $U_1, \ldots, U_{|\mathcal{G}|}.$ We need to prove that (\ref{desired-res-lemma}) holds. The proof is based on the result of \cite{smoothed}, which is given below.
    \begin{atheorem}[Theorem 1.3 in the Supplementary Material of \cite{smoothed}] \label{smooth-2}
    Let $X_1, \ldots, X_n$ be random variables on $\mathbb{R}$ or $\mathbb{Z},$ and $\mathcal{S}\subset \{1, \ldots, n\}$ such that $\{X_i: i \in \mathcal{S}\}$ are independent and identically distributed, and $\{X_i: i \in \mathcal{S}\}$ are independent of $\{X_i: i \notin \mathcal{S}\}.$
Let \begin{align*}
Y_1=g(X_{(k);\mathcal{S}}),
\end{align*}
where $g$ is a nondecreasing function,  and $X_{(k);\mathcal{S}}$ denotes the $k$-th order statistic of $\{X_i: i \in \mathcal{S}\}.$ For any $j=2, \ldots,M,$ let $$Y_j=f_j(X_1, \ldots, X_n)$$ for some nondecreasing function $f_j.$ Then $(Y_1, \ldots, Y_M)$ are PRDS with respect to $Y_1.$
    \end{atheorem}
The proof of Theorem \ref{smooth-2}, given in \cite{smoothed}, relies on the result of \cite{block1987probability}.  
Note that the assumptions of Theorem \ref{smooth-2} are satisfied for $n=m,$ $\mathcal{S}=\mathcal{G},$ and $X_j=p_j$ for $j=1, \ldots, m,$  because according to the assumptions of item (ii) of Lemma \ref{main-lemma}, $p_1, \ldots, p_m$ are independent, and $p_j$ is distributed as $\text{Un}[0,1]$ for $j\in \mathcal{G},$ since $\mathcal{G}\subseteq\mathcal{H}_0.$ According to our assumption, $T(\bm p_{\mathcal{G}})=F(G( p_{(i);\mathcal{G}})),$ where $F\circ G$ is a nondecreasing function.  Hence, by applying Theorem \ref{smooth-2} with $n=m,$ $M=m+1,$  $X_j=p_j$ for $j=1, \ldots, m,$ $k=i,$ $\mathcal{S}=\mathcal{G},$ $g=F\circ G,$ 
  $f_j(X_1, \ldots, X_m)=X_{j-1}$ for $j= 2, \ldots, m+1,$ we obtain that the vector $(T(\bm p_{\mathcal{G}}), p_{1}, \ldots, p_{m})$ is PRDS with respect to $T(\bm p_{\mathcal{G}}).$ As we showed above, this yields (\ref{desired-res-lemma}), which 
    completes the proof for combining functions satisfying part (b) of Condition 2.
    \section{Proof of Theorem \ref{thm-gen-dep}}  \label{sec:proof:gen}
The proof of items (i) and (ii) is similar to the proof of statement (d) of Theorem 1 of \cite{pfilter2} and to the proof of items (ii) and (iii) of Theorem \ref{thm:hier:SM}. Let $\mathcal{U}^*_{\beta}(\bm p) $ be the set of indices of rejected hypotheses by Weighted Focused Reshaped BH with a given reshaping function $\beta$ and weight functions $\{\hat{w}_{d,i}: d=1, \ldots, D, i=1, \ldots, |\mathcal{H}_d|\}$ satisfying Assumption 3. In the proof of each of the items, we shall use the fact that Weighted Reshaped Focused BH is self-consistent with respect to the threshold collection  $\Delta(\bm p, r, i)=\beta(r)q/(m\hat{w}_i(\bm p)).$ The proof of this fact follows from similar arguments to those that we used in the proof of Theorem 1 to show that WFBH is self-consistent with respect to the threshold collection  $\Delta(\bm p, r, i)=rq/(m\hat{w}_i(\bm p)).$ 
We use the notations from Appendix~\ref{Reshaped} and introduce some additional notations. For each depth $d\in\{1, \ldots, D\},$ let $\bm p^d$ be the vector of original $p$-values for hypotheses with depth $d$, i.e.,  $\bm p^d=(p_{d,1},\ldots, p_{d,|\mathcal{H}_d|}).$  Similarly to the proof of Proposition 1 and the proof of items (ii) and (iii) of Theorem \ref{thm:hier:SM}, for convenience, we convert the pairs of indices $(d,i)$ to the original indices in $\mathcal{H},$ according to the one-to-one correspondence between them. After conversion, we have the weight functions $\hat{w}_1, \ldots, \hat{w}_m,$ the vector $\bm p^d$ that contains the original $p$-values for hypotheses with indices in $\mathcal{H}_d,$ and the vector $\tilde{\bm p}^d$ that contains the smoothed $p$-values for hypotheses with indices in $\mathcal{H}_d,$ for $d=1, \ldots, D.$ Recall that according to Assumption 3 regarding the weight functions, for each $i\in \mathcal{H}_d,$ we have $\hat{w}_i(\bm p)= \hat{w}_i(\bm p^d)$ and $\hat{w}_i(\tilde{\bm p})= \hat{w}_i(\tilde{\bm p}^d).$ 

\paragraph{Proof of item (i)}
 Replacing smoothed $p$-values by the original $p$-values, $\mathcal{U}^*(\tilde{\bm p})$ by $\mathcal{U}_{\beta}^{*}(\bm p),$ and $|\mathcal{U}^*(\tilde{\bm p})|$ by $\beta(|\mathcal{U}_{\beta}^*(\bm p)|)$ in the expressions that lead to (\ref{thm2-term2-item2}) in the proof of items (ii) and (iii) of Theorem \ref{thm:hier:SM}, we obtain the following inequality for the FDR of Weighted Reshaped Focused BH:
\begin{align}
&\text{FDR}\leq\notag\\&\sum_{d\in \mathcal{D}_c}\sum_{i\in \mathcal{H}_d^0}\EE{\frac{\ind\{p_{i}\leq q\beta(|\mathcal{U}_{\beta}^{*}(\bm p)|)/m\}}{|\mathcal{U}_{\beta}^{*}(\bm p)|}}+\sum_{d\in \mathcal{D}_w}\sum_{i\in \mathcal{H}_d^0}\EE{\frac{\ind\{p_{i}\leq q\beta(|\mathcal{U}_{\beta}^{*}(\bm p)|)/\{m\hat{w}_{i}(\bm p^d)\}}{|\mathcal{U}_{\beta}^{*}(\bm p)|}}\label{sum-reshaped}
\end{align}
Let us first evaluate the first term in (\ref{sum-reshaped}).
\begin{align}
    \sum_{d\in \mathcal{D}_c}\sum_{i\in \mathcal{H}_d^0}\EE{\frac{\ind\{p_{i}\leq q\beta(|\mathcal{U}_{\beta}^{*}(\bm p)|)/m\}}{|\mathcal{U}_{\beta}^{*}(\bm p)|}}&=\frac{q}{m}\sum_{d\in \mathcal{D}_c}\sum_{i\in \mathcal{H}_d^0}\EE{\frac{\ind\{p_{i}\leq q\beta(|\mathcal{U}_{\beta}^{*}(\bm p)|)/m\}}{q|\mathcal{U}_{\beta}^{*}(\bm p)|/m}}\notag\\&\leq \frac{q}{m}\sum_{d\in \mathcal{D}_c}|\mathcal{H}_d^0|\leq \frac{q}{m}\sum_{d\in \mathcal{D}_c}|\mathcal{H}_d|.\label{upper-1}
\end{align}
The first  inequality in (\ref{upper-1}) follows from Proposition 3.7 of \cite{blanchard2008two}, which implies   that for each $d\in \mathcal{D}_c$ and $i\in\mathcal{H}_d^0,$\begin{align*}\EE{\frac{\ind\{p_{i}\leq q\beta(|\mathcal{U}_{\beta}^{*}(\bm p)|)/m\}}{q|\mathcal{U}_{\beta}^{*}(\bm p)|/m}}\leq 1.\end{align*}  
Let us now evaluate the second term in (\ref{sum-reshaped}). 
Since the weight functions are assumed to satisfy Assumption 3, they are nondecreasing in each coordinate of $\bm p^d$, therefore for each $i\in\mathcal{H}_d,$ 
$\hat{w}_{i}(\bm p^d_{0,i})\leq \hat{w}_{i}(\bm p^d),$ where $\bm p^d_{0,i}$ is the vector $\bm p^d$ with $p_{i}$ replaced by 0. Therefore,
\begin{align}
    &\sum_{d\in \mathcal{D}_w}\sum_{i\in \mathcal{H}_d^0}\EE{\frac{\ind\{p_{i}\leq q\beta(|\mathcal{U}_{\beta}^{*}(\bm p)|)/\{m\hat{w}_{i}(\bm p^d)\}}{|\mathcal{U}_{\beta}^{*}(\bm p)|}}\leq \notag\\&\sum_{d\in \mathcal{D}_w}\sum_{i\in \mathcal{H}_d^0}\EE{\frac{\ind\{p_{i}\leq q\beta(|\mathcal{U}_{\beta}^{*}(\bm p)|)/\{m\hat{w}_{i}(\bm p_{0,i}^d)\}}{|\mathcal{U}_{\beta}^{*}(\bm p)|}}
    \leq\notag\\& \frac{q}{m}\sum_{d\in \mathcal{D}_w}\sum_{i\in \mathcal{H}_d^0}\EE{\frac{1}{\hat{w}_{i}(\bm p^d_{0,i})}\EEst{\frac{\ind\{p_{i}\leq c_{i}\beta(|\mathcal{U}_{\beta}^{*}(\bm p)|)\}}{c_{i}|\mathcal{U}_{\beta}^{*}(\bm p)|}}{\bm p_{-i}^d}},\label{sec-term-last}
\end{align}
 where for each $d\in \mathcal{D}_w$ and $i\in\mathcal{H}_d^0,$ $\bm p_{-i}^d$ is the vector $\bm p^d$ excluding $p_{i},$ 
 and $c_{i}=q/\{m\hat{w}_{i}(\bm p^d_{0,i})\}.$ The last inequality follows from the fact that for any $d\in\mathcal{D}_w$ and $i\in \mathcal{H}_d^0,$ after fixing $\bm p_{-i}^d$  $\hat{w}_{i}(\bm p_{0,i}^d)$ is a constant.
We note that the two following facts are true for $d\in\mathcal{D}_w$ and $i\in \mathcal{H}_d^0:$
 \begin{itemize}
     \item[(a)] $c_i$ is a constant after fixing $\bm p_{-i}^d.$
     \item[(b)] The conditional distribution of $p_{i}$ given $\bm p_{-i}^d$ is the same as its marginal distribution (which is superuniform), since according to the assumption of item (i), $p_{i}$ is independent of $\bm p_{-i}^d.$ 
 \end{itemize}
Therefore, according to Proposition 3.7 of \cite{blanchard2008two}, we obtain for each $d\in\mathcal{D}_w$ and $i\in\mathcal{H}_d^0:$ 
\begin{align}\EEst{\frac{\ind\{p_{i}\leq c_{i}\beta(|\mathcal{U}_{\beta}^{*}(\bm p)|)\}}{c_{i}|\mathcal{U}_{\beta}^{*}(\bm p)|}}{\bm p_{-i}^d}\leq 1.\label{BR}
\end{align}
 Using (\ref{BR}) for each $d\in \mathcal{D}_w$ and $i\in \mathcal{H}_d^0,$ we obtain from 
 (\ref{sec-term-last}) that 
 \begin{align}
      &\sum_{d\in \mathcal{D}_w}\sum_{i\in \mathcal{H}_d^0}\EE{\frac{\ind\{p_{i}\leq q\beta(|\mathcal{U}_{\beta}^{*}(\bm p)|)/\{m\hat{w}_{i}(\bm p^d)\}}{|\mathcal{U}_{\beta}^{*}(\bm p)|}}\leq \frac{q}{m}\sum_{d\in \mathcal{D}_w}\sum_{i\in \mathcal{H}_d^0}\EE{\frac{1}{\hat{w}_{i}(\bm p_{0,i}^d)}}.\label{sec-term-2}
 \end{align}
 Since the $p$-values in the set 
 $\bm p^d$ are assumed to be valid and independent for each $d\in \mathcal{D}_w,$ we obtain from (\ref{inter-imp}) in the proof of Proposition 1  that for each $d\in \mathcal{D}_w$ the following inequality holds:
$$\sum_{i\in\mathcal{H}_d^0}\EE{\frac{1}{\hat{w}_{i}(\bm p_{0,i}^d)}}\leq |\mathcal{H}_d|.$$
 Combining this inequality with (\ref{sec-term-2}), we obtain
 \begin{align}
      \sum_{d\in \mathcal{D}_w}\sum_{i\in \mathcal{H}_d^0}\EE{\frac{\ind\{p_{i}\leq q\beta(|\mathcal{U}_{\beta}^{*}(\bm p)|)/\{m\hat{w}_{i}(\bm p^d)\}}{|\mathcal{U}_{\beta}^{*}(\bm p)|}}\leq \frac{q}{m}\sum_{d\in \mathcal{D}_w}|\mathcal{H}_d|.\label{upper-sec-term}
 \end{align}
 Now, using the upper bound for the first term in (\ref{sum-reshaped}), given in (\ref{upper-1}), and the upper bound for the second term in (\ref{sum-reshaped}), given in (\ref{upper-sec-term}), we obtain an upper bound for the FDR of the Weighted Reshaped Focused BH procedure:
 \begin{align*}
     \text{FDR}\leq \frac{q}{m}\sum_{d\in \mathcal{D}_c}|\mathcal{H}_d|+\frac{q}{m}\sum_{d\in \mathcal{D}_w}|\mathcal{H}_d|= \frac{q}{m}\sum_{d=1}^D|\mathcal{H}_d|=q.
 \end{align*}
 This completes the proof of item (i).
 \paragraph{Proof of item (ii)}
 The FDR of the Weighted Reshaped Focused BH procedure applied on the smoothed $p$-values $\tilde{\bm p}$ can be written as a sum of two terms, similarly to (\ref{sum-reshaped}):
 \begin{align}
&\text{FDR}\leq
\notag\\&\sum_{d\in \mathcal{D}_c}\sum_{i\in \mathcal{H}_d^0}\EE{\frac{\ind\{\tilde{p}_{i}\leq q\beta(|\mathcal{U}_{\beta}^{*}(\tilde{\bm p})|)/m\}}{|\mathcal{U}_{\beta}^{*}(\tilde{\bm p})|}}+\sum_{d\in \mathcal{D}_w}\sum_{i\in \mathcal{H}_d^0}\EE{\frac{\ind\{\tilde{p}_{i}\leq q\beta(|\mathcal{U}_{\beta}^{*}(\tilde{\bm p})|)/\{m\hat{w}_{i}(\tilde{\bm p}^d)\}}{|\mathcal{U}_{\beta}^{*}(\tilde{\bm p})|}}\label{sum-item2}
\end{align}
 Since it is assumed that the logical relationships in Assumption 1 hold, we obtain that for each $i\in\mathcal{H}_d^0,$  $H_i$ along with all its descendants, 
 $\{H_j, j\in Desc(i)\},$ are true null hypotheses, therefore their intersection, $H_i\cap (\cap_{j\in Desc(i)}H_j)$ is also a true null hypothesis. Hence, according to the assumption regarding the combination method which is used for obtaining  smoothed $p$-values, for each $i\in \mathcal{H}_d^0,$ the smoothed $p$-value $\tilde{p}_{i}=T(\bm p_{i, Desc(i)})$ 
 satisfies $\PP{\tilde{p}_{i}\leq x}\leq x$ for all $x\in[0,1].$ This shows that the smoothed $p$-values are valid. Therefore, by applying item (iii) of Lemma \ref{main-lemma} we obtain an upper bound for the first term in (\ref{sum-item2}):
 \begin{align}
     \sum_{d\in \mathcal{D}_c}\sum_{i\in \mathcal{H}_d^0}\EE{\frac{\ind\{\tilde{p}_{i}\leq q\beta(|\mathcal{U}_{\beta}^{*}(\tilde{\bm p})|)/m\}}{|\mathcal{U}_{\beta}^{*}(\tilde{\bm p})|}}&= \frac{q}{m}\sum_{d\in \mathcal{D}_c}\sum_{i\in \mathcal{H}_d^0}\EE{\frac{\ind\{\tilde{p}_{i}\leq q\beta(|\mathcal{U}_{\beta}^{*}(\tilde{\bm p})|)/m\}}{q|\mathcal{U}_{\beta}^{*}(\tilde{\bm p}|)/m}}\notag\\&\leq \frac{q}{m}\sum_{d\in \mathcal{D}_c}|\mathcal{H}_d^0|\leq \frac{q}{m}\sum_{d\in \mathcal{D}_c}|\mathcal{H}_d|.\label{upper-first-reshaped}
 \end{align}
 Let us now derive an upper bound for the second term in (\ref{sum-item2}), using the notations introduced in~ \ref{Reshaped}.
 \begin{align}
    & \sum_{d\in \mathcal{D}_w}\sum_{i\in \mathcal{H}_d^0}\EE{\frac{\ind\{\tilde{p}_{i}\leq q\beta(|\mathcal{U}_{\beta}^{*}(\tilde{\bm p})|)/\{m\hat{w}_{i}(\tilde{\bm p}^d)\}}{|\mathcal{U}_{\beta}^{*}(\tilde{\bm p})|}}\leq \label{first-ineq}\\&\sum_{d\in \mathcal{D}_w}\sum_{i\in \mathcal{H}_d^0}\EE{\frac{\ind\{\tilde{p}_{i}\leq q\beta(|\mathcal{U}_{\beta}^{*}(\tilde{\bm p})|)/\{m\hat{w}_{i}(\tilde{\bm p}_{0,i}^d)\}}{|\mathcal{U}_{\beta}^{*}(\tilde{\bm p})|}}=\notag\\&\sum_{d\in \mathcal{D}_w}\sum_{i\in \mathcal{H}_d^0}\EE{\EEst{\frac{\ind\{T(\bm p_{i, Desc(i)})\leq q\beta(|\mathcal{U}_{\beta}^{*}(\tilde{\bm p})|)/\{m\hat{w}_{i}(\tilde{\bm p}_{0,i}^d)\}}{|\mathcal{U}_{\beta}^{*}(\tilde{\bm p})|}}{\bm p_{-i, Desc(i)}}}\label{beinaim}
 \end{align}
 The inequality in (\ref{first-ineq}) follows from the assumption that the weight functions are coordinate-wise non-decreasing. 
 For each $d\in\mathcal{D}_w$ and $i\in \mathcal{H}_d^0,$ let us denote $\tilde{c}_{i}=q/\{m\hat{w}_{d,i}(\tilde{\bm p}_{0,i}^d)\}.$ We showed in the proof of items (ii) and (iii) of Theorem \ref{thm:hier:SM} that for each $d\in\mathcal{D}_w$ and $i\in \mathcal{H}_d^0,$ after fixing $\bm p_{-i, Desc(i)},$ 
 $\hat{w}_{i}(\tilde{\bm p}_{0,i}^d)$ is a constant;  therefore, $\tilde{c}_i$ is also a constant. 
 Using these facts, we obtain
 \begin{align}
     &\sum_{d\in \mathcal{D}_w}\sum_{i\in \mathcal{H}_d^0}\EE{\EEst{\frac{\ind\{T(\bm p_{i, Desc(i)})\leq q\beta(|\mathcal{U}_{\beta}^{*}(\tilde{\bm p})|)/\{m\hat{w}_{i}(\tilde{\bm p}_{0,i}^d)\}}{|\mathcal{U}_{\beta}^{*}(\tilde{\bm p})|}}{\bm p_{-i, Desc(i)}}}=\notag\\&\frac{q}{m}\sum_{d\in \mathcal{D}_w}\sum_{i\in \mathcal{H}_d^0}\EE{\frac{1}{\hat{w}_{i}(\tilde{\bm p}_{0,i}^d)}\EEst{\frac{\ind\{T(\bm p_{i, Desc(i)})\leq \tilde{c}_{i}\beta(|\mathcal{U}_{\beta}^{*}(\tilde{\bm p})|)\}}{\tilde{c}_{i}|\mathcal{U}_{\beta}^{*}(\tilde{\bm p})|}}{\bm p_{-i, Desc(i)}}}\leq\label{thm2-item2-first}\\&\frac{q}{m}\sum_{d\in \mathcal{D}_w}\sum_{i\in \mathcal{H}_d^0}\EE{\frac{1}{\hat{w}_{i}(\tilde{\bm p}_{0,i}^d)}}\leq \frac{q}{m}\sum_{d\in \mathcal{D}_w}|\mathcal{H}_d|.\label{thm2-item2-second}
 \end{align}
 Let us show that the inequality in (\ref{thm2-item2-first}) holds. As we noted above, for each $d\in\{1, \ldots, D\}$ and $i\in \mathcal{H}_d^0,$ the smoothed $p$-value $T(\bm p_{i, Desc(i)})$ is superuniform. Since it is assumed that for each $d\in \mathcal{D}_w$ and $i\in \mathcal{H}_d^0,$ the vector $\bm p_{i, Desc(i)}$ is independent of $\bm p_{-i, Desc(i)},$ the conditional distribution of $T(\bm p_{i, Desc(i)})$ given $\bm p_{-i, Desc(i)}$ is the same as its marginal distribution, therefore it is also superuniform. Hence, the inequality in (\ref{thm2-item2-first}) follows by applying item (iii) of Lemma \ref{main-lemma} for each $d\in\mathcal{D}_w$ and $i\in \mathcal{H}_d^0,$ with $c=\tilde{c}_{i}$ and $\mathcal{G}=\{i\}\cup Desc(i),$ which shows that the conditional expectation in (\ref{thm2-item2-first}) is upper bounded by one. 
 In order to show that the inequality in (\ref{thm2-item2-second}) holds, it is enough to show that the smoothed $p$-values, received as input to the weight functions, satisfy the assumptions of  Proposition 1, because this implies that (\ref{inter-imp}) holds for smoothed $p$-values, i.e., for each $d\in\mathcal{D}_w:$ 
$$\sum_{i\in\mathcal{H}_d^0}\EE{\frac{1}{\hat{w}_{i}(\tilde{\bm p}_{0,i}^d)}}\leq |\mathcal{H}_d|.$$ We showed above the the smoothed $p$-values are valid, therefore,  it is enough to show that for each $d\in\mathcal{D}_w,$ the $p$-values in the set $\{\tilde{p}_{d,i}, i=1, \ldots, |\mathcal{H}_d|\}$ are independent.  
The latter holds since it is assumed that $\mathcal{D}_w\subseteq \mathcal{D}_t,$ and for each $d\in \mathcal{D}_w$ and $i\in\mathcal{H}_d,$ the $p$-values in the set $\bm p_{i, Desc(i)}$ are independent of the $p$-values in the set $\bm p^{d, Desc}_{-i}.$ 
This completes the proof of the inequality in (\ref{thm2-item2-second}). 
Combining (\ref{thm2-item2-second}) and (\ref{beinaim}), we obtain an upper bound for the second term in  (\ref{sum-item2}):
\begin{align}
    & \sum_{d\in \mathcal{D}_w}\sum_{i\in \mathcal{H}_d^0}\EE{\frac{\ind\{\tilde{p}_{i}\leq q\beta(|\mathcal{U}_{\beta}^{*}(\tilde{\bm p})|)/\{m\hat{w}_{i}(\tilde{\bm p}^d)\}}{|\mathcal{U}_{\beta}^{*}(\tilde{\bm p})|}}\leq \frac{q}{m}\sum_{d\in \mathcal{D}_w}|\mathcal{H}_d|.\label{upper-second}
    \end{align}
Now, similarly to the proof of item (i), we obtain an upper bound for FDR using the upper bounds (\ref{upper-first-reshaped}) and (\ref{upper-second}) for  the first and the second term in (\ref{sum-item2}), respectively:
\begin{align*}
    \text{FDR}\leq  \frac{q}{m}\sum_{d\in \mathcal{D}_c}|\mathcal{H}_d|+\frac{q}{m}\sum_{d\in \mathcal{D}_w}|\mathcal{H}_d|=
    \frac{q}{m}\sum_{d=1}^D|\mathcal{H}_d|=q.
\end{align*}
This completes the proof of item (ii).
\section{Theoretical results for intersection DAGs}\label{sec:intersec}
We consider the setting addressed in Section 3.1 in \cite{filtering}. Assume that we have a set of $K$ items, indexed $1, \ldots, K,$ and a DAG with $m$ nodes, so that each node $i\in\{1, \ldots, m\}$ is associated with a set of items $\mathcal{A}_i\subseteq\{1, \ldots, K\}.$ Assume that if $H_i$ is a parent of $H_j,$ then $\mathcal{A}_j\subseteq \mathcal{A}_i.$ Each item $j\in\{1, \ldots, K\}$ is associated with an item-level hypothesis $H_j^{item},$ and 
\begin{align*}H_i=\cap_{j\in \mathcal{A}_i}H_j^{item},\label{inters}\end{align*}
that is, $H_i$ is the intersection of  the item-level hypotheses for items in $\mathcal{A}_i,$ for $i=1, \ldots, m.$ \cite{filtering} describe this setting in the context of the example we considered in the Introduction: a gene expression study with hypotheses addressing the gene sets which have the GO structure. In this example, the items are the genes, each node is annotated with a gene set so that child nodes are annotated to subsets of their parent nodes, and the item-level null hypothesis for each gene states that it is not associated with the disease. The self-contained hypothesis (\cite{goeman2007analyzing}) for each node is the intersection of its item-level null hypotheses. 
A DAG where each parent hypothesis is an intersection of all its leaf descendants also satisfies the structure we describe above, with leaf hypotheses as the item-level hypotheses.

The vector of $K$ $p$-values for the item-level hypotheses is denoted by $\bm p^{item}.$ For $i=1, \ldots, m,$  the $p$-value for $H_i,$ denoted by $p_i,$ is obtained by combining the $p$-values for its item-level hypotheses, $\{p_j^{item}, j\in\mathcal{A}_i\},$ using a combining method which gives a valid  $p$-value for $\cap_{j\in \mathcal{A}_i} H_j^{item}$ under the dependency between the $p$-values in the set $\{p_j^{item}, j\in\mathcal{A}_i\}.$ Recall that such combining methods are used for smoothing the $p$-values in DAGs satisfying Assumption 1. Our theoretical results regarding FDR control of WFBH when applied on smoothed $p$-values can be easily adjusted for addressing intersection DAGs. We introduce several notations and then present theoretical results for intersection DAGs. 

Let $\mathcal{D}^{int}_t\subseteq\{1, \ldots, D\}$ be the set of depths defined as follows:
\begin{align*}
    \mathcal{D}^{int}_t=\{d\in\{1, \ldots, D\}: \forall i,j\in \mathcal{H}_d, \mathcal{A}_i\cap \mathcal{A}_j=\emptyset\}.
\end{align*}
So, for each depth $d\in \mathcal{D}^{int}_t$ and any two nodes with depth $d,$ the sets of items associated with these nodes are non-overlapping. For each $d\in \mathcal{D}_t^{int}$ and $i\in \mathcal{H}_d,$ let $\bm p^{item}_{\mathcal{A}_i}$
be the set of item-level $p$-values for items belonging to $\mathcal{A}_i,$ i.e., 
 $\bm p^{item}_{\mathcal{A}_i}=\{p_j^{item}: j\in \mathcal{A}_i\},$ and let $\bm p^{item}_{d, -\mathcal{A}_i}$  be the item-level $p$-values for the remaining nodes with depth $d,$ i.e., $\bm p^{item}_{d, -\mathcal{A}_i}=(\cup_{j\in \mathcal{H}_d} \bm p_{\mathcal{A}_j}^{item})\setminus \bm p_{\mathcal{A}_i}^{item}.$ Denote by $\mathcal{H}_0^{item}\subseteq\{1, \ldots, K\}$ the set of indices of true null item-level hypotheses.  
\begin{atheorem}\label{thm-inters}
    Consider an intersection DAG as defined above, with valid item-level $p$-values $\bm p^{item}.$ We address below FBH, WFBH,  and Weighted Reshaped Focused BH when applied on $\bm p,$ with a filter $\F,$  and  weight functions satisfying Assumption 3 with a set $\mathcal{D}_w$ for WFBH and Weighted Reshaped Focused BH. 
 \begin{itemize}
        \item[(i)] Assume that the item-level $p$-values are independent;  $p_i$ is obtained by combining $\bm p_{\mathcal{A}_i}^{item}$ using Simes' combining function for $i=1, \ldots, m;$  $\mathcal{D}_w\subseteq \mathcal{D}_t^{int};$  $\F$ is monotonic. Then WFBH controls the FDR at level $q.$
        \item[(ii)] Assume that the item-level $p$-values are independent; for each $j\in \mathcal{H}_0^{item},$ $p_j^{item}\sim\text{Un}[0,1];$  $p_i$ is obtained by combining $\bm p_{\mathcal{A}_i}^{item}$ using any combining function satisfying Condition 2 for $i=1, \ldots, m;$ $\mathcal{D}_w\subseteq \mathcal{D}_t^{int};$  $\F$ is monotonic. Then WFBH controls the FDR at level $q.$
        \item[(iii)] Assume that the item-level $p$-values are PRDS on $\mathcal{H}_0^{item};$  $p_i$ is obtained by combining $\bm p_{\mathcal{A}_i}^{item}$ using Simes' combining function for $i=1, \ldots, m;$ $\F$ is monotonic. Then FBH   controls the FDR at level $q.$
        \item[(iv)] Assume that $\mathcal{D}_w\subseteq \mathcal{D}_t^{int};$ for each $d\in \mathcal{D}_w$ and $i\in \mathcal{H}_d,$ the item-level $p$-values in the set $\bm p^{item}_{\mathcal{A}_i}$ are independent of the item-level $p$-values in the set $\bm p^{item}_{d, -\mathcal{A}_i};$  $p_i$ is obtained using any combination method which gives a valid $p$-value for $H_i=\cap_{j\in \mathcal{A}_i} H_j^{item}$ under the dependency structure among the $p$-values in the set $\bm p^{item}_{\mathcal{A}_i};$ $\F$ is arbitrary. Then Weighted Reshaped Focused BH controls the FDR at level $q.$
    \end{itemize}
\end{atheorem}
Item (iii) gives the result of Theorem 3 of \cite{filtering}. The proofs of items (i), (ii), and (iv) are obtained by making the following changes in the proofs of items (ii) and (iii) of Theorem \ref{thm:hier:SM} and in the proof of item (ii) of Theorem \ref{thm-gen-dep}, respectively: $\bm p$ is replaced by $p^{item};$ for each $d\in \{1, \ldots, D\}$ and $i\in \mathcal{H}_d,$ $\bm p_{i, Desc(i)}$ is replaced by $\bm p^{item}_{\mathcal{A}_i},$
so $\tilde{p}_i=T(\bm p_{i, Desc(i)})$ is replaced by $p_i=T(\bm p^{item}_{\mathcal{A}_i}),$ and $\bm p_{-i}^{d, Desc}$ is replaced by $\bm p^{item}_{d, -\mathcal{A}_i}.$ While in the proofs of items (ii) and (iii) of Theorem \ref{thm:hier:SM} and of item (ii) of Theorem \ref{thm-gen-dep} we used Assumption 1 to obtain that for $i\in \mathcal{H}_0,$ $\{i\}\cup Desc(i)\subseteq \mathcal{H}_0,$ and applied Lemma \ref{main-lemma}  with $\mathcal{G}=\{i\}\cup Desc(i)$ and the original $p$-values, in the proofs of items (i), (ii), and (iv) of Theorem \ref{thm-inters} we obtain that for each $i\in \mathcal{H}_0,$ $\mathcal{A}_i\subseteq \mathcal{H}_0^{item},$ and apply Lemma \ref{main-lemma} with $\mathcal{G}=\mathcal{A}_i$ for the item-level $p$-values. Note that when the DAG is a tree  with independent leaf $p$-values, and each parent hypothesis is an intersection of all its leaf descendants, then: (1) $\mathcal{D}^{int}_t=\{1, \ldots, D\},$ so the assumption that $\mathcal{D}_w\subseteq \mathcal{D}^{int}_t$ holds for any $\mathcal{D}_w;$ (2) the assumption in item (iv), that for each $d\in \mathcal{D}_w$ and $i\in \mathcal{H}_d,$ the item-level $p$-values in the set $\bm p^{item}_{\mathcal{A}_i}$ are independent of the item-level $p$-values in the set $\bm p^{item}_{d, -\mathcal{A}_i},$ holds for any $\mathcal{D}_w.$

\section{Further Simulation Studies}
\begin{figure}
\centering
  \includegraphics[width=1\linewidth]{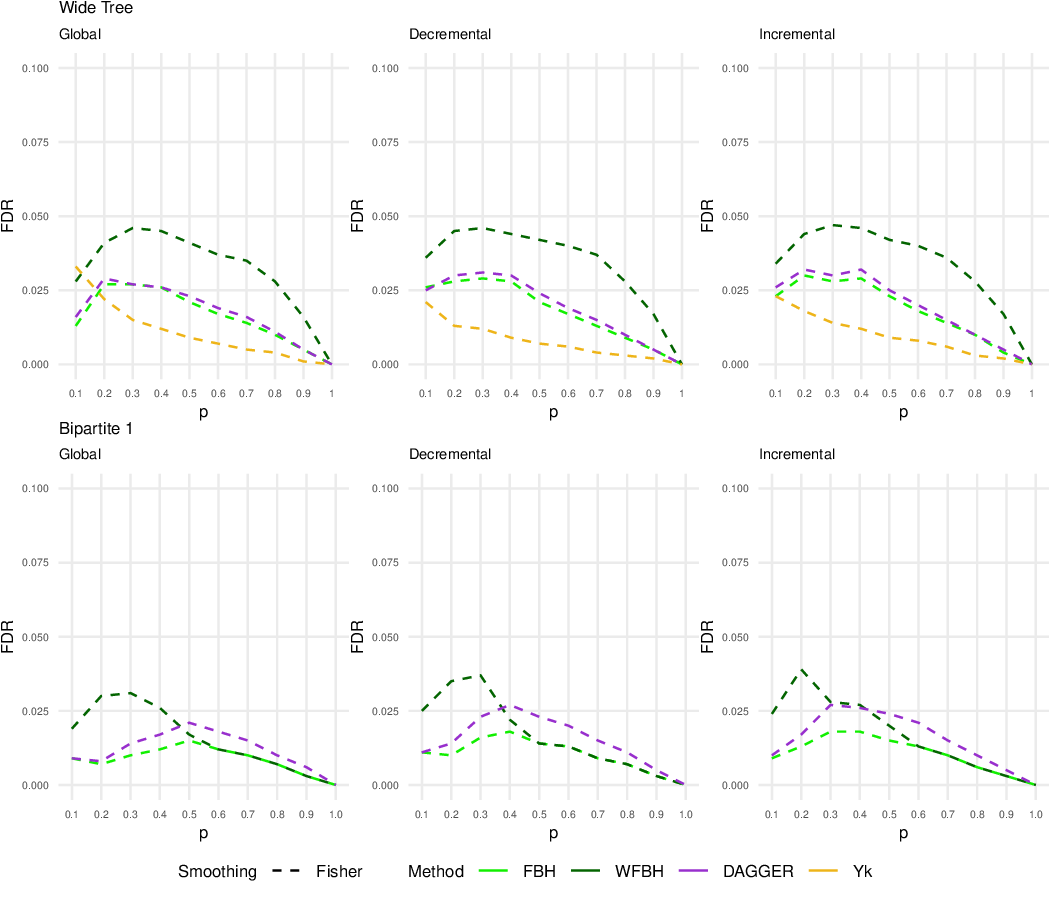}
\caption{Estimated FDR for wide tree and bipartite 1 setups, when original $p$-values are independent, and the methods are applied to Fisher's smoothed $p$-values (the settings are described in Section 4 in the main manuscript).  All the methods target FDR control at level $q=0.05;$ $\lambda=0.5$ for WFBH.}
\label{fig_fdr_ind_tree_bip2}
\end{figure}

We show additional simulation results for the methods addressed in Section 4, under both independence and positive dependence. 
We first address the settings described in Section 4. As noted there, in these settings Yekutieli's method applied to Fisher's smoothed $p$-values for the wide tree and WFBH applied to the same $p$-values for the bipartite graph 1 do not have theoretical FDR control guarantees. Figure \ref{fig_fdr_ind_tree_bip2} shows the estimated FDR (based on 200 iterations) of the FBH, WFBH, DAGGER, and Yekutieli's procedures  when applied to Fisher's smoothed $p$-values, at varying levels of the proportion of non-null leaves $p$, under the settings of Section 4. 
We observe that all methods control FDR conservatively at $q=0.05$ under the global, decremental, and incremental setups. 
We further consider two more graph structures given below along with the choices of $\mathcal{D}_w$ for the weights of the WFBH method:
\begin{enumerate}[label=(A\arabic*)] 
\item[(G1)] Deep tree: 
A DAG with $555$ nodes distributed over three depths, $5$ nodes are roots, each of them has $10$ children at depth 2; each node at depth 2 has $10$ children at depth 3. Each node at depths 2 and 3 has one parent. For WFBH, we set $\mathcal{D}_w=\{1,2,3\},$ i.e., all the depths are assigned with data-dependent weights.
\item[(G2)] Bipartite graph 2: 
A DAG of $551$ nodes, with $61$ roots and $490$ leaves. Each root has $10$ children and $370$ leaves have one parent each, and $120$ leaves have two parents. For WFBH, we set $\mathcal{D}_w=\{1,2\},$ i.e., all the depths are assigned with data-dependent weights.
\end{enumerate}
For these graphs, null/non-null status of each hypothesis and the strength of signals for non-null hypotheses is determined as described in Section 4, addressing three setups - global, incremental, and decremental. The $p$-values are also obtained as described in Section 4.  

\begin{figure}
\centering
  \includegraphics[width=1\linewidth]{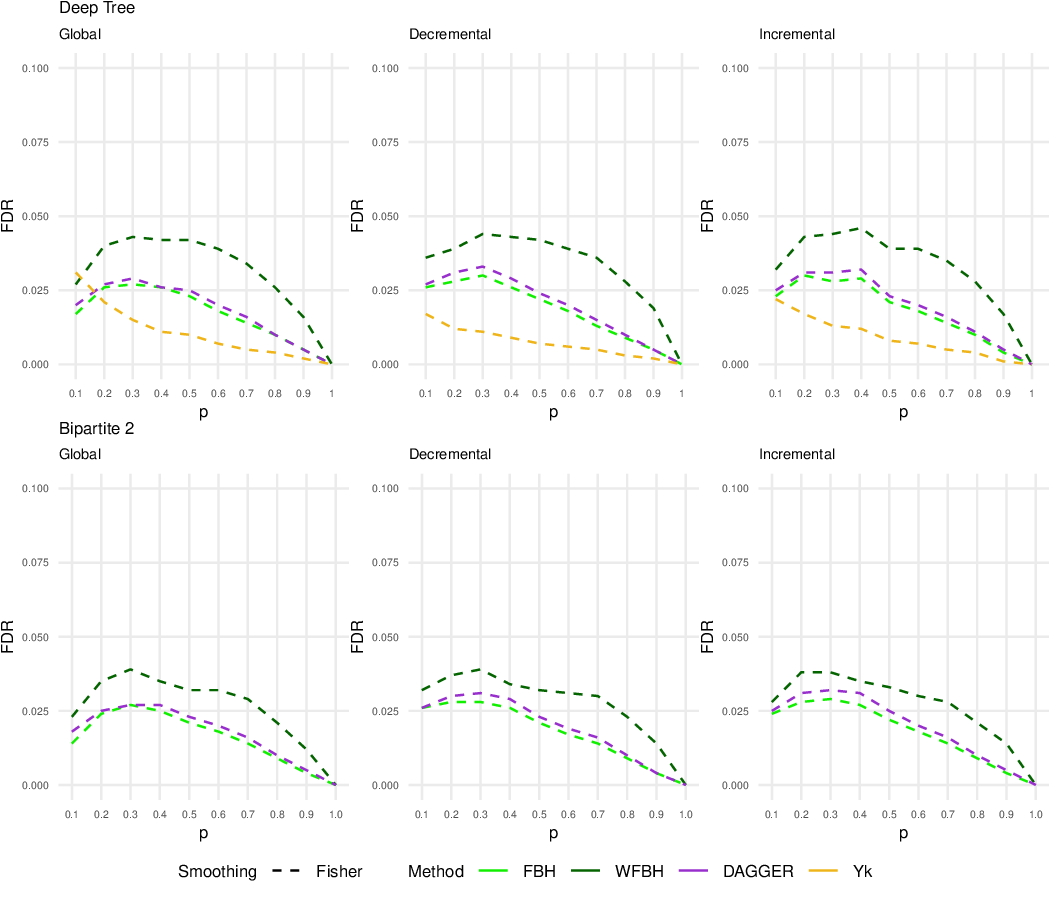}
\caption{Estimated FDR for deep tree and bipartite 2 setups,  when original $p$-values are independent, and the methods are applied to Fisher's smoothed $p$-values. All the methods target FDR control at level $q=0.05;$ $\lambda=0.5$ for WFBH. }
\label{fig_fdr_deeptree_bipv2}
\end{figure}

Figure \ref{fig_fdr_deeptree_bipv2} shows that at varying non-null proportion of leaves $p$, FDR is controlled by all the methods 
for the deep tree and bipartite 2 graph structures, when applied to Fisher's smoothed $p$-values obtained from independent original $p$-values. 
This provides additional empirical evidence of FDR control for WFBH  applied to Fisher's smoothed $p$-values for a DAG that is not a tree, as well as for Yekutieli's method applied to Fisher's smoothed $p$-values for 
 a tree (recall that these are the only two methods addressed in Figure \ref{fig_fdr_deeptree_bipv2} for which we do not have theoretical FDR control guarantees). 

\begin{figure}
\centering
  \includegraphics[width=1\linewidth]{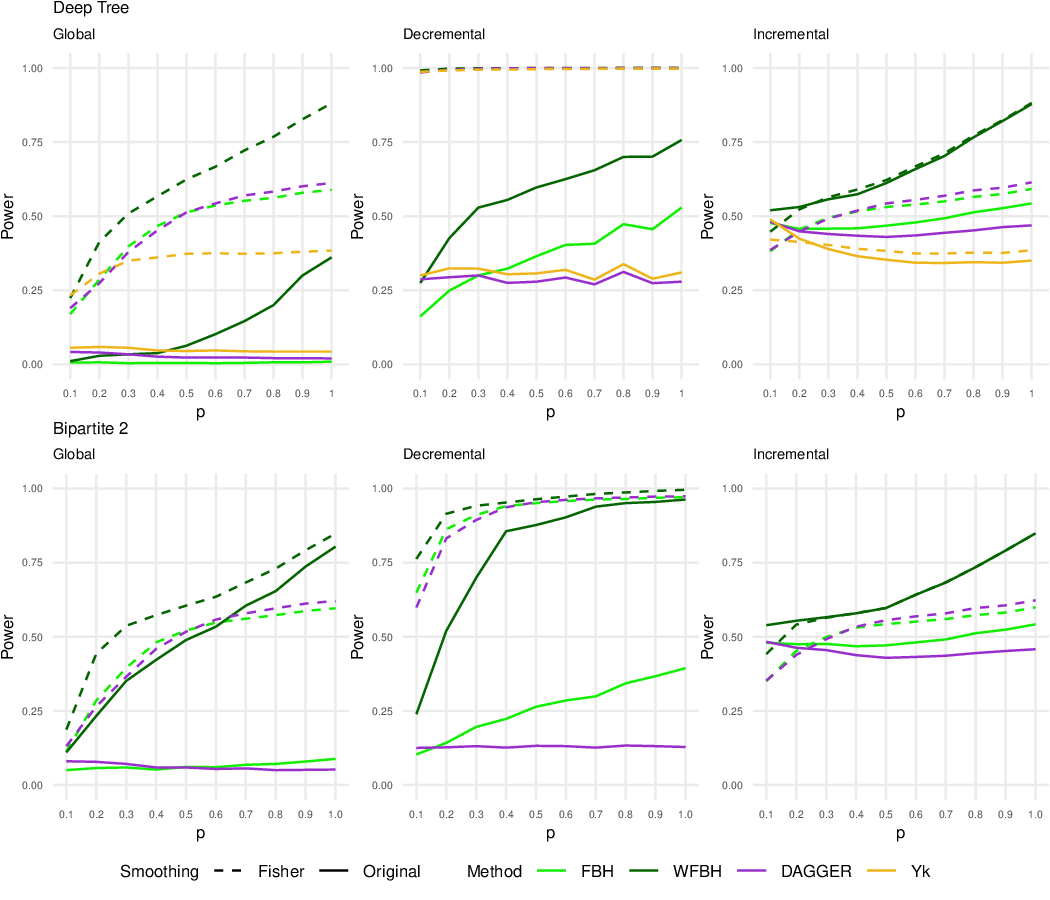}
\caption{Estimated average power, for deep tree and bipartite 2 setups, when the original $p$-values are independent; the methods are applied to the original and Fisher's smoothed $p$-values. All the methods target FDR control at level $q=0.05;$  $\lambda=0.5$ for WFBH. 
}
\label{fig_pow_ind_dpt_bip2}
\end{figure}

Figure \ref{fig_pow_ind_dpt_bip2} shows the average power (estimated over 200 iterations) for the deep tree and the bipartite 2 graph structures, when the WFBH, FBH, DAGGER and Yekutieli's procedures are applied to the original independent $p$-values and their Fisher's smoothed versions. Yekutieli's procedure is applied only to the tree structure, and its level is adjusted to target full-tree FDR control, as described in Section 4.  Power is compared for varying $p$, the proportion of non-null leaves in each graph, across the three above mentioned setups.
In all configurations, WFBH is more powerful than FBH and the other competitors when applied to the original $p$-values. 
All-descendant smoothing using Fisher's method increases the power significantly for all the methods and graphs in the global and decremental setups, and its effect is the strongest in the decremental setup. Smoothing is especially advantageous for the power of WFBH in the deep tree setting. The high power advantage for the decremental setup in natural, as noted in Section 4, because in this setup the combined parent $p$-values for the non-null hypotheses are expected to be smaller than their original $p$-values, especially for high $p,$ due to the stronger signal strengths of their descendant non-null hypotheses.  In the incremental setup, for small $p$, smoothing harms the power of all the methods, for all graphs. Recall that in incremental setup, in contrast to the decremental setup, the signal strengths for the parent non-null hypotheses are stronger than those of their child non-null hypotheses. When $p$ is small, each non-null parent is expected to have a small proportion of non-null children. Due to these facts, in the incremental setup, when $p$ is small, combining the parent $p$-values with their descendant $p$-values  may be harmful, possibly resulting in higher combined $p$-values than the original $p$-values. Note that due to our data generation process, for the same value of $p,$ the overall proportion of non-null hypotheses in each graph can be different, therefore the comparisons of power performances across different graphs for the same value of $p$ are difficult.  

In the following, we demonstrate the simulation results for positively dependent $p$-values, obtained as follows. For every $d\in\{1, \ldots, D\}$ and $i\in\{1, \ldots, |\mathcal{H}_d|\},$  $p_{d,i}=1-\Phi(X_{d,i}),$ where $X_{d,i}$ is generated as follows: 
\begin{equation} \label{data-gen}
X_{d,i} = \mu_{d,i} + (1-\rho)Z_{d,i} + \rho Z_0,
\end{equation}
where $Z_{d,i}, d=1, \ldots, D, \,i=1, \ldots, |\mathcal{H}_d|$ and $Z_0$ are independent standard normal variables, and $\rho\in(0,1)$ determines the constant correlation among the test statistics. The value of $\mu_{d,i}$ is determined according to the global, incremental, and decremental setups, as described in Section 4. In this case, we consider the performance of the methods when applied to original $p$-values and to smoothed $p$-values obtained by all-descendant smoothing using Simes' method (\cite{Simes1986}). Under this setting, the $p$-values satisfy the PRDS property on each one from $\mathcal{H}_0,$ therefore Focused BH and DAGGER have theoretical FDR control guarantees when applied on either original or smoothed $p$-values (according to \cite{filtering} and \cite{DAGGER}, respectively). However,  WFBH and Yekutieli's methods do not have theoretical FDR control guarantees under positive dependence. In simulations with positively dependent $p$-values, we set $\lambda = q = 0.05$ for WFBH,   motivated by the simulation results of \cite{blan2009} for the adaptive BH procedure of \cite{SetS04} (which we refer to as Storey-BH), and the simulation results of \cite{nandi1, nandi2} for their data-adaptive weighted BH procedures. These simulation results show that incorporating Storey's estimator with $\lambda=q$ in the above procedures results in their empirical FDR control under positive dependence. 

\begin{figure}
\centering
  \includegraphics[width=1\linewidth]{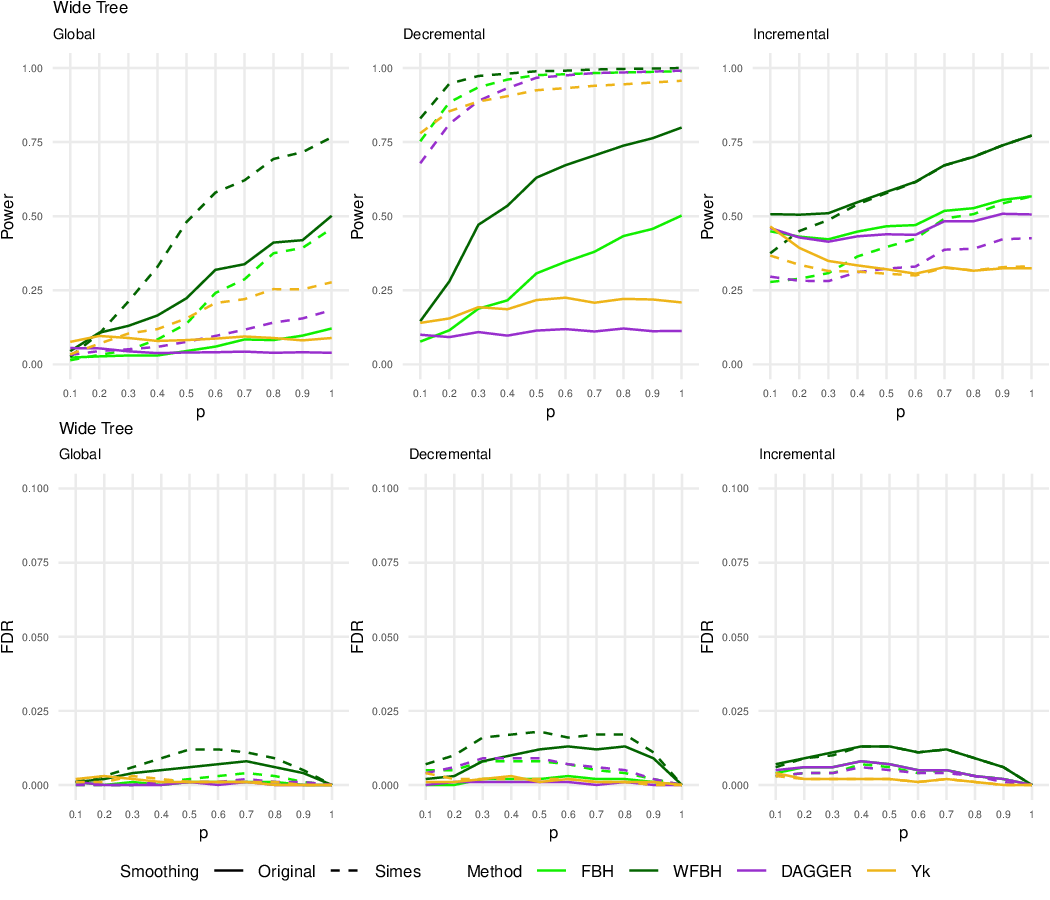}
\caption{Estimated average power and FDR, when the methods are applied to positively dependent original $p$-values (obtained from (\ref{data-gen}) with $\rho=0.2$), and corresponding Simes' smoothed $p$-values, for the wide tree setup. The target FDR level is $q=0.05$ for all the methods; $\lambda=q$ for WFBH.} 
\label{fig_fdrpow_prds_widetree}
\end{figure}

\begin{figure}
\centering
  \includegraphics[width=1\linewidth]{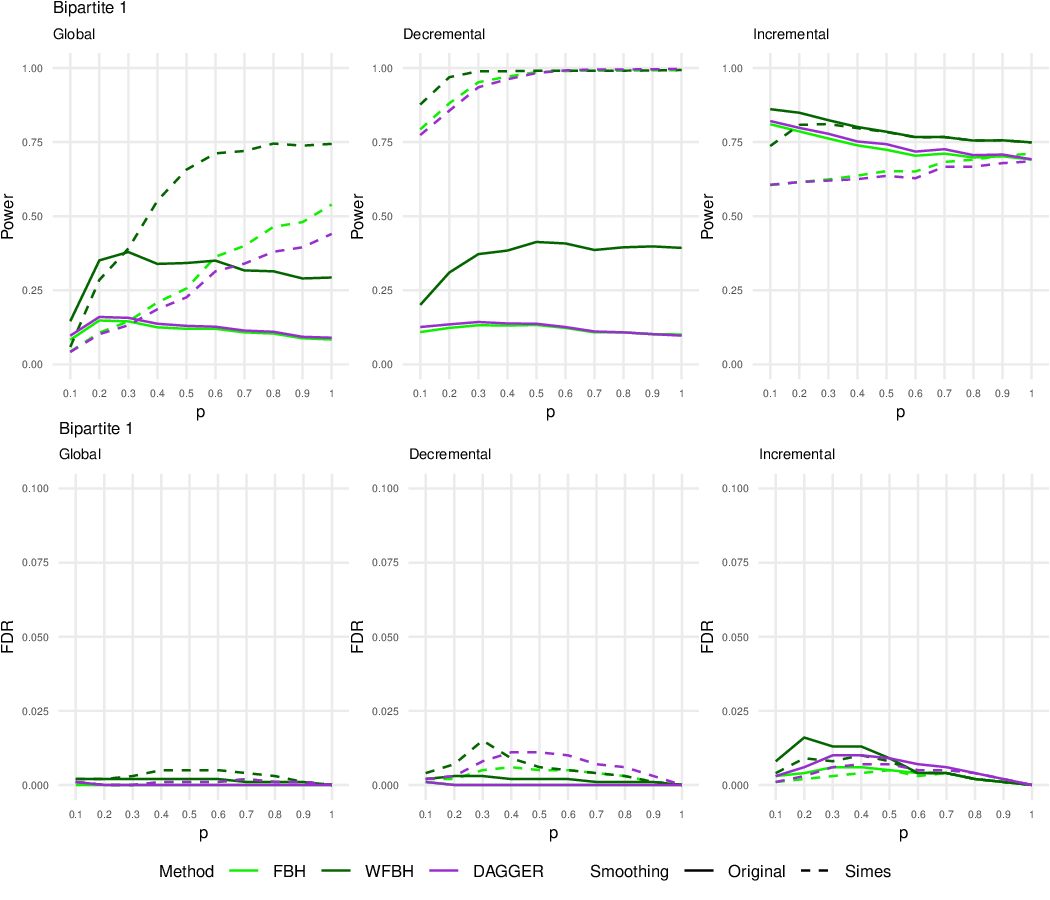}
\caption{Estimated average power and FDR, when the methods are applied to positively dependent original $p$-values (obtained from (\ref{data-gen}) with $\rho=0.2$), and corresponding Simes' smoothed $p$-values, for the bipartite 1 setup. The target FDR level is $q=0.05$ for all the methods; $\lambda=q$ for WFBH.}
\label{fig_fdrpow_prds_bip2}
\end{figure}

Figures \ref{fig_fdrpow_prds_widetree}  and  \ref{fig_fdrpow_prds_bip2} show FDR and power comparisons for varying non-null leaf proportion $p$ and $\rho=0.2$ in (\ref{data-gen}) individually for the four different graph structures - wide tree and bipartite 1, described in Section 4, as well as  deep tree (G1) 
and bipartite 2 (G2). 
The figures provide empirical evidence that all competitive methods control FDR when applied to original $p$-values that are positively dependent and when applied to their respective Simes' smoothed versions.  
The power performance of the methods applied to the original $p$-values is qualitatively similar to that under independence. It is noticeable that the power of WFBH with $\lambda=q$ applied to the original positively dependent $p$-values is lower than the power of WFBH with $\lambda=0.5$ applied to the original independent $p$-values; however, WFBH with $\lambda=q$ is still the most powerful under positive dependence.  
The graphs also show that, across all the considered DAG structures, 
though Simes' smoothing enhances the power for all methods in the global and decremental setups (except for small $p$ in the global setup for bipartite graph 1), it is unable to boost the power of the methods in the incremental setup. Moreover, in this setup, Simes' smoothing is detrimental for the power of all the methods for small values of $p,$ and for DAGGER and FBH, it harms the power for all $p<1.$ This phenomenon is natural in the incremental setup, as discussed above, however, for Simes' smoothing and positively dependent $p$-values it is much more pronounced than for Fisher's smoothing and independent $p$-values. In the global setup, the power of the methods applied to Fisher's smoothed $p$-values in the independent case is higher than when the methods are applied to Simes' smoothed $p$-values in the case of positive dependence; however, in the decremental setup, both types of smoothing lead to similar good power performances for all the methods.   

\begin{figure}
\centering
  \includegraphics[width=1\linewidth]{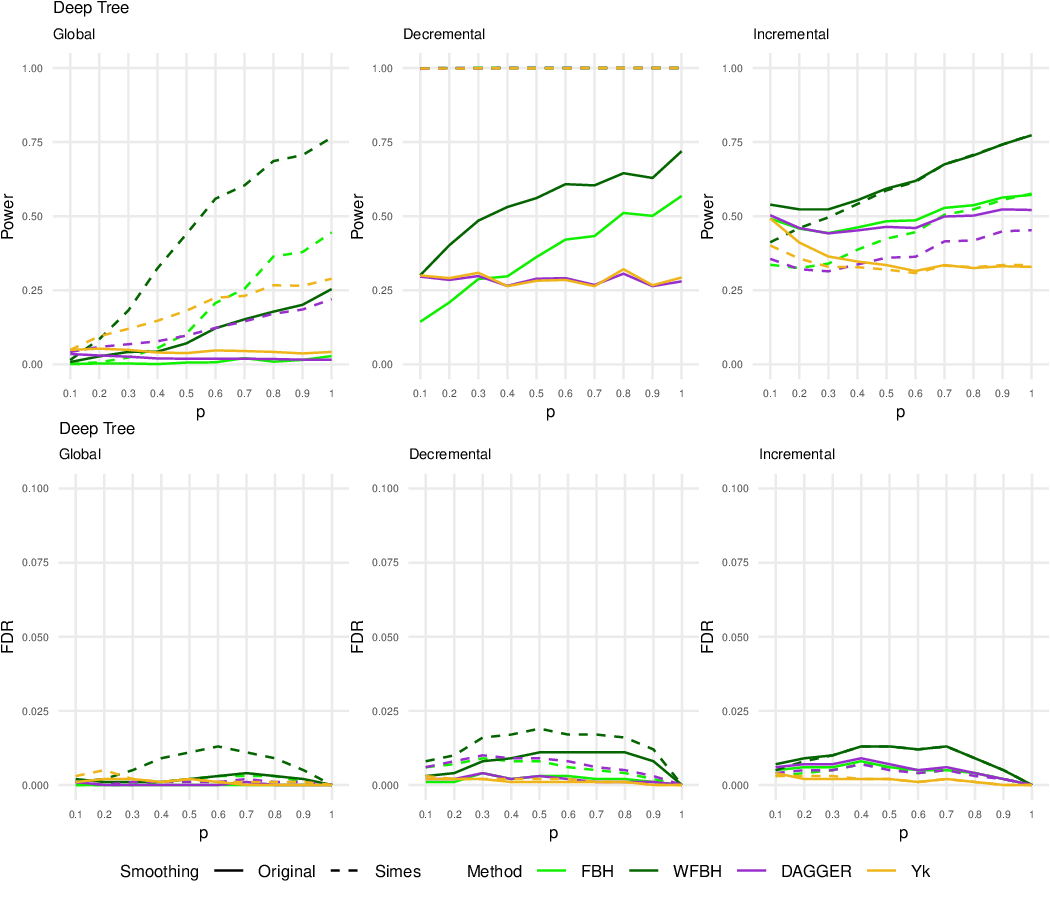}
\caption{Estimated average power and FDR, when the methods are applied to positively dependent original $p$-values (obtained from (\ref{data-gen}) with $\rho=0.2$), and corresponding Simes' smoothed $p$-values, for the deep tree setup. The target FDR level is $q=0.05$ for all the methods; $\lambda=q$ for WFBH.
}
\label{fig_fdrpow_prds_deeptree}
\end{figure}

\begin{figure}
\centering
  \includegraphics[width=1\linewidth]{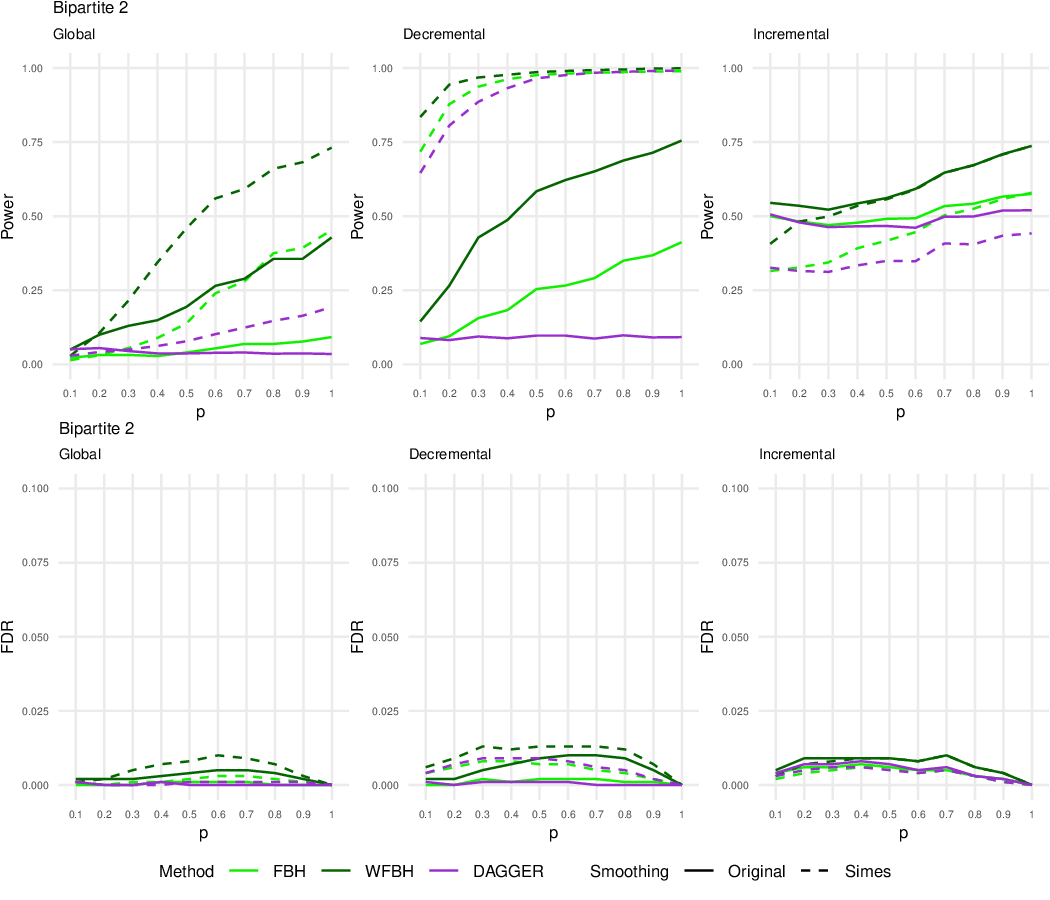}
\caption{Estimated average power and FDR, when the methods are applied to positively dependent original $p$-values (obtained from (\ref{data-gen}) with $\rho=0.2$), and corresponding Simes' smoothed $p$-values, for the bipartite 2 setup. The target FDR level is $q=0.05$ for all the methods; $\lambda=q$ for WFBH.}
\label{fig_fdrpow_prds_bip1}
\end{figure}

\begin{figure}
\centering
  \includegraphics[width=1\linewidth]{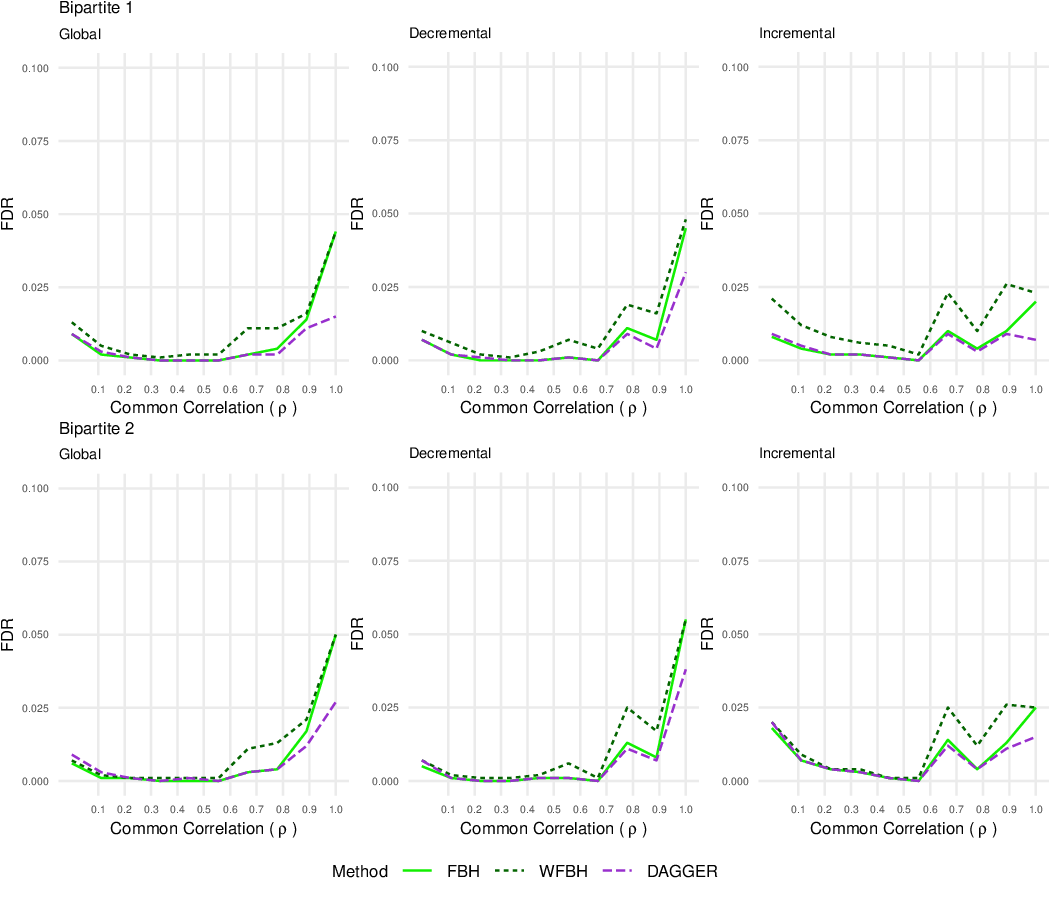}
\caption{Estimated FDR, when the methods are applied to positively dependent original $p$-values for varying strengths of common correlation $\rho$, for bipartite 1 and bipartite 2 setups. The target FDR level is $q=0.05$ for all the methods; $\lambda=q$ for WFBH.}
\label{fig_fdr_rho_bipartite}
\end{figure}

The above figures provide empirical evidence that WFBH with $\lambda=q$ controls FDR under positive dependence with $\rho=0.2$ in (\ref{data-gen}). To address various strengths of correlation among test statistics, we explore the empirical FDR of WFBH with $\lambda=q$ applied to positively dependent statistics for $\rho$ increasing from $0$ to $1$. 
Figure \ref{fig_fdr_rho_bipartite} illustrates that WFBH with $\lambda=q$ appears to control the FDR when applied to positively dependent $p$-values corresponding to all $\rho<1,$ for the two bipartite graph structures. 
Similar results hold for the wide and deep tree structures. 

\section{Further Data Analysis}
\subsection*{Analysis of phylogenetic tree}
In the main manuscript we analyze a tree structure of hypotheses corresponding to the phylum Chlamydiae, which is a subset of the data from the study of \cite{Caporaso2011} on the abundance of various microbes across different environments. Here we consider a different subset of the same data set concerning the phylum  Actinobacteria, which consists of $1631$ microbes. The taxonomic hierarchy within this phylum yields a tree structure of $3261$ nodes at $39$ depths, each representing a null hypothesis. Similarly to the Chlamydiae phylum, the null hypothesis in this case states that each microbial taxonomic classification is uniformly distributed across all environments. To test each null, an F-statistic is used. 

Table \ref{tab1} summarizes the number of discoveries made by each of the FBH, WFBH (both with DAG-structured filter), DAGGER, Storey-BH, BH, and Yekutieli's procedures applied to the original 
p-values. We  set $q=0.05$ as the target FDR level for all the methods, and set $\lambda=q$ in Storey's null-proportion estimator incorporated in the weights of WFBH (obtained according to Assumption 3 with $\mathcal{D}_w=\{1, \ldots, 39\}$), as well as in Storey-BH, because the $p$-values are dependent.  The highest number of discoveries is made by Storey-BH and BH, however, their rejection sets do not have a form of a rooted sub-tree of the original tree, in contrast to the rejection sets of their competitors, which respect the strong heredity principle. 
Comparing the results in Table \ref{tab1} with the analysis of the tree structure corresponding to the Chlamydiae phylum, we observe several similar phenomena: incorporating weights in FBH leads to several additional rejections, and both WFBH and BH make more rejections  than DAGGER. 
However, in contrast to the tree structure corresponding to the Chlamydiae phylum, Yekutieli's procedure for Actinobacteria phylum results in more rejections than FBH and WFBH. This can be explained by the fact that 
FBH and WFBH correct for multiplicity of all the hypotheses in the tree, 
whereas Yekutieli's procedure corrects for multiplicity of hypotheses within each group that is tested (at a reduced level of $q/2$). 
 Given that the Actinobacteria tree is significantly larger than the Chlamydiae tree, and both trees have only two hypotheses within each group, FBH and WFBH 
lead to fewer rejections compared to Yekutieli's procedure in the Actinobacteria tree, as opposed to the Chlamydiae tree. 
\begin{table}[ht]
	\centering
	\begin{tabular}{ccccccc}
		\hline
		& DAGGER & WFBH & FBH & Yekutieli & Storey-BH & BH \\ 
		\hline
		Number of discoveries &  86 & 133 & 123 & 243 & 1231 & 1013 \\ 
		\hline\\
	\end{tabular}
    \caption{Number of discoveries made by the methods applied to original  $p$-values obtained from the data on the Actinobacteria phylum, with target full-tree FDR level $q=0.05,$ and $\lambda=q$ for WFBH and Storey-BH. Yekutieli's procedure is applied at level   $q/2$, according to the adjustment addressed in Section 5. Only $230$ rejections made by BH, and  $259$ rejections made by Storey-BH remain if the DAG-structured filter is applied on their rejection sets.} 
    \label{tab1}
\end{table}

\subsection*{Analysis of GO graph}
\begin{figure}
	\centering
       \begin{minipage}{0.5\textheight}
           \centering
            \includegraphics[width=1\linewidth]{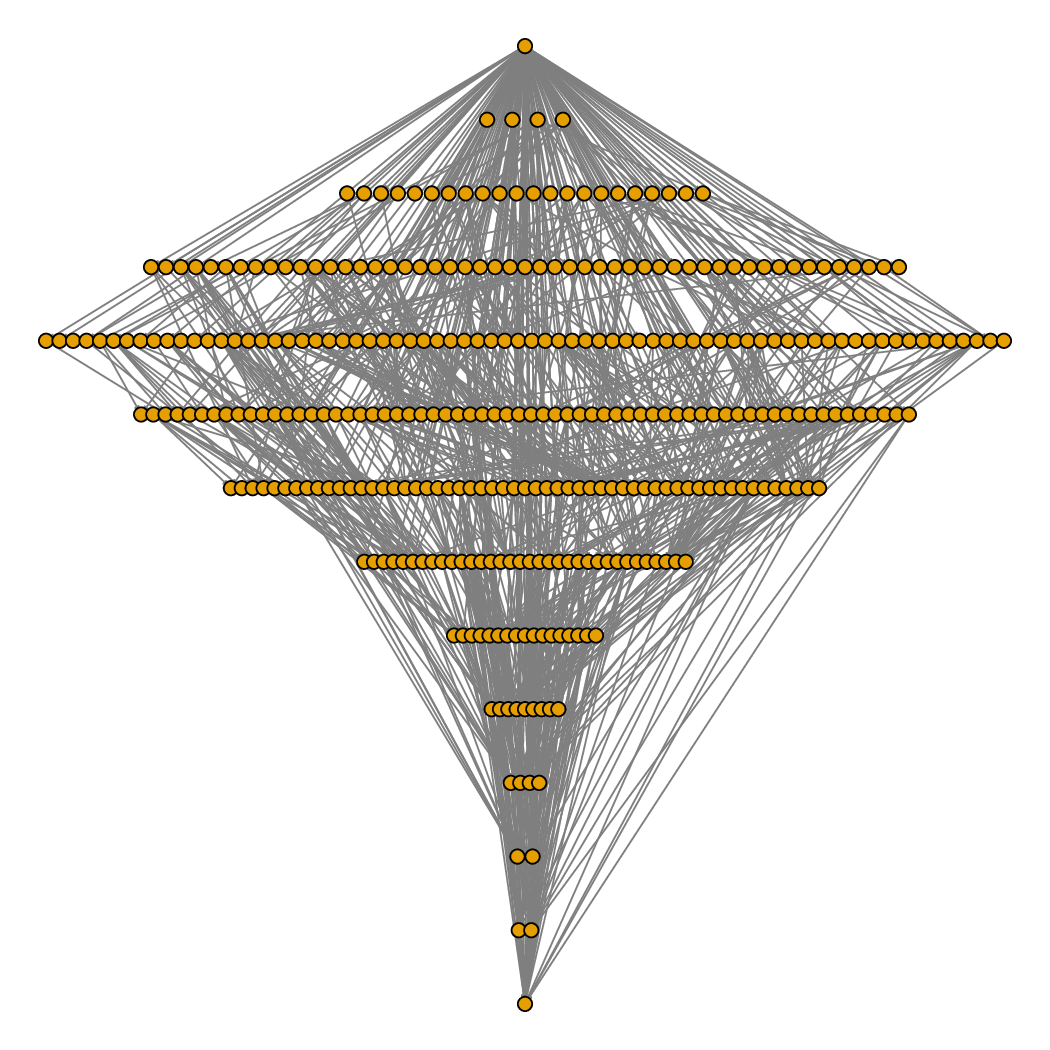}
        \caption{The partial GO graph under study for the Golub dataset.}\label{gographdata}
       \end{minipage}\\
       \begin{minipage}{0.4\textheight}           
            \centering
             \includegraphics[width=1\linewidth]{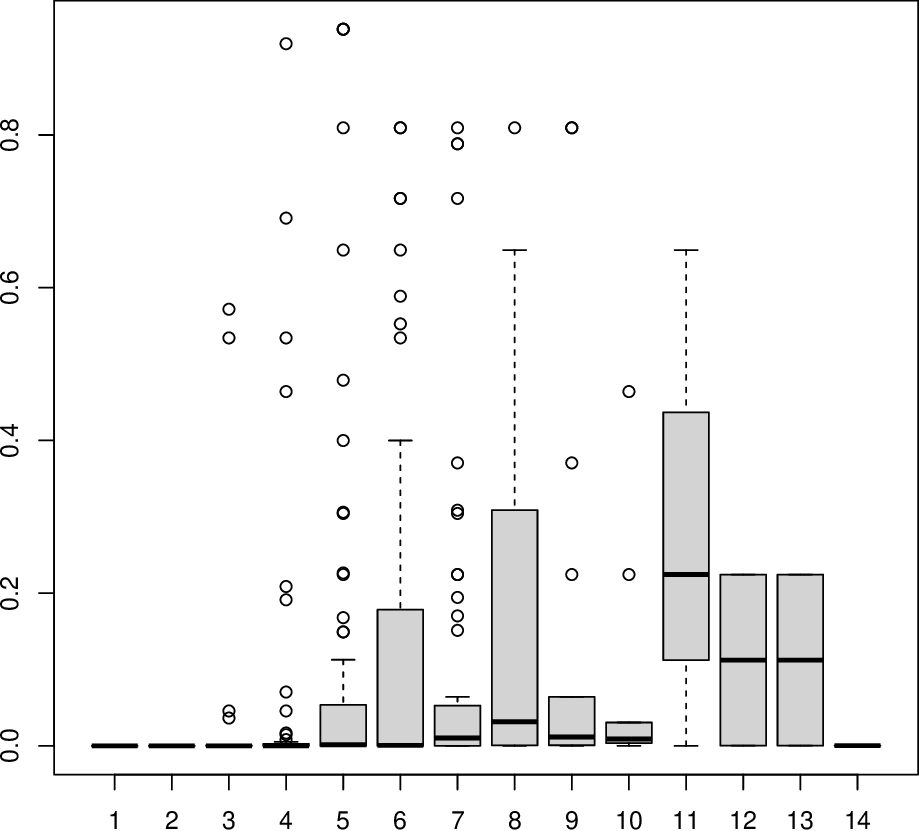}
        \caption{Boxplots showing the spread of $p$-values at all 14 depths of the partial GO graph under consideration.}\label{gographboxplot}
        \end{minipage}		
	\end{figure}

We consider a subset of the Golub dataset \citep{golubEsets, Golub1999-kh}, which contains data on the expression of 7129 genes from a sample of 47 patients with acute lymphoblastic leukemia (ALL) and 25 patients with acute myeloid leukemia (AML). We focus on a part the GO hierarchy, a DAG consisting of 338 nodes, that includes the biological process cell cycle term GO:0007049 and its descendants.  
See Figure \ref{gographdata} for an illustration.  
A similar dataset was analyzed in \cite{DAGGER}. 

	
Each node represents a GO term, and is associated with 
a global null hypothesis, which is 
an intersection of individual null hypotheses corresponding to the genes that share that particular GO term. The global null hypothesis at each node states that none of the individual genes involved are differentially expressed between AML and ALL. 

The tests are performed in R using the GlobalANCOVA package by comparing a full model consisting of all parameters and a reduced model that satisfies the null hypothesis. 
The subsequent $p$-values have complex dependencies, so in this example, we address  methods that have theoretical FDR control guarantees under arbitrary dependence: the BH procedure with a conservative adjustment for handling arbitrary dependence, which we refer to as BY (\cite{BY01}), reshaped FBH (\cite{filtering}), and reshaped DAGGER (\cite{DAGGER}). The latter two methods are conservative variants of FBH and DAGGER, respectively, with an adjustment that incorporates a general reshaping function $\beta,$ defined in (\ref{beta}). We address the variants of reshaped FBH and reshaped DAGGER with $\beta(r)=r/(\sum_{i=1}^m1/i),$ i.e., a reshaping function which is incorporated in the BY procedure. We refer to these variants as FBH.BY and DAGGER.BY, respectively. All three methods are applied to the original $p$-values and smoothed $p$-values, obtained by all-descendant smoothing using Bonferroni's combining method (which outputs the minimum of all the combined $p$-values multiplied by the number of combined $p$-values if the result is smaller than one, and outputs one otherwise). This method gives a valid global null $p$-value under arbitrary dependence among the combined $p$-values, therefore BY, FBH.BY, and DAGGER.BY have theoretical FDR control when applied to Bonferroni's smoothed $p$-values, as well as when applied to the original $p$-values, for arbitrary dependencies among the original $p$-values. For comparative purposes, we also apply the BH, FBH and DAGGER procedures to the original and Bonferroni's smoothed $p$-values.

\begin{figure}
		\centering
\includegraphics[width=0.8\textwidth, keepaspectratio]{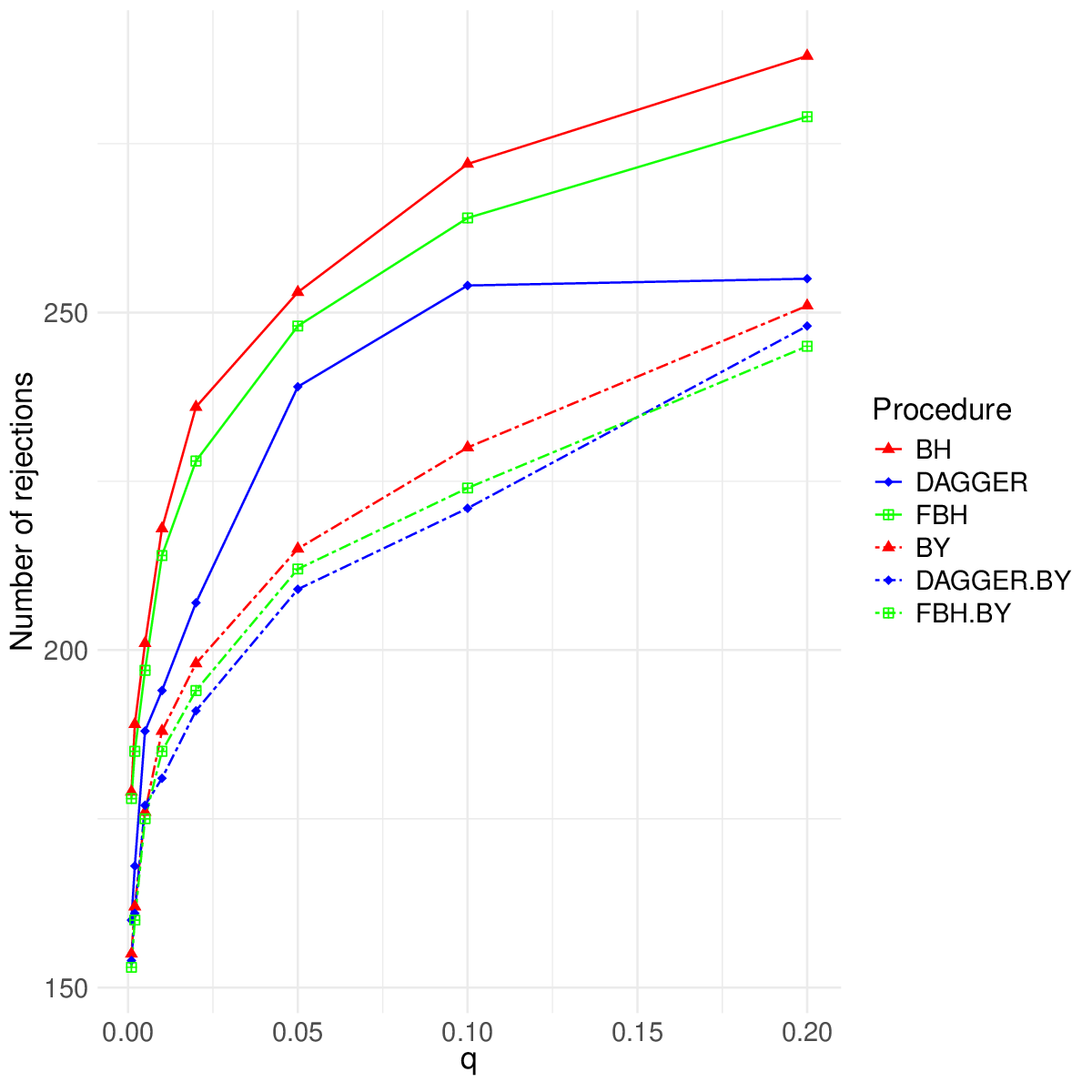}
        \caption{Number of rejections made by the different procedures when applied to original $p$-values, for varying values of target FDR level $q$.}
        \label{GO:original}
	\end{figure}

Figure \ref{gographboxplot} illustrates the spread of $p$-values at each depth of the DAG. Figure \ref{GO:original} shows the number of discoveries made by different methods when applied to the original $p$-values, for varying target FDR level $q.$  Table \ref{tab2} shows the number of discoveries of the methods with target FDR level $q=0.05,$ when applied to the original or Bonferroni's smoothed $p$-values. 
\begin{figure}
		\centering
\includegraphics[width=1\textwidth, keepaspectratio]{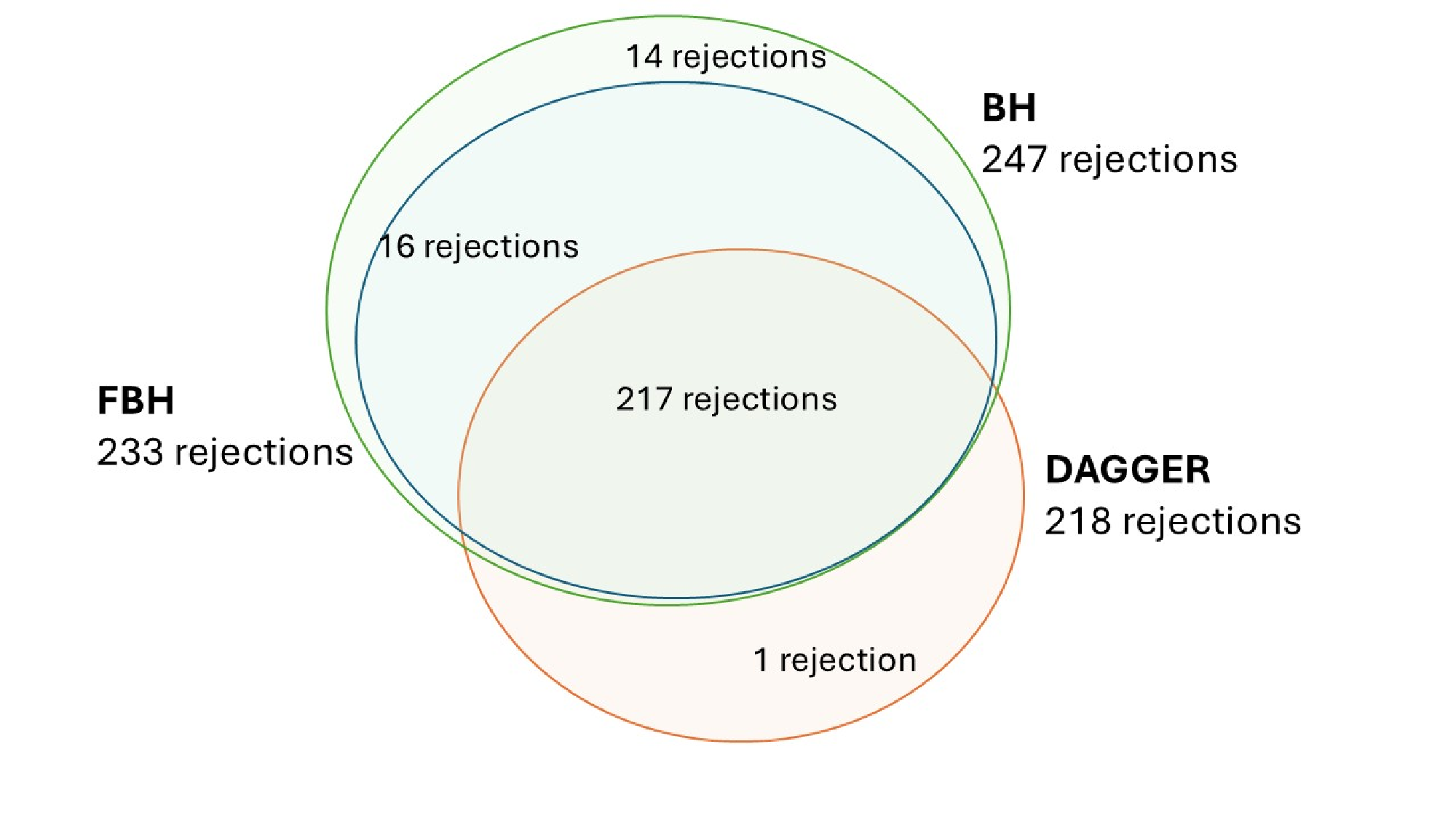}
        \caption{Venn diagram showing the overlapping and unique discoveries made by the BH, FBH and DAGGER procedures when applied to Bonferroni's smoothed $p$-values at FDR level $q=0.05$.}
        \label{gographplots}
	\end{figure}

  \begin{table}
	\centering
	\begin{tabular}{|p{2.2cm}|c|c|c|c|c|c|}
		\hline
       & \multicolumn{2}{c|}{DAGGER} & \multicolumn{2}{c|}{FBH} & \multicolumn{2}{c|}{BH}\\
        \cline{2-7}
     & Original & Reshaped & Original & Reshaped & Original & Reshaped \\
    \hline
		Original &  239 & 209 & 248 & 212 & 253 & 215  \\ 
		Bonferroni & 218 & 196 & 233 & 191 & 247 & 203\\
		\hline
	\end{tabular}
    \caption{Number of discoveries made by BH, FBH, DAGGER, and their reshaped versions, at level $q=0.05,$ when applied to original and Bonferroni's smoothed $p$-values obtained from the data on the GO graph. Reshaped BH is the BY procedure. 
    All the additional discoveries made by BH, beyond those made by FBH, have non-rejected ancestors; the same holds for BY (reshaped BH) and reshaped FBH.
    }
    \label{tab2}
\end{table}
 As expected, reshaped variants of the procedures lead to a smaller number of rejections
than their original versions. Since the BH and BY procedures do not respect the strong heredity principle, in contrast to FBH and DAGGER, they lead to a higher
number of rejections when applied either to the original or to the smoothed $p$-values. 
Figure \ref{gographplots} shows the Venn diagram for the number of rejections of BH, FBH and DAGGER with target FDR level $q=0.05$ when applied to Bonferroni's smoothed $p$-values. All the additional 14 discoveries that are made by BH and are not made by FBH have non-rejected ancestors.
The number of rejections of DAGGER.BY and FBH.BY are similar for
all values of $q$ considered in Figure \ref{GO:original}. According to Table \ref{tab2}, each of the methods that we addressed results
in a lower number of rejections when applied to the Bonferroni's smoothed $p$-values 
than when applied to the original $p$-values. This can be explained by the fact that, as we observe in Figure \ref{gographboxplot}, most of the $p$-values for hypotheses at low depths are small, while many of the $p$-values for hypotheses at higher depths are much higher. Therefore, when we combine the parent $p$-values at low depths with all their descendants using Bonferroni's method, we may obtain combined $p$-values which are higher than the original $p$-values, which is harmful for the power of all the methods. We observed a similar phenomenon in our simulation studies: in the incremental setup, where the signals at low depths are stronger than the signals at higher depths, the power of the methods applied to smoothed $p$-values is often lower than their power when they are applied to the original $p$-values, especially when the proportion of signals is low, the $p$-values are positively dependent, and smoothing is performed using Simes' method. 

\bibliographystyle{plain}  
\bibliography{Mybib}  

\end{document}